\documentclass[%
 reprint,
superscriptaddress,
nofootinbib,
 amsmath,amssymb,
 aps,
onecolumn,
notitlepage,
leqno,
noarxiv,
accepted=2020-06-30]{quantumarticle}

\pdfoutput=1

\usepackage{graphicx}% Include figure files
\usepackage{dcolumn}% Align table columns on decimal point
\usepackage{bm}% bold math
\usepackage{physics}
\usepackage{multirow}
\usepackage{amsthm}
\usepackage{apptools}
\usepackage{chngcntr}
\usepackage{mathtools}
\usepackage{subcaption}
\usepackage{color}
\usepackage{accents}
\usepackage{hhline}
\AtAppendix{\counterwithin{theorem}{section}}
\usepackage[margin=1in]{geometry}
\theoremstyle{definition}
\newtheorem{theorem}{Theorem}
\newtheorem{Theorem}[theorem]{Theorem}

\newtheorem{Definition}[theorem]{Definition}
\makeatletter
\def\blfootnote{\xdef\@thefnmark{}\@footnotetext}
\makeatother

\newcommand{\vect}[1]{\accentset{\rightharpoonup}{#1}}

\begin{document}

\preprint{APS/123-QED}

\title{Generating Fault-Tolerant Cluster States \newline from Crystal Structures}% Force line breaks with \\

\author{Michael Newman}
\email{mgnewman@google.com}
\affiliation{Departments of Electrical and Computer Engineering, Chemistry, and Physics, Duke University, Durham, NC, 27708, USA}
\author{Leonardo Andreta de Castro}
\affiliation{Departments of Electrical and Computer Engineering, Chemistry, and Physics, Duke University, Durham, NC, 27708, USA}
\affiliation{Present address: Q-CTRL Pty Ltd, Sydney, NSW, Australia}
\author{Kenneth R. Brown}
\affiliation{Departments of Electrical and Computer Engineering, Chemistry, and Physics, Duke University, Durham, NC, 27708, USA}

\date{July 13, 2020}% It is always \today, today,
             %  but any date may be explicitly specified
\begin{abstract}
Measurement-based quantum computing (MBQC) is a promising alternative to traditional circuit-based quantum computing predicated on the construction and measurement of cluster states.  Recent work has demonstrated that MBQC provides a more general framework for fault-tolerance that extends beyond foliated quantum error-correcting codes.  We systematically expand on that paradigm, and use combinatorial tiling theory to study and construct new examples of fault-tolerant cluster states derived from crystal structures.  Included among these is a robust self-dual cluster state requiring only degree-$3$ connectivity.  We benchmark several of these cluster states in the presence of circuit-level noise, and find a variety of promising candidates whose performance depends on the specifics of the noise model.  By eschewing the distinction between data and ancilla, this malleable framework lays a foundation for the development of creative and competitive fault-tolerance schemes beyond conventional error-correcting codes.
\end{abstract}

\maketitle

\section{Introduction}

Fault-tolerant quantum computation is possible, provided that error processes are sufficiently weak and uncorrelated \cite{Aliferis:2006, Knill:1996b,Aharonov:1997,terhal2005fault, aliferis2005fault}.  However, quantum systems are inherently delicate, and so it is important maximize the robustness of a fault-tolerance strategy relative to the specific noise patterns a quantum processor experiences \cite{knill2005quantum, Dennis:2002, raussendorf2007fault, wang2003confinement, ohno2004phase, Wang:2009, fowler2009high, fowler2012towards, Suchara:2014, wang2011surface, stephens2013high}.  

Typically, fault-tolerance schemes are founded on constructing and maintaining a quantum error-correcting code \cite{shor1996fault,Shor:1995b,calderbank1996good,steane1996error,steane1997active, Gottesman:1997} on which abstracted logical circuits can be implemented with high fidelity.  However, there is an alternative to circuit-based quantum computation (CBQC) that is built on the adaptive measurement of highly-entangled resource states \cite{raussendorf2003measurement,briegel2009measurement}.  This alternative has been appropriately named measurement-based quantum computation (MBQC), and has blossomed within the framework of fault-tolerance \cite{raussendorf2007fault, raussendorf2007topological, raussendorf2006fault, raussendorf2005long}.

Most proposals for fault-tolerant MBQC are based on a topological approach, whereby quantum circuits are woven into resource states via adaptive measurement choices \cite{raussendorf2007topological, raussendorf2006fault}.  The ambient space on which this topological pattern occurs is commonly referred to as the $3$D cluster state vacuum, and represents the error-resilient stitching that protects the computation.  The essential property of this state is that it can both propagate highly non-local correlations, which carry the logical information being manipulated, while supporting local correlations, which act as consistency checks to catch errors.  This permits the sharing of long-range entanglement in the presence of noise, which is essential for any fault-tolerance scheme \cite{raussendorf2005long}.

Early on, a fundamental connection was observed between $3$D cluster states and surface codes \cite{raussendorf2007fault}, which was later extended to CSS codes \cite{bolt2016foliated, bolt2018decoding} and eventually all stabilizer codes \cite{brown2018universal}.  This connection has been termed foliation, which according to its name, maps a stabilizer code to a cluster state representing the evolution of that code in time.  In particular, each data qubit of the code is replaced by a time-like $1$D cluster state, so that the foliated code maintains the error-correction properties of the base code in a naturally phenomenological error model.

Recently, pioneering work by Nickerson and Bomb{\'i}n studied cluster states that cannot be realized as the foliation of any quantum code \cite{nickerson2018measurement}.  The central insight is that fault-tolerant cluster states need not behave as time-like error-correcting codes to maintain their essential property: the coexistence of long-range logical correlations with local consistency checks.  At the circuit level, this manifests as constantly measuring and reinitializing all of the qubits, so that each plays a role in both propagating logical information and catching errors.  This stands in contrast to a quantum error-correcting code, in which the logical information is held statically in data qubits that remain unmeasured until the end of the computation, while only the ancilla qubits search for errors. Relaxing this division of labor opens up a new design space to explore, and \cite{nickerson2018measurement} employed a splitting technique reminiscent of approaches in network theory \cite{nguyen1994three, parhami2001unified} to construct new cluster states that were resilient to both dephasing and erasure noise \cite{barrett2010fault}.

\begin{figure}[htb!]
\includegraphics[width=\linewidth]{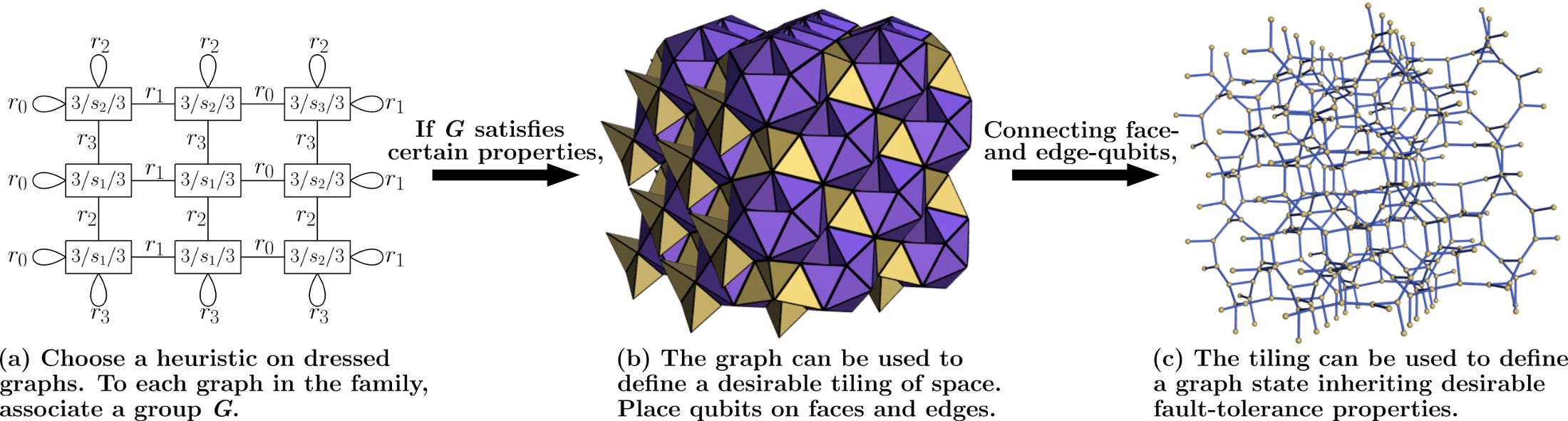}
\caption{Finding a degree-$3$ self-dual fault-tolerant cluster state, which necessarily cannot be realized as the foliation of any code. This is optimally local, as any degree-$2$ chain cannot fill space, and so cannot correct any type of error.  More generally, a cluster state inherits properties affecting fault-tolerance from the chosen graph heuristic.}
\label{main}
\end{figure}

In this work, we use tools from combinatorial tiling theory \cite{dress1987presentations} to study and construct a broad range of robust fault-tolerant cluster states.  The central idea is to use a discrete analogue of orbifold covers to turn the geometric problem described in \cite{nickerson2018measurement} into an algebraic problem on groups \cite{delgado2001recognition}. Although we focus on cluster states that are built from periodic crystal structures, we also discuss how these techniques can apply more generally.  To illustrate the power of this algebraic approach, we construct a self-dual fault-tolerant cluster state that is optimally local, requiring only a $3$-regular underlying graph state.  See Figure \ref{main} for a high-level description of the main idea.

In addition, we benchmark these cluster states in the presence of circuit-level noise with varying relative parameters.  We do so using a minimum-weight perfect matching decoder, which is less efficient than the union-find decoder \cite{delfosse2017almost, huang2020fault} utilized in \cite{nickerson2018measurement}, but which exhibits higher performance and relates to the zero-temperature phase transition of associated statistical mechanical models with quenched disorder \cite{wang2003confinement, chubb2018statistical}.
We find a variety of competitive cluster states whose robustness depends on the specifics of the noise parameters.

The paper is organized as follows.  In Section \ref{background}, we discuss the general formulation of fault-tolerant cluster states in terms of chain complexes, and introduce their relevant circuit-level noise models.  In Section \ref{tilings}, we review the foundations of combinatorial tiling theory, and use it to identify interesting fault-tolerant cluster states based on crystal structures.  In Section \ref{cs_numerics}, we provide numerics on examples drawn from an extensive classification of self-dual tilings as well as our numerical search to both benchmark promising candidates, and to probe the effects of circuit-level noise. Finally, we discuss avenues for future development in Section \ref{final_conclusions}.

\section{Fault-Tolerant Cluster States}\label{background}

\subsection{CSS codes as chain complexes}

To understand fault-tolerant cluster states, it is instructive to first review quantum CSS codes.  In particular, there is a well-established interpretation of CSS codes as chain complexes \cite{bombin2007homological,kitaev2003fault,freedman2002z2}.  Throughout, we implicitly work with coefficients in $\mathbb{Z}_2$, reflecting the use of qubits.

\begin{Definition}
A \emph{chain complex} $\mathcal{C}$ is a sequence of vector spaces $\{C_i\}$ and linear maps between them $\{\partial_i\}$ called boundaries: $$0 \xrightarrow{} C_\ell \xrightarrow{\partial_\ell} C_{\ell-1}\xrightarrow{} \ldots \xrightarrow{}C_1 \xrightarrow{\partial_1}  C_0 \xrightarrow{} 0$$ satisfying $\partial_{i-1}\circ\partial_{i} = 0$.  Every such complex has an associated \emph{cochain complex} $\mathcal{C}^*$ comprised of vector spaces $\{C^i \coloneqq \text{Hom}(C_i,\mathbb{Z}_2)\}$ and linear maps between them $\{\partial^i(\cdot) \coloneqq (\cdot) \circ \partial_i\}$ called coboundaries: $$0 \xleftarrow{} C^\ell \xleftarrow{\partial^\ell} C^{\ell-1}\xleftarrow{} \ldots \xleftarrow{}C^1 \xleftarrow{\partial^1}  C^0 \xleftarrow{} 0$$ satisfying $\partial^{i+1} \circ \partial^{i} = 0.$  We sometimes refer to $\mathcal{C}^*$ as the \emph{dual} of $\mathcal{C}$, and we call $\ell$ the \emph{length} of the chain complex.
\end{Definition}

  One important characteristic of a chain complex relevant to the associated quantum code is its homology. This measures the failure of each sequence of boundary maps to be exact.

\begin{Definition}
The \emph{homology groups} of the chain complex $\mathcal{C}$ are the quotient spaces $H_k(\mathcal{C}) \coloneqq \ker(\partial_k)/\text{im}(\partial_{k+1})$. The \emph{cohomology groups} of the chain complex $\mathcal{C}$ are the quotient spaces $H^k(\mathcal{C}) \coloneqq \ker(\partial^{k+1})/\text{im}(\partial^k)$.  Representatives of elements in these quotient spaces are referred to as \emph{cycles} and \emph{cocycles}, respectively.
\end{Definition}

It turns out that chain complexes and their homology provide a natural language for describing and building CSS codes.

\begin{Definition} \label{correspondence}
An $[[n,k]]$ \emph{CSS code} can be identified with a length-$2$ chain complex $\mathcal{C}$: $$C_2 \xrightarrow{\partial_2} C_1 \xrightarrow{\partial_1} C_0$$ where $n = \dim(C_1)$ and $k = \dim(H_1(\mathcal{C}))$.
\end{Definition}

In particular, we have the following correspondence:
\[
\begin{matrix}
C_2 & \xrightarrow{\hspace{.1cm} \partial_2 \hspace{.1cm}} & C_1 & \xrightarrow{\hspace{.1cm} \partial_1 \hspace{.1cm}} & C_0,\\
\updownarrow & & \updownarrow & & \updownarrow\\
\text{\{$Z$-stabilizers\}} & & \text{\{$Z$-errors\}} & & \text{\{$X$-syndromes\}} \\ \\ \\
C^2 & \xleftarrow{\hspace{.1cm} \partial^2 \hspace{.1cm}} & C^1 & \xleftarrow{\hspace{.1cm} \partial^1 \hspace{.1cm}} & C^0.\\
\updownarrow & & \updownarrow & & \updownarrow\\
\text{\{$Z$-syndromes\}} & & \text{\{$X$-errors\}} & & \text{\{$X$-stabilizers\}}
\end{matrix}
\]
The boundary condition $\partial_1 \circ \partial_2 = 0$ ensures the commutativity of the $X$- and $Z$-type stabilizers.  By construction, the class of $Z$-type ($X$-type) logical operator representatives can be identified with nontrivial cycles (cocycles) in $H_1(\mathcal{C})$ ($H^1(\mathcal{C})$).

Logical qubits can be identified with pairs of $X$- and $Z$-type logical operators $\{\overline{X}_i, \overline{Z}_i\}$, and so it should be the case that $\dim(H_1(\mathcal{C})) = \dim(H^1(\mathcal{C}))$.  Furthermore, these logical operators should satisfy canonical anti-commutativity relations $\overline{X}_i\overline{Z}_j = (-1)^{\delta_{ij}}\overline{Z}_j\overline{X}_i$.  This correspondence manifests as a structure theorem on chain complexes \cite{Hatcher:478079}.

\begin{Theorem}[Universal Coefficient Theorem] \label{UCT}
There is a well-defined isomorphism
\[
\begin{matrix}
\phi: & H^i(\mathcal{C}) &\longrightarrow & \text{Hom}(H_i(\mathcal{C}),\mathbb{Z}_2)
\end{matrix}
\] 
given by $\phi([f])([x]) = f(x)$.
\end{Theorem}
Thus, for any nontrivial cycle $x \in C_1$ representing a logical $Z$-type operator, there is a nontrivial cocycle $f \in C^1$ representing a logical $X$-type operator that intersects $x$ with odd parity when identified with its dual in $C_1$.  Furthermore, this element will intersect any equivalence class of cycles $[y] \neq [x]$ with even parity.  Consequently, the factorization of the logical subspace into logical qubit subsystems is captured by the structure of the chain complex.

\begin{figure}[htb!]
\centering
\includegraphics[width=.95\linewidth]{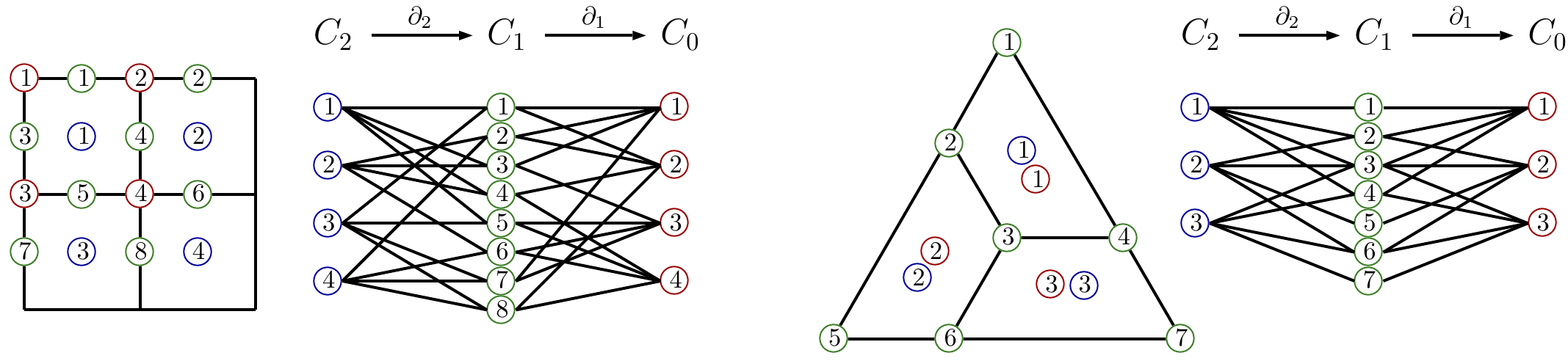}
\caption{A toric code (left) and color code (right) regarded as chain complexes.  For the toric code, the boundary maps between chain complexes are given by the geometric boundary maps in the periodic lattice.  For the color code, the boundary maps between chain complexes are given by a pair of abstract Tanner graphs, which are not defined by geometric boundary maps.}
\label{examples}
\end{figure}

This perspective has been extremely profitable in several ways, including the construction of codes with improved parameters \cite{kitaev2003fault, freedman2002z2, bravyi2014homological, hastings2016quantum, londe2017golden, audoux2015tensor}, the study of single-shot error-correction \cite{campbell2019theory, bombin2015single}, and the design of fault-tolerant gates \cite{jochym2019fault, krishna2019fault}.  We detail two foundational examples of this correspondence in Figure \ref{examples}.

In the language of chain complexes, error-correction in the code capacity model can be rephrased as the following problem:
\begin{align} \label{problem1}
\begin{split}
\text{given $\left(\partial_1(\vect{z}),\partial^2(\vect{x})\right) \in C_0 \oplus C^2$}&, \text{ determine $\left(\vect{c}_z, \vect{c}_x\right) \in C_1 \oplus C^1$}\\ \text{so that $\left([\vect{z} + \vect{c}_z], [\vect{x} + \vect{c}_x]\right)$ } & \text{$= (0,0) \in H_1(\mathcal{C}) \oplus H^1(\mathcal{C})$.} 
\end{split}
\end{align}
Colloquially, this reads: given $X$- and $Z$-type syndromes, determine corrections that, taken together with the errors, apply the logical identity.  For many error models, this problem can be treated independently in each coordinate without significant loss in performance \cite{bombin2012strong,Dennis:2002}.  In the presence of depolarizing noise, it is then optimal to choose code families for which these two problems are symmetric, as the threshold will be determined by the worse of the two.  These include families like the toric and color codes, while asymmetric families like Shor's code \cite{Shor:1995b} will be bottlenecked by one type of error \cite{li20192d}.  

The chain complex formulation of error-correction can be extended to phenomenological noise as well.  In particular, recent work has studied redundant syndromes checks in the context of length-$4$ chain complexes in order to characterize single-shot error-correction \cite{campbell2019theory, bombin2015single}.  It is then apt that fault-tolerant cluster states, which eschew the distinction between data and ancilla, meet in the middle between length-$2$ and length-$4$ chain complexes.

\subsection{Fault-tolerant cluster states as chain complexes}

Having described CSS codes in the chain complex framework, we turn to describing fault-tolerant cluster states.  Despite their similarities, fault-tolerant cluster states afford an extra degree of freedom that is useful in the construction of robust quantum processes \cite{nickerson2018measurement}.  As with most measurement-based schemes, our fault-tolerant cluster states will be built from a particular class of stabilizer states associated to graphs.

\begin{Definition}
For a graph $G = (V,E)$, the associated \emph{graph state} $\ket{G}$ is the unique state stabilized by the set of operators $$\{X_v\prod\limits_{(v,w) \in E} Z_w: v \in V\}.$$ Informally, \emph{cluster states} refer to particular families of graph states that are used as a resource for MBQC.
\end{Definition}

Defining a complete fault-tolerant MBQC scheme involves several considerations, including details describing the implementation of a universal gate set.  For existing proposals, this information requires the specification of additional structures, such as input/output surfaces and their relation to defects \cite{raussendorf2007topological, brown2018universal}.  

In this work, we focus on the error-resilience of the vacuum.  Consequently, we consider a distilled problem: the preservation of long-range correlations in the presence of errors.  As was noted in \cite{nickerson2018measurement}, the fault-tolerance schemes we discuss can be readily adapted to universal fault-tolerant MBQC using the topological techniques in \cite{raussendorf2007topological}.  However, this simplified perspective allows for a concise description of error-resilient vacuums.

\begin{Definition}
A \emph{fault-tolerant cluster state} can be identified with a length-$3$ chain complex $\mathcal{C}$: $$C_3 \xrightarrow{\partial_3} C_2 \xrightarrow{\partial_2} C_1 \xrightarrow{\partial_1} C_0.$$ The underlying graph state is given by the bipartite graph $(C_2,C_1)$ with biadjacency matrix $\partial_2$.  We refer to qubits indexed by $C_2$ as \emph{dual}, and qubits indexed by $C_1$ as \emph{primal}.
\end{Definition}
We emphasize that the underlying graph state is entirely determined by the map $\partial_2$, which encodes a single bipartite graph; the maps $\partial_3$ and $\partial_1$ serve to distinguish the correlations used for error-correction. 

There is an intrinsic asymmetry in cluster states, indicated by the presence of $X$-type symmetries and absence of $Z$-type symmetries, introduced by the asymmetric definition of a graph state.  This asymmetry also manifests in the construction and processing of a fault-tolerant cluster state.  Following the prescription in \cite{raussendorf2007topological}, for a cluster state on $n$ qubits with underlying graph state $G = (V,E)$, this proceeds in three steps.
\begin{itemize}
\itemsep0em
    \item[$(i)$] Prepare $\ket{+}^{\otimes n}$.
    \item[$(ii)$] For each $(v,w) \in E$, apply CZ$_{v,w}$ in some specified order.
    \item[$(iii)$] Measure each qubit in the $X$-basis.
\end{itemize}
In particular, we have the following correspondence:
\[
\resizebox{\linewidth}{!}{$
\begin{matrix}
C_3 & \xrightarrow{\hspace{.1cm} \partial_3 \hspace{.1cm}} & C_2 & \xrightarrow{\hspace{.1cm} \partial_2 \hspace{.1cm}} & C_1 & \xrightarrow{\hspace{.1cm} \partial_1 \hspace{.1cm}} & C_0,\\
\updownarrow & & \updownarrow & & \updownarrow & & \updownarrow\\
\text{\{dual $X$-stabilizers\}} & & \text{\{dual $X$-symmetries\}} & & \text{\{primal $Z$-errors\}} & & \text{\{primal syndromes\}} \\ \\ \\
C^3 & \xleftarrow{\hspace{.1cm} \partial^3 \hspace{.1cm}} & C^2 & \xleftarrow{\hspace{.1cm} \partial^2 \hspace{.1cm}} & C^1 & \xleftarrow{\hspace{.1cm} \partial^1 \hspace{.1cm}} & C^0.\\
\updownarrow & & \updownarrow & & \updownarrow & & \updownarrow\\
\text{\{dual syndromes\}} & & \text{\{dual $Z$-errors\}} & & \text{\{primal $X$-symmetries\}} & & \text{\{primal $X$-stabilizers\}}
\end{matrix}$}\vspace{.25cm}
\]
For any $\vect{v} \in C_2$ indexing dual qubits, there is a corresponding symmetry of the form $$\prod\limits_{i: \vect{v}_i = 1} X_i \prod\limits_{j: \partial_2(\vect{v})_j = 1} Z_j,$$ where the $X_i$ are supported on dual qubits and the $Z_j$ are supported on primal qubits.  The reverse relation holds for $\vect{v} \in C^1$ in the cochain complex.  Thus, when we identify $C_2$ ($C^1$) with dual (primal) $X$-symmetries, we mean symmetries whose restriction to dual (primal) qubits is of $X$-type.

Just as the structure of a length-$2$ chain complex can be interpreted naturally in terms of the properties of a CSS code, so too can the structure of a length-$3$ chain complex be interpreted naturally in terms of the properties of a fault-tolerant cluster state.\footnote{Note that primal logical symmetries and dual logical symmetries correspond, respectively, to dual correlation surfaces and primal correlation surfaces as described in \cite{raussendorf2007topological}.  For a geometrically defined cluster state, a primal (dual) logical symmetry can be visualized as a closed surface in the dual (primal) lattice.}  Note that this chain complex is also related to the ungauging complex introduced for quantum codes, and more specifically for $3$D surface codes when the cluster state is built from a tiling \cite{kubica2018ungauging}.  In the table below, the left-hand side represents either an object associated to or property of the chain complex. The right-hand side represents how that object or property plays a role in the protected computation within the MBQC model.

\begin{table}[htb!]
\centering
\begin{tabular}{|c|l|}
\hline
\textbf{Chain Complex}            & \multicolumn{1}{c|}{\textbf{Fault-tolerant Cluster State}}        \\ \hline
$\partial_2 \circ \partial_3 = 0$ & Dual $X$-stabilizers are accessible via $X$-measurements.         \\ \hline
$H_2(\mathcal{C})$                & Indexes equivalence classes of dual logical symmetries.           \\ \hline
$\partial_1 \circ \partial_2 = 0$ & Primal $Z$-errors equivalent to dual $X$-errors are undetectable. \\ \hline
$H_1(\mathcal{C})$                & Indexes equivalence classes of primal logical $Z$-errors.                 \\ \hline
$\partial^2 \circ \partial^1 = 0$ & Primal $X$-stabilizers are accessible via $X$-measurements.       \\ \hline
$H^1(\mathcal{C})$                & Indexes equivalence classes of primal logical symmetries.         \\ \hline
$\partial^3 \circ \partial^2=0$   & Dual $Z$-errors equivalent to primal $X$-errors are undetectable. \\ \hline
$H^2(\mathcal{C})$                & Indexes equivalence classes of dual logical $Z$-errors.                   \\ \hline
\end{tabular}
\end{table}
Equivalence classes of dual (primal) logical symmetries are defined up to dual (primal) stabilizer equivalence, while equivalence classes of undetectable dual (primal) $Z$-errors are defined up to primal (dual) $X$-error equivalence.  Such errors are trivial, as they commute with our $X$-measurements. Theorem \ref{UCT} ensures that there are canonical isomorphisms
\begin{align*}
    H^2(\mathcal{C}) &\leftrightarrow H_2(\mathcal{C}) \\
    H^1(\mathcal{C}) &\leftrightarrow H_1(\mathcal{C})
\end{align*}
so that for every logical symmetry, there is a corresponding undetectable error that negates it.  

One subtlety is that, without specifying input/output surfaces, there is no explicit correspondence between primal and dual symmetries emulating a logical qubit \cite{raussendorf2005long, brown2018universal}.  This correspondence is explicit for foliated codes, as measuring the bulk of the lattice generates an encoded long-range Bell pair between these surfaces \cite{bolt2016foliated}.  We focus on fault-tolerant cluster states realized by periodic cellulations of $3$-space, for which the notions of input/output surfaces, defects, and topologically protected gates can be introduced directly \cite{raussendorf2007topological,nickerson2018measurement}.  However, fault-tolerant cluster state families need not be geometrically defined, as the chain complex structure need not stem from cellulations of a manifold.  Foliations of more general quantum low-density parity check code families are examples of such states, and fault-tolerant cluster states can be more general still.

Having established the parallels between length-$3$ chain complexes and fault-tolerant cluster states, error-correction in the fault-tolerant MBQC model can be rephrased as the following problem:
\begin{align}\label{problem2}
\begin{split}
\text{given $\left(\partial_1(\vect{z}_p),\partial^3(\vect{z}_d)\right) \in C_0 \oplus C^3$}&, \text{ determine $\left(\vect{c}_p, \vect{c}_d\right) \in C_1 \oplus C^2$}\\ \text{so that $\left([\vect{z}_p + \vect{c}_p], [\vect{z}_d + \vect{c}_d]\right)$ } & \text{$= (0,0) \in H_1(\mathcal{C}) \oplus H^2(\mathcal{C})$.} 
\end{split}
\end{align}
Once again, this can be read colloquially as: given primal and dual syndromes, determine primal and dual corrections that, taken together with the errors, preserve the primal and dual logical symmetries.  Mirroring CSS codes, this problem can be treated independently in each coordinate, and so it is often optimal to choose a fault-tolerant cluster state for which these two problems are symmetric.  In particular, both primal and dual symmetries are required to perform universal fault-tolerant computation \cite{raussendorf2007topological}.

\begin{figure}[htb!]
\centering
    \begin{subfigure}[b]{0.4\textwidth}
        \centering
        \includegraphics[width=\linewidth]{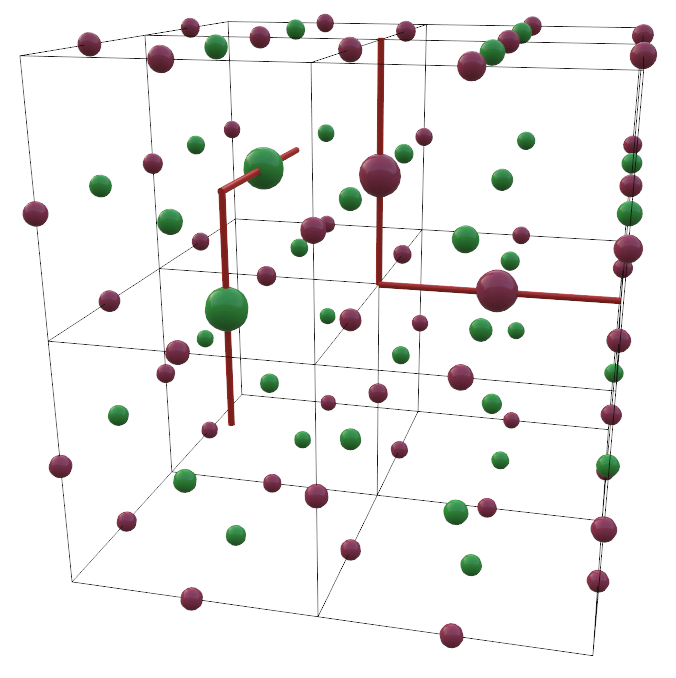}
        \caption{The $3$D cluster state.}
    \end{subfigure}%
    \hspace{2.5cm}
    \begin{subfigure}[b]{0.4\textwidth}
        \centering
        \includegraphics[width=\linewidth]{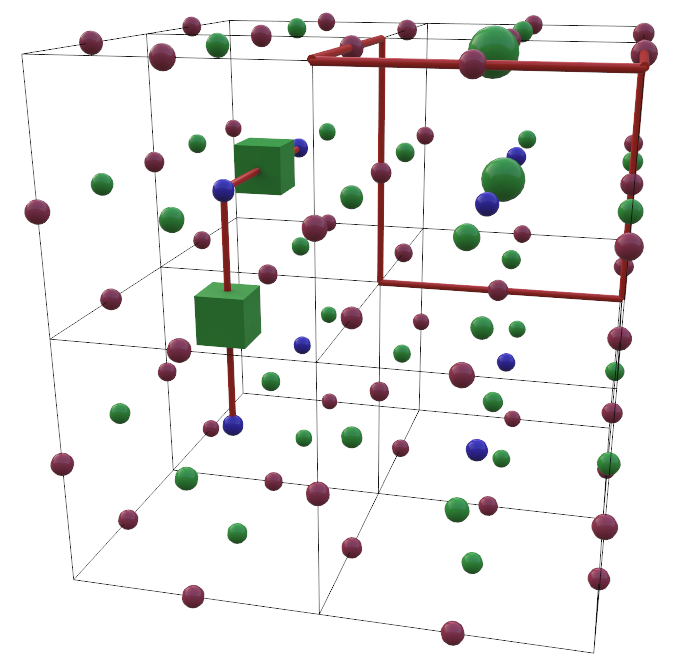}
        \caption{The $3$D surface code.}
    \end{subfigure}
    \caption{A comparison of error-correction in the superficially similar (a) $3$D cluster state and (b) $3$D surface code on a cubic lattice.  $Z$-errors ($X$-errors) are denoted by enlarged circles (cubes), and their syndromes are represented as red lines. (a) In the $3$D cluster state, only $Z$-errors are damaging, as $X$-errors commute with the $X$-measurements.  Each qubit is measured, and syndromes are processed as correlations between outcomes.  $Z$-errors on dual (primal) qubits are detected by correlations about cubes (vertices).  (b) In the $3$D surface code, there are code qubits (green), $Z$-ancilla (blue), and $X$-ancilla (red).  $X$-errors ($Z$-errors) are caught by $Z$-ancilla ($X$-ancilla), producing point-like (string-like) syndromes.  However, measurements themselves can fail, which is captured in (a) but not (b).  In particular, vertices play a role in (a) but not (b), highlighting the difference between the underlying length-$3$ and length-$2$ chain complexes.}
\label{sc_vs_cs}
\end{figure}

Although the problems of error-correction in CSS codes and error-correction in fault-tolerant cluster states appear similar, comparing problems (\ref{problem1}) and (\ref{problem2}), there are key differences between them (see Figure \ref{sc_vs_cs}).  There is a mapping between these two problems that associates to any stabilizer code a fault-tolerant cluster state called the \emph{foliation} of that code.  The key insight made in \cite{nickerson2018measurement} is that this mapping is too restrictive; there is an extra degree of freedom in problem (\ref{problem2}) that can be exploited to construct more robust fault-tolerant protocols.  To highlight this difference, we review foliation in the case of CSS codes, which has also been extended to all stabilizer codes \cite{brown2018universal}.

\subsection{Foliated quantum codes}

Foliated quantum codes are not codes per se, but rather fault-tolerant cluster states that are closely related to codes.  In particular, foliation is a map that takes any CSS (or more generally stabilizer) code and constructs a corresponding fault-tolerant cluster state that emulates that code evolving in time for some number of time steps \cite{bolt2016foliated}.  The most famous example is the $3$D cluster state on a cubic lattice, which can be seen as a foliated $2$D surface code.

\begin{figure}[htb!]
\includegraphics[width=\linewidth]{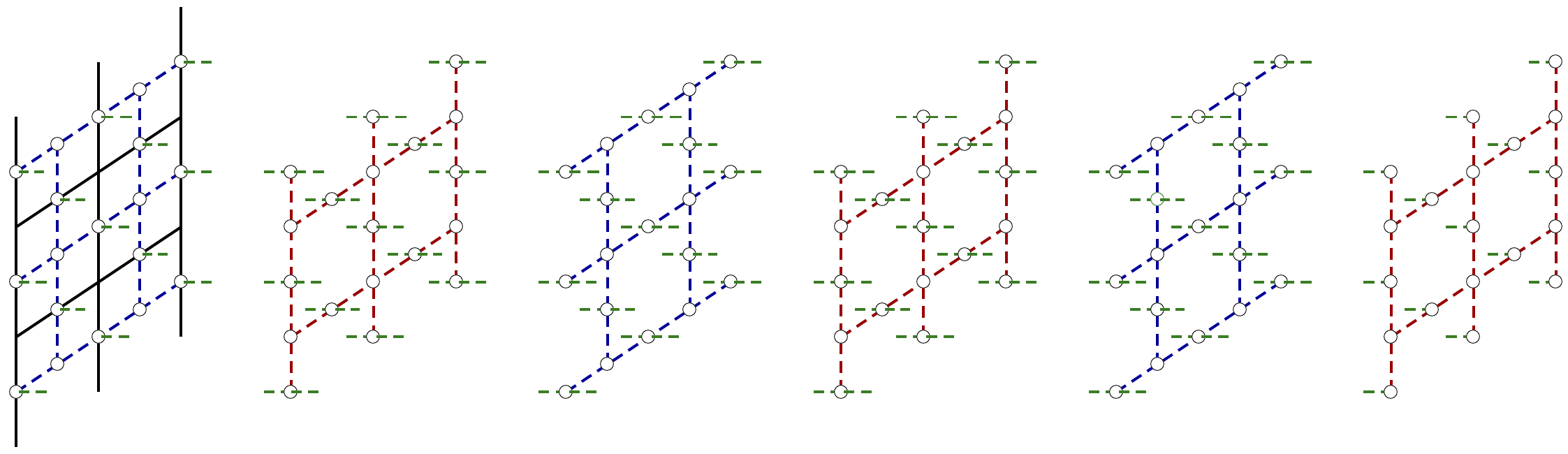}
\caption{A $3$D cluster state regarded as a foliated surface code.  Dotted green lines (abbreviated for clarity) represent time-like $1$D cluster states, dotted blue lines represent $Z$-type ancilla interactions, and dotted red lines represent $X$-type ancilla interactions.  The time-like $1$D cluster states can be regarded as Hadamard clocks, with $Z$- and $X$-type information available at alternating time steps, with time moving from left to right.}
\label{surface_foliated}
\end{figure}

\begin{figure}[htb!]
\centering
    \begin{subfigure}[b]{0.3\textwidth}
        \centering
        \includegraphics[width=\linewidth]{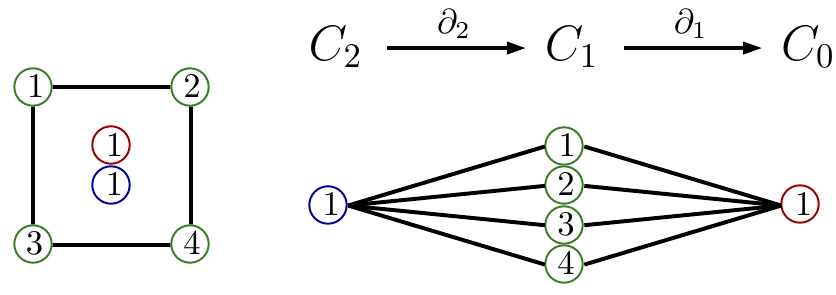} \vspace{1.5cm}
        \caption{The [[4,2,2]] code.}
    \end{subfigure}%
    \hspace{1.5cm}
    \begin{subfigure}[b]{0.58\textwidth}
        \centering
        \includegraphics[width=\linewidth]{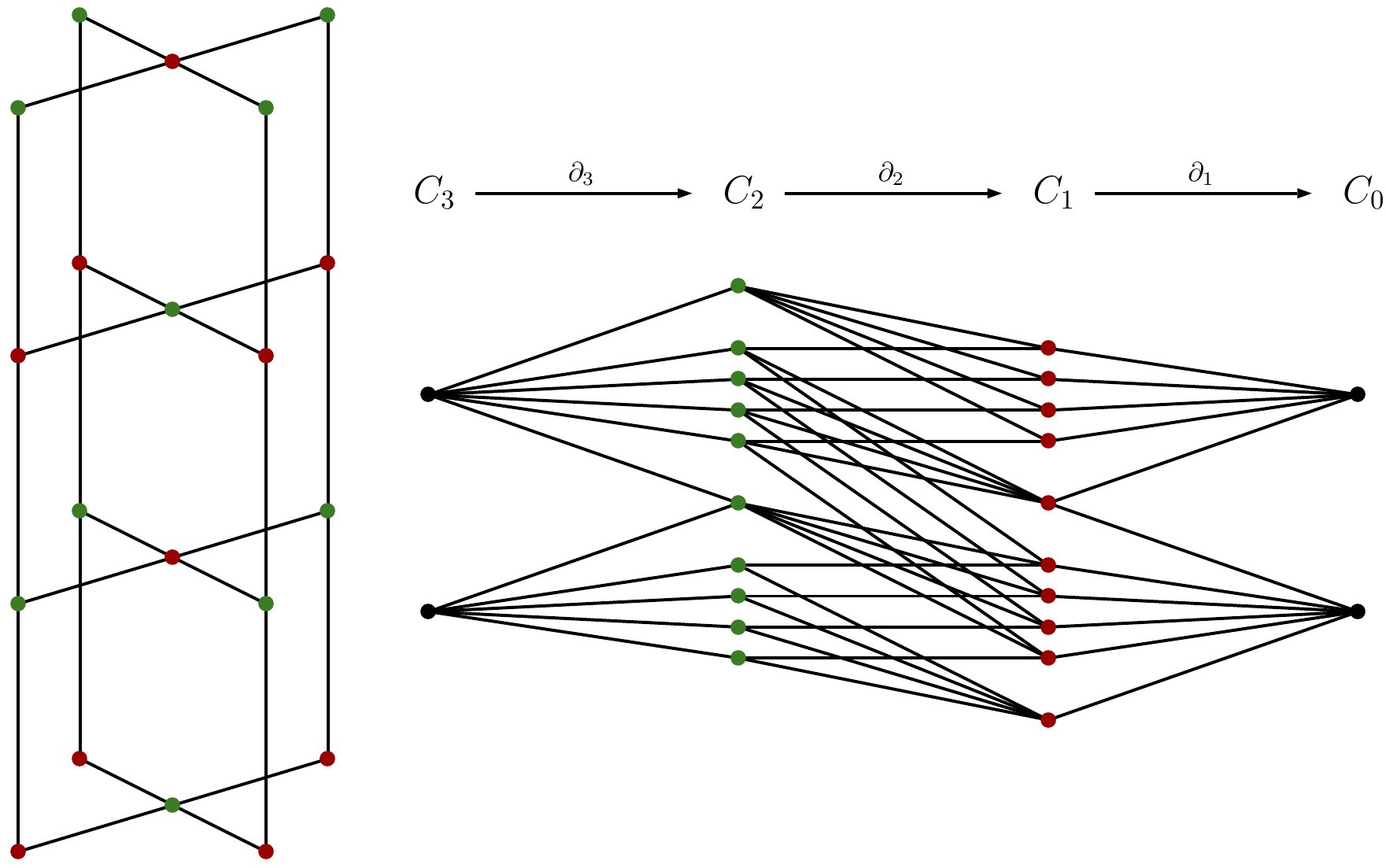}
        \caption{The foliated [[4,2,2]] code.}
    \end{subfigure}
    \caption{(a) A standard representation of the $[[4,2,2]]$ code and its associated length-$2$ chain complex.  (b) The underlying graph state of the $2$-time-step foliated $[[4,2,2]]$ code and its associated length-$3$ chain complex. Dual (primal) qubits are colored green (red).  For clarity, the four data qubits in a particular time-slice are grouped together.  Each time-slice is further subdivided into two steps, which is encoded by the location of the group of four in either $C_2$ or $C_1$.}
\label{422}
\end{figure}

Given a CSS code presented as a length-$2$ chain complex $$C_2 \xrightarrow{\partial_2} C_1 \xrightarrow{\partial_1} C_0$$ and a number of time-steps $T$, its \emph{$T$-time-step foliation} is the following fault-tolerant cluster state, presented as a length-$3$ chain complex:
$$C_2^{\oplus T} \xrightarrow{\delta_3} (C_2 \oplus C_1)^{\oplus T} \xrightarrow{\delta_2} (C_0 \oplus C_1)^{\oplus T} \xrightarrow {\delta_0} C_0^{\oplus T}.$$  Let $\text{id}^V$ denote the identity on $V$, and for $f:V \rightarrow W$, let $f_{i,j}: V^{\oplus T} \rightarrow W^{\oplus T}$ denote the map that applies $f$ from $V_i$ to $W_j$ and is zero everywhere else.  Then the corresponding boundary maps are
\begin{align*}
    \delta_3 &= \text{id}^{C_2}_{T,T} +  \partial_{2_{T,T}} + \sum\limits_{i = 1}^{T-1} \text{id}^{C_2}_{i,i} + \text{id}^{C_2}_{i, i+1} + \partial_{2_{i,i}}, \\
    \delta_2 &= \text{id}^{C_1}_{T,T} + \partial_{2_{T,T}} + \partial_{1_{T,T}} + \sum\limits_{i=1}^{T-1} \text{id}^{C_1}_{i,i} + \text{id}^{C_1}_{i,i+1} + \partial_{2_{i,i}} + \partial_{1_{i,i}}, \\
    \delta_1 &= \text{id}^{C_0}_{T,T} + \partial_{1_{T,T}} + \sum\limits_{i=1}^{T-1} \text{id}^{C_0}_{i,i} + \text{id}^{C_0}_{i,i+1} + \partial_{1_{i,i}}.
\end{align*}
Intuitively, the above prescription assembles $1$D cluster states of length $2T$ for each data qubit.  Then, $Z$-type and $X$-type ancilla qubits are introduced in alternating layers, and connected via CZ gates to the data qubits in their associated stabilizers.  Commutativity of the stabilizers ensures that this chain complex is well-defined, and the corresponding fault-tolerant cluster state inherits the properties of the underlying CSS code \cite{bolt2016foliated, brown2018universal}.  This picture recovers the $3$D cluster state as a foliated surface code, depicted in Figure \ref{surface_foliated}. We include the $2$-time-step foliated $[[4,2,2]]$ code, along with its associated chain complexes, as another leading example in Figure \ref{422}.

Foliated quantum codes are closely related to their corresponding CSS codes in a phenomenological model; indeed the space-time decoding processes for each can be nearly identical \cite{Dennis:2002,raussendorf2007fault}.  Foliated codes also yield a prescription for building fault-tolerant cluster states: a priori, building length-$3$ chain complexes with desirable properties like high distance, high encoding rate, or locality can be a difficult task.   Generally, foliation allows one to port existing constructions of quantum error-correcting codes to the MBQC framework, making it a powerful tool.

However, it is also restrictive.  There are chain complexes corresponding to fault-tolerant cluster states that cannot be realized as the foliation of any quantum code \cite{nickerson2018measurement}.  The broadened framework of length-$3$ chain complexes permits a wide variety of robust cluster states that often outperform their foliated cousins.  In some sense, this generalizes the notion of quantum error-correction in a phenomenological error model.  Mirroring the construction of the toric code, one good prescription for building sparse chain complexes corresponding to robust cluster states is by cellulating manifolds.  However, before building cluster states from cellulations, it will be helpful to first interpret the effects of different types of noise on such states.  In what follows, we describe a circuit-level error model and identify which cluster states should be robust to which parameter regimes of noise.

\subsection{Circuit-based error model and its effects}\label{simple}

We consider a simple circuit-level Pauli error model that captures the correlated errors that may occur due to gate failures. Decoding will be performed independently on primal and dual qubits, and so we describe errors that manifest as $Z$-type on primal qubits; errors on dual qubits can be treated analogously. We consider three types of errors.
\begin{itemize}
\itemsep0em
    \item[($i$)] Measurement errors occur with probability $p_m$. This can also account for preparation errors.
    \item[($ii$)] $X$-type errors occur on the dual qubit support of a CZ gate with probability $p_X$ after the gate.
    \item[($iii$)] $Z$-type errors occur on the primal qubit support of a CZ gate with probability $p_Z$ after the gate.
\end{itemize}
Although $X$-errors commute with our measurements, $X$-errors occurring on dual qubits during the construction of the cluster state will result in correlated $Z$-errors on primal qubits.  Note that this is \emph{not} the standard depolarizing model, in which each two-qubit gate would ordinarily be followed by a depolarizing channel on its support.  For example, we do not include idling errors, and each two-qubit gate has two independent chances to fail, so that the probability of having no errors is $(1-p_X)(1-p_Z)$ rather than $(1-p)$.  This atypical model is more damaging, and leads to lower thresholds than the standard error model. However, it allows us to sample errors directly as edges on the decoder graph, which simplifies numerical simulations over a wide range of complicated lattices significantly.  In order to draw some comparison between this model and the standard model, we include benchmarks of the usual $3$D cluster state as a point of reference \cite{raussendorf2007fault}.

\begin{figure}[htb!]
\centering
\includegraphics[width=.95\linewidth]{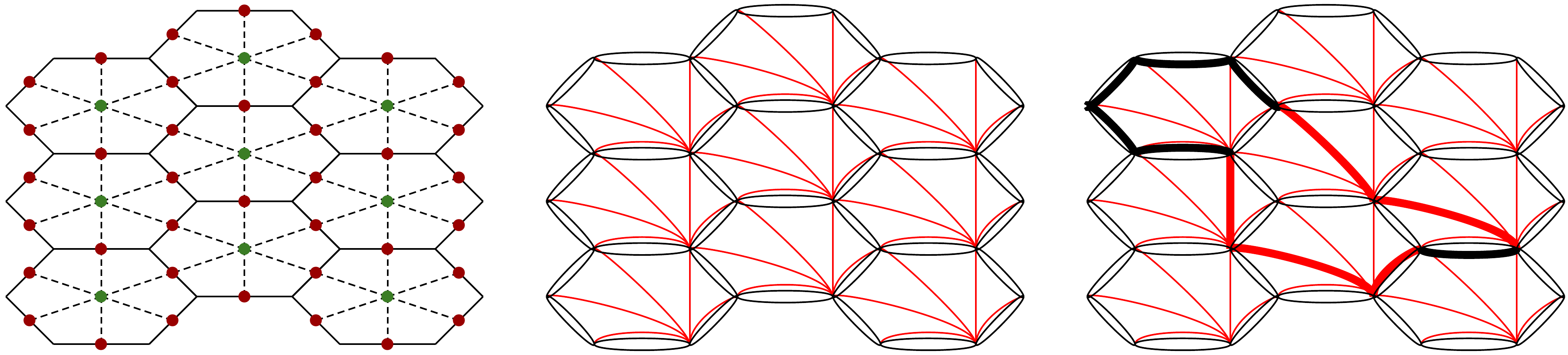}
\caption{On the left, part of a cluster state, with dual (primal) qubits in green (red).  The cell complex is defined by solid lines, while the underlying graph state is defined by dashed lines. The cluster state is constructed by applying CZ gates in each face beginning with the bottom edge and proceeding clockwise.  In the center, each black edge represents a $Z$-error occurring on a primal qubit immediately following a CZ gate.  Each red edge represents a correlated $Z$-error resulting from an $X$-error on a dual qubit immediately following a CZ gate.  Measurement errors are omitted for clarity.  On the right, a collection of highlighted gate errors resulting in a trivial cycle.  Like the foliated repetition code, this $2$D cluster has no checks on dual qubits.  However, this partial protection extends to more general cluster phases defined on $2$D lattices supporting one-form symmetries \cite{daniel2019computational}.}
\label{2D}
\end{figure}

For fault-tolerant cluster states constructed as cellulations, we build the underlying graph state face-by-face. In each face, the CZ gates are applied in consecutive order about the face so that the resulting correlated errors are string-like.\footnote{There is a subtlety here: the error-correction performance may depend on this ordering, and optimal orderings for primal and dual qubits may not be compatible.}   An augmented $1$-skeleton of the cellulation will then serve as the decoder graph, see Figure \ref{2D}. We elaborate on the unit cell of each lattice, along with the decoder graph, and the gate scheduling in Appendix \ref{others}.

There are three broad design principles discussed in \cite{nickerson2018measurement} relevant to the construction of fault-tolerant cluster states built as cellulations of manifolds:
\begin{itemize}
\itemsep0em
    \item[($i$)] Low average vertex degree: heuristically, lower degree vertices yield more information about the surrounding errors, resulting in more robust error-correction.  In the limiting case, the repetition code in a code capacity model can be thought of as having a degree-$2$ decoder graph.  This is particularly important when $p_m$ dominates, as measurement errors are indifferent to the construction cost of the cluster state.  Consequently, choosing the most robust error-correction is optimal.
    \item[($ii$)] Few edges in each face: a face with many edges may result in long-range correlated errors.  This is particularly important when $p_X$ dominates, as $X$-errors on dual qubits cause such errors.
    \item[($iii$)] Few faces incident to each edge: to first order in $p_Z$, the effective error rate each qubit experiences is given by the number of CZ gates it participates in.  This is determined by the number of faces its corresponding edge belongs to, and so is particularly important when $p_Z$ dominates.
\end{itemize}
 As correlated errors on two-qubit gates tend to commute with the gates themselves \cite{aliferis2009fault, PhysRevA.78.052331,trout2018simulating}, we generally expect $p_Z > p_X$ with $p_m$ varying; however, we consider a variety of noise models.  Additionally, for self-dual complexes, principles ($ii$) and ($iii$) become identical.  A final heuristic worth following is increasing the symmetry of the underlying graph states, as homogeneous structures tend to perform better.  Broadly, we expect that given different relative error strengths, different fault-tolerant cluster states appropriately balancing robust error-correction with locality become viable.
 
\section{Combinatorial tiling theory}\label{tilings}

Building quantum codes from manifolds has been enormously profitable, resulting in some of the best known local codes \cite{freedman2002z2,kitaev2003fault,hastings2016quantum, londe2017golden, breuckmann2017hyperbolic, breuckmann2016constructions, conrad2018small}.  However, we do not focus on the topological properties of a fault-tolerant cluster state, but rather the way in which its vacuum is stitched together.  Correspondingly, the objects of interest are no longer manifolds but rather certain \emph{orbifolds},\footnote{A space locally modeled on Euclidean space modulo a linear finite group action, which may not be free.} and in particular, their underlying simplicial structures.  These admissible triangulations are encoded into dressed graphs known as \emph{extended Schl{\"a}fli symbols}, and their study is broadly known as \emph{combinatorial tiling theory} \cite{dress1985regular, dress1987presentations, dress1987tilings, grunbaum1987tilings, huson1993generation, balke1996two, friedrichs2003three,  delgado2001recognition, friedrichs1999systematic, friedrichs1997orbifold, thurston1997geometry, hyde2006towards}.  Put simply for our purposes, it is the algebraic study of ways to periodically tile space.

Following \cite{nickerson2018measurement}, we focus on building fault-tolerant cluster states from such tilings, forming a length-$3$ chain complex in the usual way.  The underlying graph state is constructed by placing qubits on faces and edges, with graph state edges determined by $\partial_2$.  The $2$-skeleton of the tiling will determine the robustness of primal error-correction, with a decoder graph that is an augmented version of the $1$-skeleton.  The same holds for dual qubits in the dual tiling, and so we focus on self-dual complexes to realize symmetric thresholds.

\subsection{Constructing tilings} \label{constructions}
We next review the foundations of combinatorial tiling theory, following the prescription of \cite{delgado2001recognition} towards the construction and recognition of tilings.

\begin{Definition}
A \emph{tiling} is a regular CW-complex\footnote{Closure-finite, weak topology.} $\mathcal{C}$ realizing a simply connected $d$-dimensional manifold. Its \emph{symmetry group} $\Gamma$ is the group of all automorphisms that respect the induced CW-filtration. 
\end{Definition}

Informally, a tiling is a partition of space, with each tile glued face-to-face in such a way that the entire space is filled. It determines a group of symmetries that respect the partition.  These are reflections, rotations, and translations that map the tiles back to themselves in such a way that the new tiling is the same as the old.  There is a notion of equivalence between tilings that depends on both their topological structure and their symmetry group.  This will determine the level of granularity at which we will consider tilings, with two equivalent tilings sharing the same combinatorial structure.

\begin{Definition}
Two tilings $\mathcal{C}_1, \mathcal{C}_2$ with respective symmetry group $\Gamma_1,\Gamma_2$ are \emph{equivariantly equivalent} if there exists a homeomorphism $f: \mathcal{C}_1 \rightarrow \mathcal{C}_2$ respecting the filtration that induces an isomorphism $\Gamma_1 \rightarrow \Gamma_2$ given by $\gamma_1 \mapsto f\circ\gamma_1\circ f^{-1}$.
\end{Definition}

To study tilings, it is useful to break them down into elemental combinatorial objects called flags.  These will be defined in terms of objects of different dimensions.  We refer to such objects $C_i$ as cells, where $i$ denotes the dimension of the cell.  For example, $C_0$ is a point, $C_1$ an edge, and so on.  The set of all cells of dimension $\leq i$ is called the $i$-skeleton, and when $i=d$, it represents the entire CW-complex.
\begin{Definition}
A \emph{flag} is an increasing sequence of cells $(C_0, C_1, \ldots, C_d)$ with $C_i \subseteq C_{i+1}$ and where each $C_i$ belongs to the $i$-skeleton, but not the $(i-1)$-skeleton.
\end{Definition}

Geometrically, flags can be visualized as $d$-simplices obtained in the following way.  For each $i$-cell in a complex, associate a vertex at its barycenter.  Then, connect the vertices of the cells comprising a flag to form its corresponding $i$-simplex.  For example, a square has $8$ flags corresponding to triangles obtained from the barycentric subdivision of its faces and edges, while a cube has $48$ flags corresponding to tetrahedra.

Let $\mathcal{F}$ denote the set of all flags.  A central insight of \cite{dress1985regular} is to study two relevant group actions on $\mathcal{F}$. The first is induced by $\Gamma$, as each automorphism respects the filtration. The second is by the free Coxeter group. 
\begin{Definition}
The \emph{free Coxeter group} generated by $d+1$ reflections is the group $$\mathcal{R} \coloneqq \langle \{r_i\}_{i=0}^d : \{r_i^2 = 1\}_{i=0}^d \rangle.$$ 
\end{Definition}
For each flag $F = (C_i)_i$ and each index $j \in \{0, \ldots, d\}$, there is a unique flag $F' = (C_i')_i$ satisfying $C_i = C_i'$ for all $i \neq j$ and $C_j \neq C_j'$.  This yields a well-defined action given by $r_j(F) \coloneqq F'$.  Geometrically, this action can be visualized as reflecting the $d$-simplex corresponding to $F$ about the face opposite the vertex corresponding to $C_j$.  This new $d$-simplex corresponds to $F'$. 

The actions of $\Gamma$ and $\mathcal{R}$ commute, and so the $\mathcal{R}$-action descends to an action on the set of flag orbits $\{\Gamma \cdot F: F \in \mathcal{F}\}.$  Specifying the free Coxeter group action, in terms of its generators and relations, on each flag orbit carries all relevant information about a tiling.  The objects representing this information are known as \emph{extended Schl{\"a}fli symbols} (or \emph{Delaney symbols}, see \cite{blatov2010vertex}). 

\begin{Definition} \label{main_def}
A $d$-dimensional \emph{extended Schl{\"a}fli symbol} is a set $S$ equipped with a transitive $\mathcal{R}$-action along with, for each $s \in S$ and for all $0 \leq i , j \leq d$, positive natural numbers $m_{ij}(s)$ satisfying
\begin{itemize}
\itemsep0em
    \item[($i$)] $m_{ii}(s) = 1$,
    \item[($ii$)] $(r_ir_j)^{m_{ij}(s)} \cdot s = s$,
    \item[($iii$)] $m_{ij}(s) = m_{ji}(s) = m_{ij}(s')$ for any $s' \in \langle r_i,r_j \rangle \cdot s$.
\end{itemize}
The symbol is further called \emph{regular} if $m_{ij} = 2$ for all $|i-j| > 1$.
\end{Definition}

We will generally refer to these as symbols $S$, leaving implicit the extra information carried by the set. There is also a notion of maps between symbols called \emph{coverings}, analogous to covering maps between orbifolds.

\begin{Definition}
A \emph{covering} of a $d$-dimensional symbol $S$ by another $d$-dimensional symbol $S'$ is an $\mathcal{R}$-equivariant map $f:S'\rightarrow S$ satisfying $m_{ij}(f(s')) = m_{ij}(s')$ for all $s' \in S'$.  The covering is an isomorphism if $f$ is bijective.
\end{Definition}

Informally, coverings are simply maps between symbols that preserve the structure of the symbol, namely the labeled adjacency relations and the vertex labels. We will often be concerned with self-dual cluster states, as our thresholds will be limited by error-correction on the worse of the two lattices.  Fortunately, the corresponding notion of duality on symbols is quite simple.
\begin{Definition}
The \emph{dual} of a $d$-dimensional extended Schl{\"a}fli symbol is the symbol obtained by interchanging $r_i \leftrightarrow r_{d-i}$ and $m_{ij} \leftrightarrow m_{d-i,d-j}$.  A symbol is called \emph{self-dual} if it is isomorphic to its dual.
\end{Definition}
Thus, for  the purposes of building fault-tolerant cluster states, we will mostly be concerned with self-dual symbols.  For a $d$-dimensional tiling $\mathcal{C}$ with symmetry group $\Gamma$, we can associate a $d$-dimensional symbol in the following way.
\begin{itemize}
\itemsep0em
    \item[($i$)] Let $S \coloneqq \{\Gamma \cdot F: F \in \mathcal{F}\}$ with corresponding transitive $\mathcal{R}$-action inherited from $\mathcal{F}$.
    \item[($ii$)] For each flag orbit $\Gamma \cdot F$, for any flag $F' \in \Gamma \cdot F$, and for any $0\leq i , j \leq d$, let $m_{ij}$ denote the order of the action about $r_ir_j$, namely $m_{ij} \coloneqq \min \{k: (r_ir_j)^kF' = F'\}$.  This is independent of the choice of representative $F'$.
\end{itemize}
Note that for a tiling, the $i$th cell of $r_i(F)$ depends only on the $(i-1)$- and $(i+1)$-cells of $F$.  Consequently, any symbol associated to a tiling must be regular, as the corresponding $r_i$,$r_j$ must commute.

\begin{figure}[htb!]
\centering
    \begin{subfigure}[b]{0.45\textwidth}
        \centering
        \includegraphics[width=\linewidth]{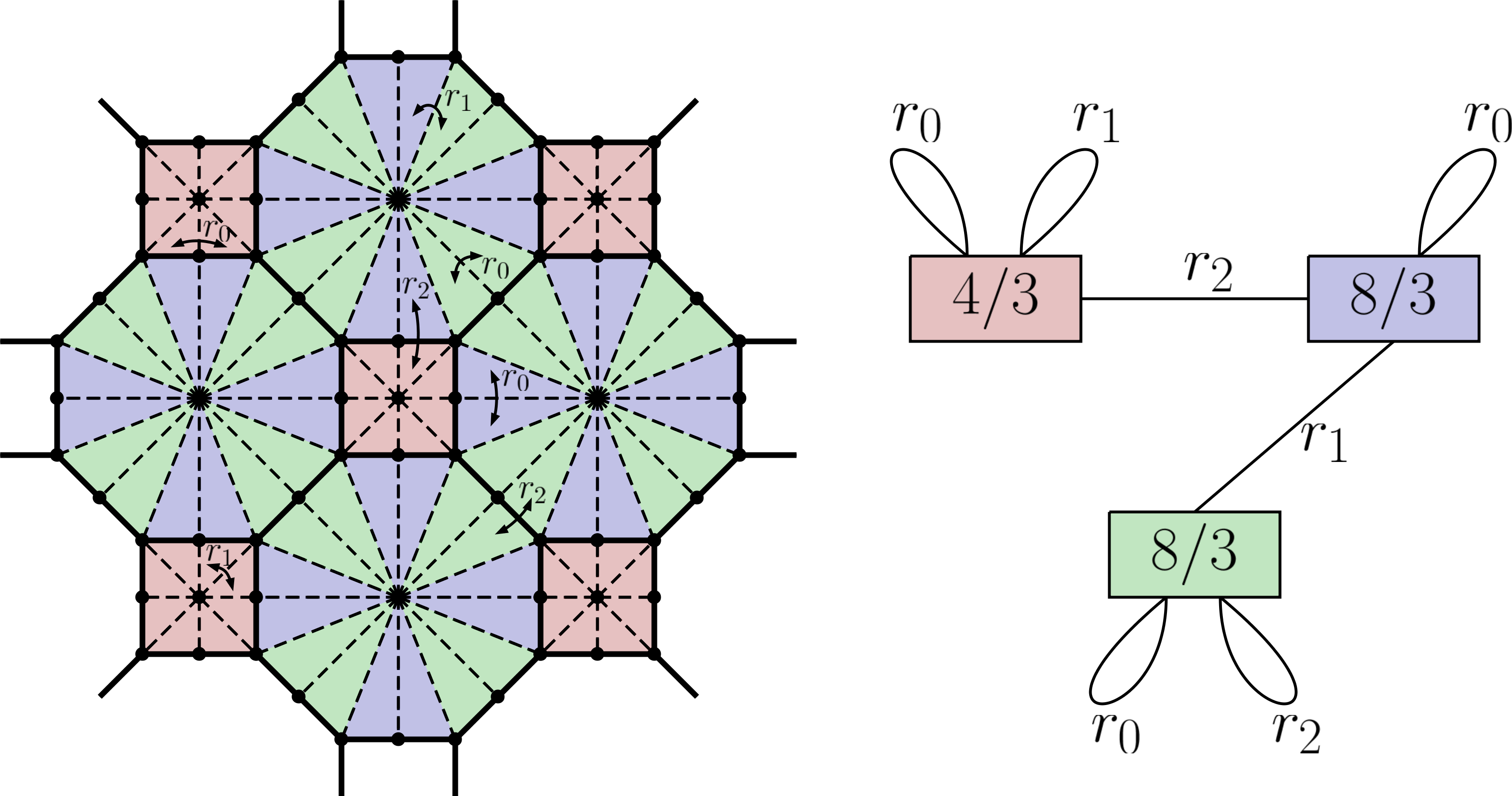}
        \caption{The 4.8.8 tiling.}
    \end{subfigure}%
    \hspace{0.5cm}
    \begin{subfigure}[b]{0.47\textwidth}
        \centering
        \includegraphics[width=\linewidth]{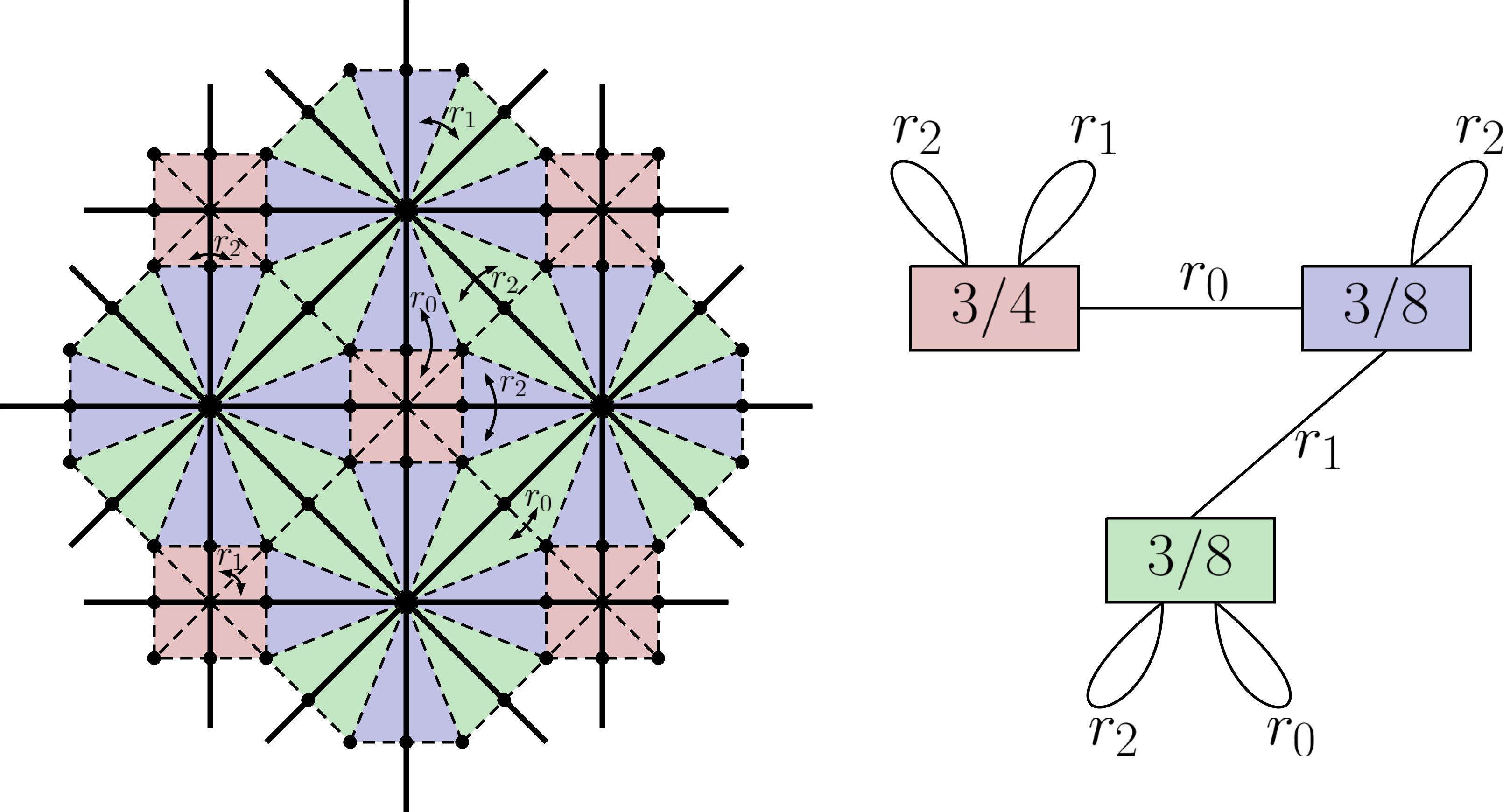}
        \caption{The dual of the 4.8.8 tiling.}
    \end{subfigure}
\caption{The $4.8.8$ tiling and its corresponding extended Schl{\"a}fli symbol, represented as a dressed graph, along with  their duals.  The edges of the original lattice are solid, while edges introduced by the barycentric subdivision are dashed, with the added vertices shown.  Up to symmetry, there are three unique flags colored red, blue, and green.  Each triangle represents a flag, with its vertices corresponding to a face, edge, and vertex.  The action of $r_i$ corresponds to reflecting about the facet opposite the vertex representing an $i$-cell.  Examples of each reflection generator appearing in the symbol are shown.  Dissimilar to a regular tiling, not all of these reflections are themselves symmetries.  For example, any edge corresponding to a reflection mapping between flags of different colors cannot be a symmetry of the tiling, as it maps between two symmetry inequivalent flags.  Each flag orbit is labeled with $m_{01}/m_{12}$; in dimension $2$, this corresponds to the number of edges in the face containing the flag, and the degree of the vertex incident to the flag, respectively.  The nodes of the dressed graph are colored for clarity.  As the $4.8.8$ tiling is not self-dual, the two symbols are not isomorphic.}
\label{symbol}
\end{figure}

To summarize, a symbol has a vertex for each unique flag, up to symmetries of the tiling.  For example, the symbol for a regular tiling like that of the plane by squares has a single vertex. In contrast, the $4.8.8$ tiling of the plane has three vertices corresponding to flags (triangles) within squares, flags that border two octagons, and flags that border both a square and an octagon; see Figure \ref{symbol}. Extended Schl{\"a}fli symbols provide a sharp invariant for characterizing equivariant equivalence classes of tilings.

\begin{Theorem}[\cite{dress1985regular}]
Two tilings are equivariantly equivalent if and only if their associated extended Schl{\"a}fli symbols are isomorphic.
\end{Theorem}

Symbols are often presented as dressed graphs \cite{delgado2001recognition}.  The vertex set is given by $S$, with each vertex labeled by its corresponding $m_{i,j}$.  For regular symbols associated to tilings, the order of the necessarily commuting reflections is typically omitted, so that each vertex is only labeled by $m_{i,i+1}$.  The edge set is determined by the $\mathcal{R}$-action, with each edge labeled by the corresponding reflection generator.  For tilings, these can be visualized by barycentrically subdividing the tiling into $d$-simplices representing flags, with the $\mathcal{R}$-action represented by reflections about facets; see Figure \ref{symbol}.

Just as we can create a symbol from a tiling, so too can we construct the underlying space \textbf{Top}$(S)$ of an orbifold encoded by a symbol $S$.  This is done by associating to each $s \in S$ a simplex with $(d+1)$-facets, and then identifying the $i$th facets of two simplices if the corresponding elements of $S$ are exchanged by $r_i$.\footnote{The information specifying the singular locus of the orbifold is lost in this mapping.}  Coverings can also be realized as continuous maps that identify simplices.

To summarize, we can map from tilings to symbols, and from symbols to topological spaces, where the equivariant equivalence class of a tiling is determined by its associated symbol.\footnote{The composition of these maps need not recover the tiling, but rather encodes an orbifold covered by the tiling.}  We would like to construct tilings from symbols directly in order to turn the former geometric problem into a (dressed) graph problem.  To do so, we must determine which symbols can be used to realize tilings of Euclidean $3$-space.  

In order to construct tilings from symbols, we can use a covering space theory of symbols inherited from the covering space theory of orbifolds \cite{delgado2001recognition,thurston1997geometry}.  This will further transform the problem on dressed graphs into an algebraic problem on their associated fundamental groups.

\begin{Definition}
Let $S$ be an extended Schl{\"a}fli symbol with basepoint $s$. Up to isomorphism, there is a unique symbol $\overline{S}$ with basepoint $\overline{s}$ and a covering $u_S:\overline{S} \rightarrow S$ with $u_S(\overline{s}) = s$ such that for any other covering $f:S'\rightarrow S$ and any $s' \in f^{-1}(s)$, there is a unique $u_{S'}:\overline{S} \rightarrow S'$ through which $u_S$ factors: $u_S = f \circ u_{S'}$ with $u_{S'}(\overline{s}) = s'$ \cite{thurston1997geometry}.  We call $\overline{S}$ the \emph{universal cover} of $S$.
\end{Definition}

Informally, a symbol is an encoding of a topological space with some symmetry `packed away'.    The larger the symbol, the more flags it has, the more triangles glued together it represents, and the less of its symmetry is packed away.  To come back to the square tiling example, a $2$-torus corresponds to a square, which can be constructed from a symbol with eight vertices. There is an implicit translational symmetry in the $x$- and $y$-directions.  The square tiling is its universal cover, obtained from `unfurling' these symmetries; this is constructed from an infinite symbol that has no symmetries packed away.  Each of its triangles is explicitly defined by the symbol, rather than implicitly defined by a symmetry.  These implicit symmetries of a symbol are described by the \emph{fundamental group} of the symbol, and to obtain the fundamental group of a symbol, we must construct its universal cover. 

An explicit presentation for the universal cover of $S$ can be constructed according to the prescription in \cite{delgado2001recognition}.  Namely, fix a basepoint $s \in S$ and identify $\overline{S}$ with cosets in $\mathcal{R}$ of the normal subgroup $$U_s \coloneqq \langle r^{-1} (r_ir_j)^{m_{ij}(r\cdot s)}r: r \in \mathcal{R}, 0 \leq i , j \leq d\rangle,$$  where the $\mathcal{R}$-action is inherited from the group, and define $m_{ij}(r\cdot U) \coloneqq m_{ij}(r \cdot s).$  As $U_s \subseteq \text{Stab}_\mathcal{R}(s)$, this defines a valid cover with covering map $r \cdot U \mapsto r \cdot s$, where the information about $S$ is passed to $U_s$ through the group action in the exponent of $r_ir_j$.  In fact, opposite to the universal cover $\overline{S}$ of $S$, there also exists a \emph{final} (or minimal) \emph{cover} $\underline{S}$.  Up to basepoints, this is defined as the unique symbol with covering $f_S:S \rightarrow \underline{S}$ such that, for any other covering $f: S \rightarrow S'$, there is a unique $f_{S'}: S' \rightarrow \underline{S}$ through which $f_S$ factors: $f_S = f_{S'} \circ f.$ 

Given the universal cover of a symbol $S$, we can define the fundamental group of $S$ as the group of deck transformations in the usual way.

\begin{Definition} \label{second_def}
Let $S$ be an extended Schl{\"a}fli symbol with basepoint $s$, and let $\overline{S}$ be its universal cover with covering map $u_S$.   Then the \emph{fundamental group} of $S$ is given by the subgroup of automorphisms of $\overline{S}$ annihilated by $u_S$, $$\pi_1(S,s) \coloneqq \{\phi \in \text{Aut}(\overline{S}) :  u_S \circ \phi = u_S\}.$$
\end{Definition}
Following the construction of the universal cover, the fundamental group can be constructed explicitly as $\pi_1(S,s)= \text{Stab}_\mathcal{R}(s)/U_s$ with a finite presentation \cite{balke1996chamber}. If this symbol is covered by a tiling, then the fundamental group corresponds to a subgroup of the symmetries of that tiling.  For example, only reflections $r_i$ that map between symmetry equivalent flags are themselves symmetries of the tiling.  Correspondingly, these are exactly the reflections that belong to $\text{Stab}_\mathcal{R}(s)$.  More generally, we have the usual correspondence between subgroups $G \subseteq \pi_1(S,s)$ and covers of $S$, given by $G \leftrightarrow \overline{S}/G$ and pictured below.

\begin{center}
\includegraphics[width=0.75\linewidth]{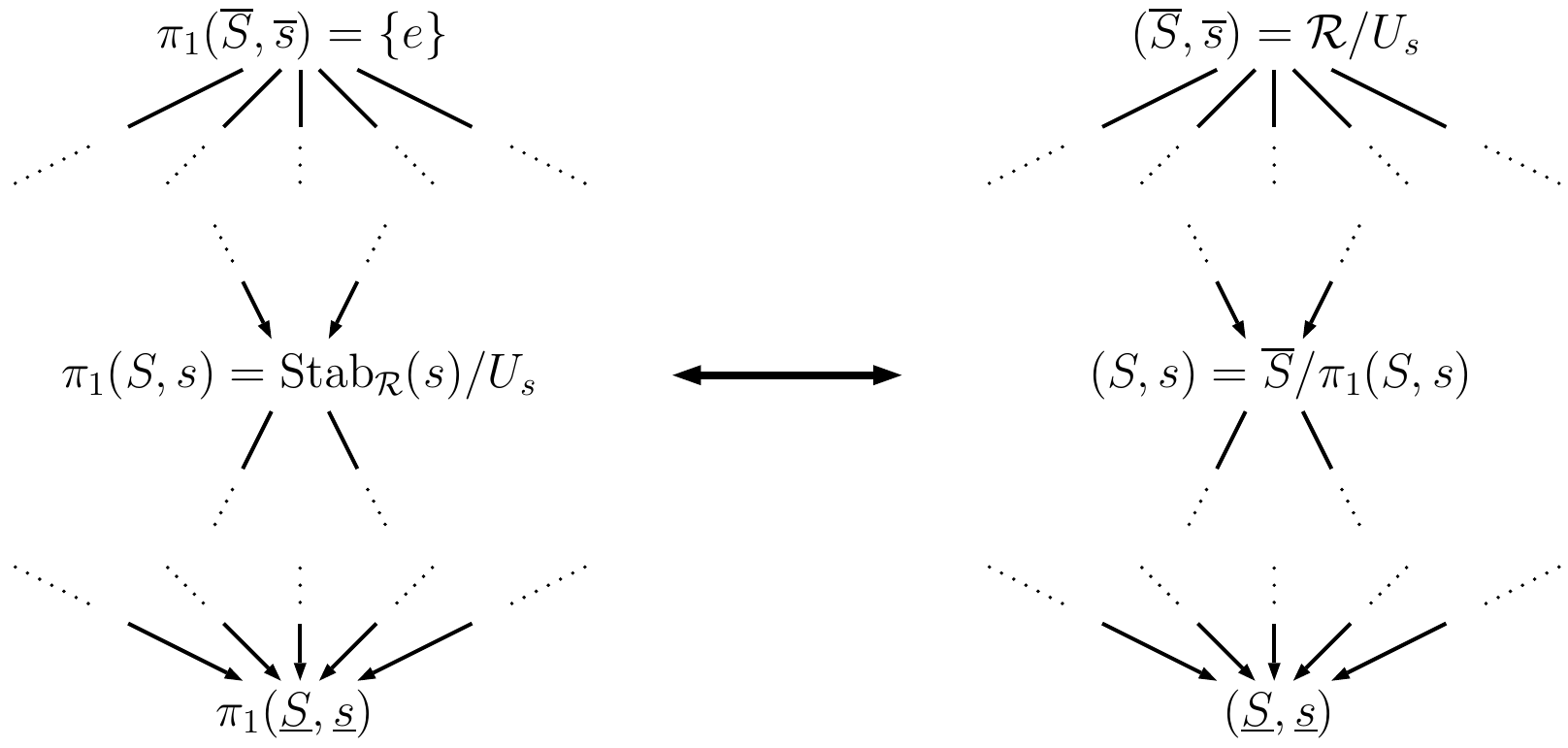}
\end{center}

On the left, we have a sequence of inclusions of groups, and on the right, a corresponding sequence of coverings of symbols.  From top to bottom, the sizes of the objects on the left increase, while the sizes of the objects on the right decrease. We can regard moving up the diagram as unfurling the symmetries of a symbol; the final cover has a maximal group of symmetries $\pi_1(\underline{S},\underline{s})$ and the universal cover is a complete unfurling of those symmetries.  In particular, the final cover is the symbol we associate to a tiling, while the universal cover is the symbol that we extract in order to construct a tiling.\footnote{In particular, the symbols we associated to tiling were actually their final covers, corresponding to a realization of space by a minimal set of simplices, but expanded by a maximal set of symmetries.}

For now, we are interested in determining which symbols $S$ correspond to (unfurled) periodic tilings of Euclidean $3$-space.  Such an $S$ should be covered by $\mathbb{R}^3$ with deck transformations realizing Euclidean isometries, up to homeomorphism.  As symbols also carry the underlying orbifold structure, we also insist that the $\mathbb{R}^3$ cover carry a trivial orbifold structure free of local group actions.  Correspondingly, the universal cover must satisfy $$(r_ir_j)^k(\Gamma \cdot F) \neq (\Gamma \cdot F)$$ for all $k < m_{ij}(\Gamma \cdot F)$.\footnote{This property, called \emph{semi-good} in \cite{delgado2001recognition}, is not necessary for the construction of the chain complex; we only care about the underlying manifold structure.  However, it is helpful in not double-counting symbols that may correspond to the same underlying tiling.}  Symbols satisfying these two properties are called \emph{flat}, and it suffices to test flatness on any symbol covered by $S$.

\begin{Theorem}[\cite{delgado2001recognition}] \label{restrict}
If $S$ and $S'$ are $3$-dimensional symbols with $S$ covering $S'$, then $S$ is flat if and only if $S'$ is flat.
\end{Theorem}
This implies that no matter which symbol in a sequence of coverings we choose, we can check for flatness directly on that symbol.
This is important because, by design, we are looking for infinite universal covers (corresponding to a tiling).  Fortunately, by Theorem \ref{restrict}, we can instead restrict our attention to finite symbols $S$ corresponding to compact orbifolds that are covered by tilings. Any finite flat symbol must use an infinite translational symmetry group of finite index $\mathbb{Z}^d \cong H \subseteq \pi_1(S,s)$ to fill $\mathbb{R}^d$ \cite{bieberbach1911bewegungsgruppen, bieberbach1912bewegungsgruppen}.  Thus, one can enumerate over finite index subgroups of $\pi_1(S,s)$ until identifying one isomorphic to $\mathbb{Z}^3$, and then construct the corresponding finite cover.  If the symbol is free of local group actions, then one can check algorithmically whether its simplicial topological realization is homeomorphic to a $3$-torus \cite{hemion1992classification}.  If so, then the symbol encodes a tiling by unfurling these final translational symmetries.  For a complete description of how to perform this check efficiently, see \cite{delgado2001recognition}.  In practice, we perform these tests using the open-source software Gavrog \cite{Gavrog}, but have included a summary of the internal sequence of checks below for completeness.
\subsection*{Checking symbols}
\begin{enumerate}
    \item Write down valid regular symbol according to Definition \ref{main_def}, chosen so that the corresponding graph state (if it exists) has desirable properties like locality, self-duality, etc.
    \item Check if locally Euclidean \cite{delgado2001recognition}. If yes,
    \item Construct orbifold graph and orbifold invariants \cite{friedrichs1997orbifold} to determine space group.  If it corresponds to one of the $219$ classified space groups,
    \item From fundamental group constructed according to Definition \ref{second_def} and \cite{balke1996chamber}, enumerate over finite index subgroups.  Test if the next group isomorphic to $\mathbb{Z}^3$.  If yes,
    \item Construct the corresponding (possibly toroidal) cover of the symbol as quotient of universal cover by that subgroup.
    \item Perform symbol reductions \cite{delgado2001recognition} to simplify symbol.  Once simplified,
    \item Check if the reduced symbol belongs to a known list of simplified symbols.  If yes, declare success; the tiling and its corresponding desired graph state exist.  If no, continue from Step 4.  If there are no more subgroups isomorphic to $\mathbb{Z}^3$, declare failure.
\end{enumerate}

In summary, in order to construct a fault-tolerant cluster state with desirable properties like self-duality and locality, one begins by constructing valid symbols with those desired properties.  There are several necessary conditions one can impose to narrow this search, such as demanding regularity and positive curvature on tiles, as they themselves must be spherical. Then, one can follow the above prescription to construct the associated fundamental group explicitly, and search for a subgroup of spanning translational symmetries.  If found, the corresponding finite cover encodes a tiling realizing the desired cluster state, which can be built from translations of the unit cell represented by this symbol.

It is worth noting that, instead of building fault-tolerant cluster states based on tilings of $3$-space, we could instead try and build CSS codes based on tilings of $2$-space.  In the $2$D setting, the Euler characteristic provides a simple invariant for detecting the zero-curvature of a flat $2$D symbol $S$, which must satisfy $\sum_s m_{01}(s) + m_{12}(s) = (\#s)/2$. However, this limits the average degree of a self-dual tiling to $4$, which also determines the locality of the code, and so the toric code is essentially optimal with symmetric noise.  The central insight in \cite{nickerson2018measurement} is that this limitation is relaxed in $3$D (where the Euler characteristic is zero) due to the increased length of the chain complex.  However, fault-tolerant cluster states still incur a tradeoff between the graph state degree and decoder graph degree in a self-dual lattice, as we will see shortly.

\subsection{Generating self-dual fault-tolerant cluster states with local underlying graph states} \label{local}

\subsubsection{An optimally local fault-tolerant cluster state}
Before detailing a large class of self-dual fault-tolerant cluster states, presented in terms of their corresponding symbols, we begin with a single fully-realized and optimally local example.  Broadly, we have discussed two design principles from \cite{nickerson2018measurement} for building a robust fault-tolerant cluster state: low average degree of the underlying $1$-skeleton, contributing to robust error-correction, and underlying graph-state locality, contributing to a low effective error-rate and minimal correlated errors.

\begin{figure}[htb!]
\centering
    \begin{subfigure}[b]{0.46\textwidth}
        \centering
        \includegraphics[width=\linewidth]{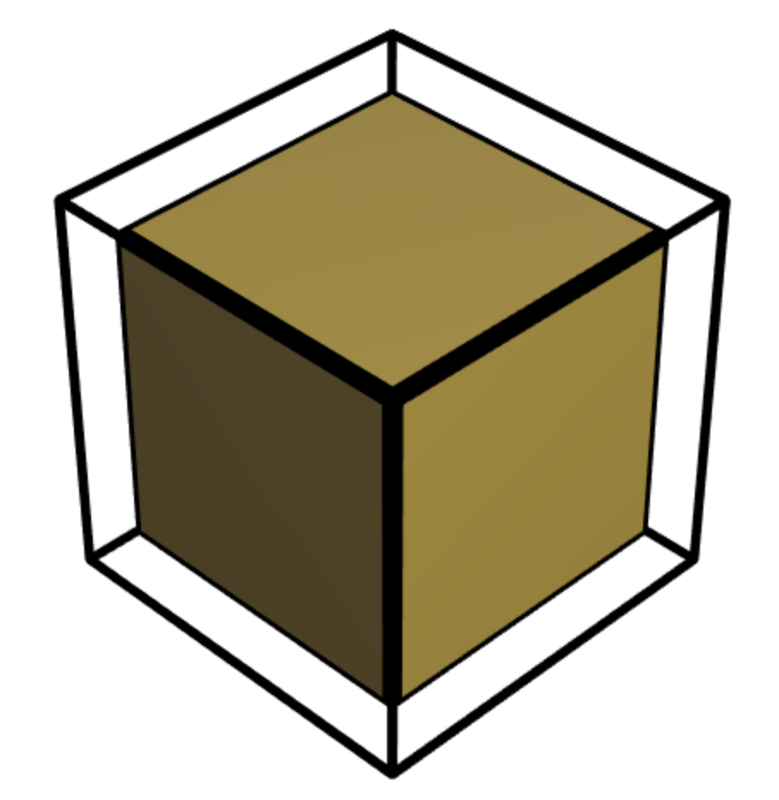}
        \caption{$3$-regular graph state, $6$-regular decoder graph.}
    \end{subfigure}%
    \hspace{1.25cm}
    \begin{subfigure}[b]{0.44\textwidth}
        \centering
        \includegraphics[width=\linewidth]{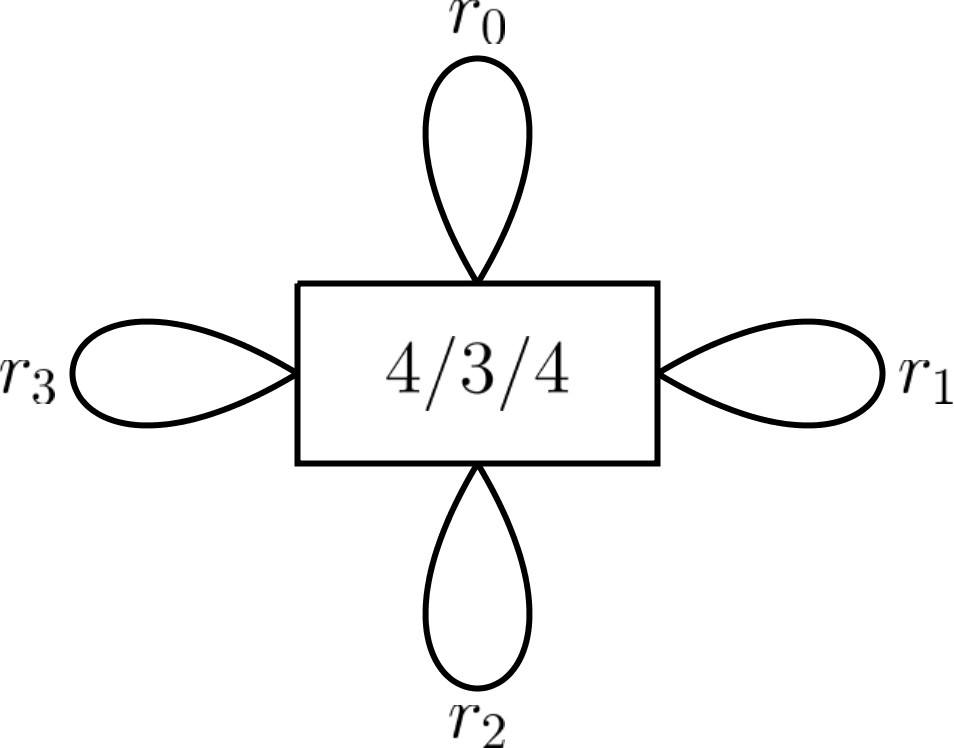}
        \vspace{.5 cm}
        \caption*{The duality isomorphism is the identity.}
    \end{subfigure}
    \begin{subfigure}[b]{0.46\textwidth}
        \centering
        \includegraphics[width=\linewidth]{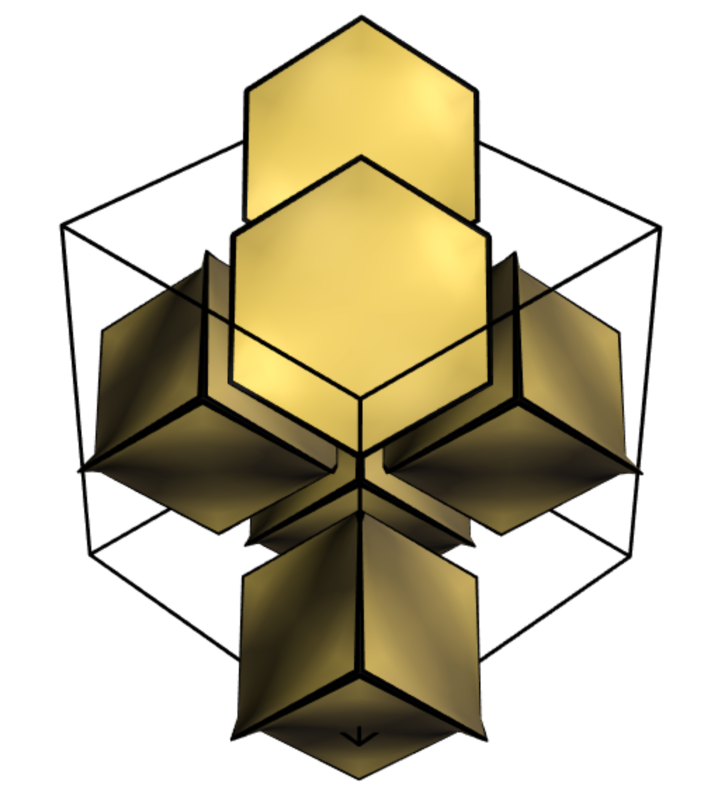}
        \caption{$6$-regular graph state, $4$-regular decoder graph.}
    \end{subfigure}%
    \hspace{1.25cm}
    \begin{subfigure}[b]{0.44\textwidth}
        \centering
        \includegraphics[width=\linewidth]{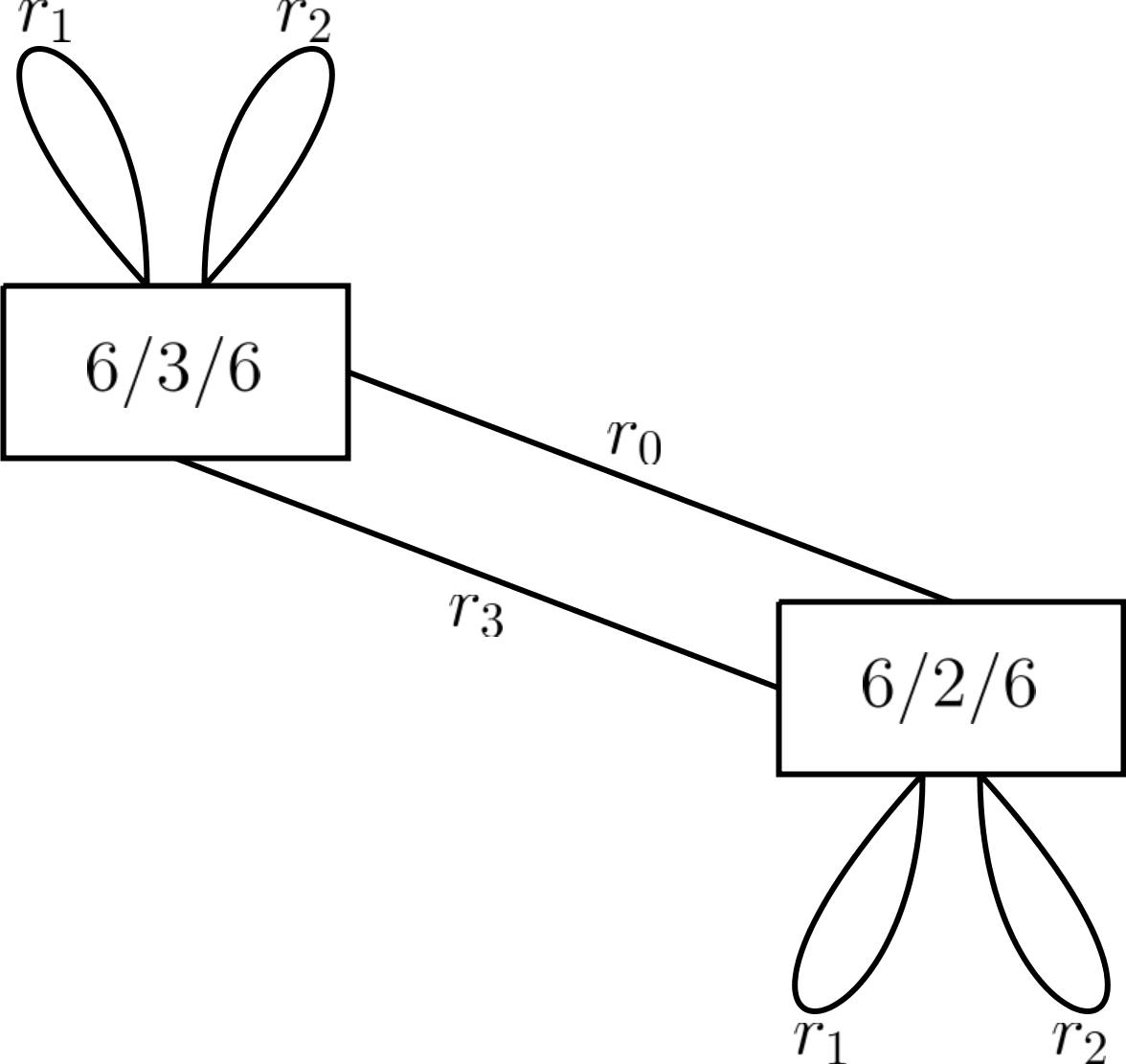}
        \vspace{.2cm}
        \caption*{The duality isomorphism is the identity.}
    \end{subfigure}
    \caption{Two self-dual fault-tolerant cluster states  that balance locality of the decoder graph and the graph state, built from the well-studied cubic and diamond lattices, and their corresponding final covers. The first represents the usual $3$D cluster state originally introduced in \cite{raussendorf2007topological}, while the second was studied in \cite{nickerson2018measurement} (figures rendered with Gavrog \cite{Gavrog}).}
\label{cubic_vs_diamond_1}
\end{figure}

\begin{figure}[htb!]
\centering
    \begin{subfigure}[b]{0.45\textwidth}
        \centering
        \includegraphics[width=\linewidth]{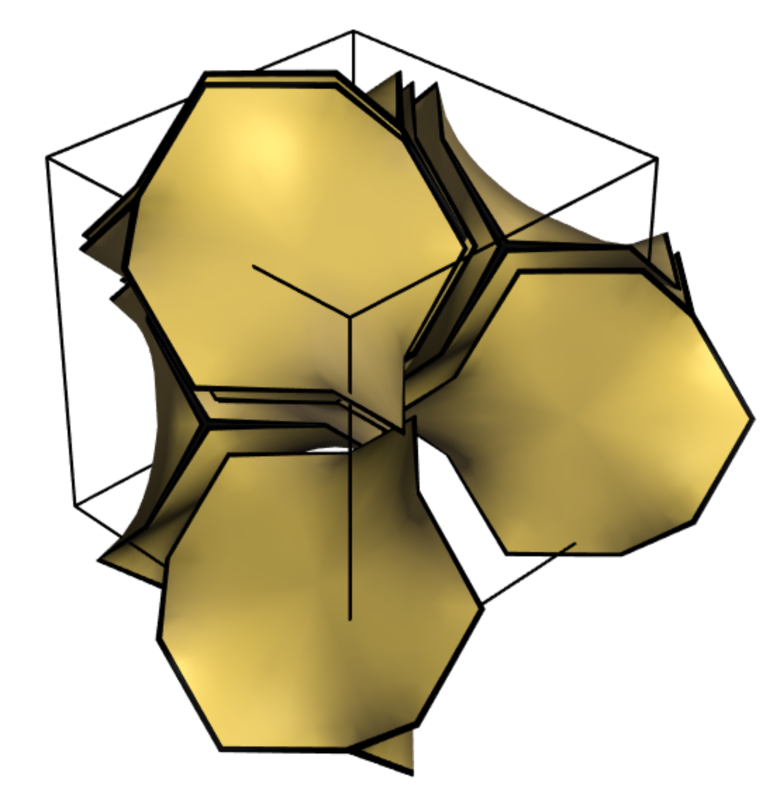}
        \caption{$10$-regular graph state, $3$-regular decoder graph.}
    \end{subfigure}%
    \hspace{1.25cm}
    \begin{subfigure}[b]{0.45\textwidth}
        \centering
        \includegraphics[width=\linewidth]{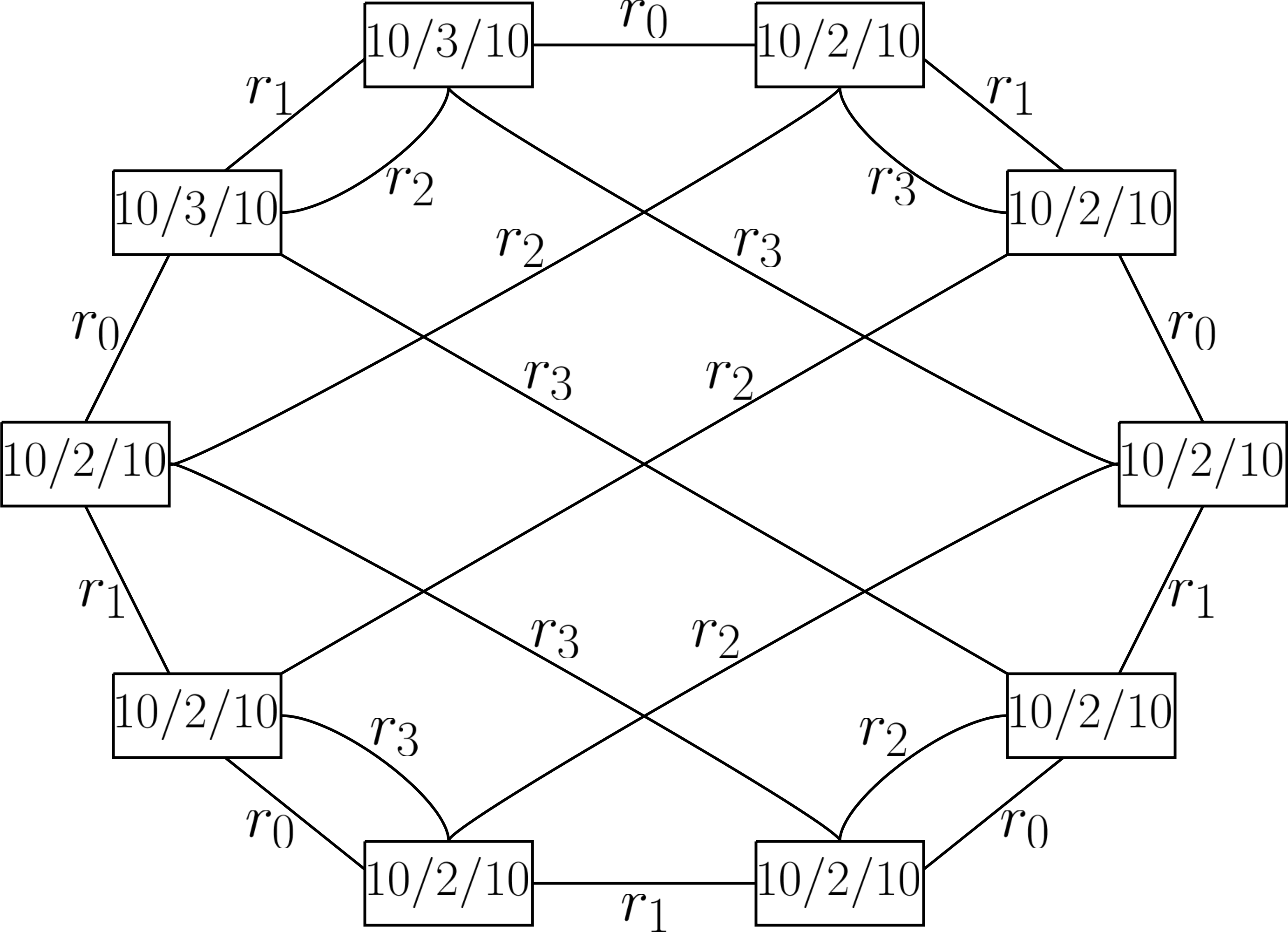}
        \vspace{.3 cm}
        \caption*{The duality isomorphism swaps pairs in each corner and the pair opposite each other along the
        equator.}
    \end{subfigure}
    \begin{subfigure}[b]{0.45\textwidth}
        \centering
        \includegraphics[width=\linewidth]{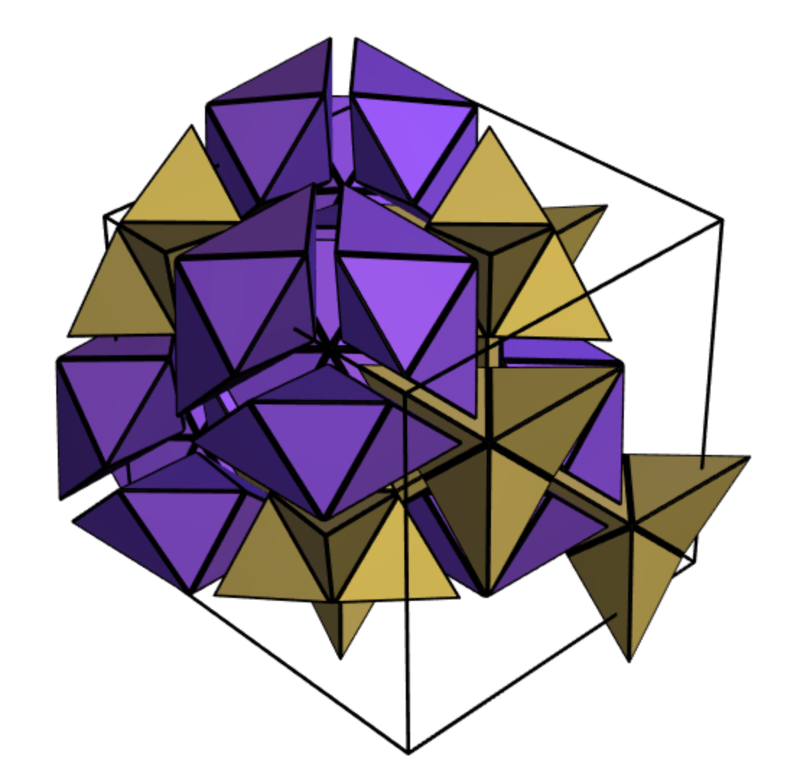}
        \caption{$3$-regular graph state, $12$-regular decoder graph.}
    \end{subfigure}%
    \hspace{1.25cm}
    \begin{subfigure}[b]{0.45\textwidth}
        \centering
        \includegraphics[width=\linewidth]{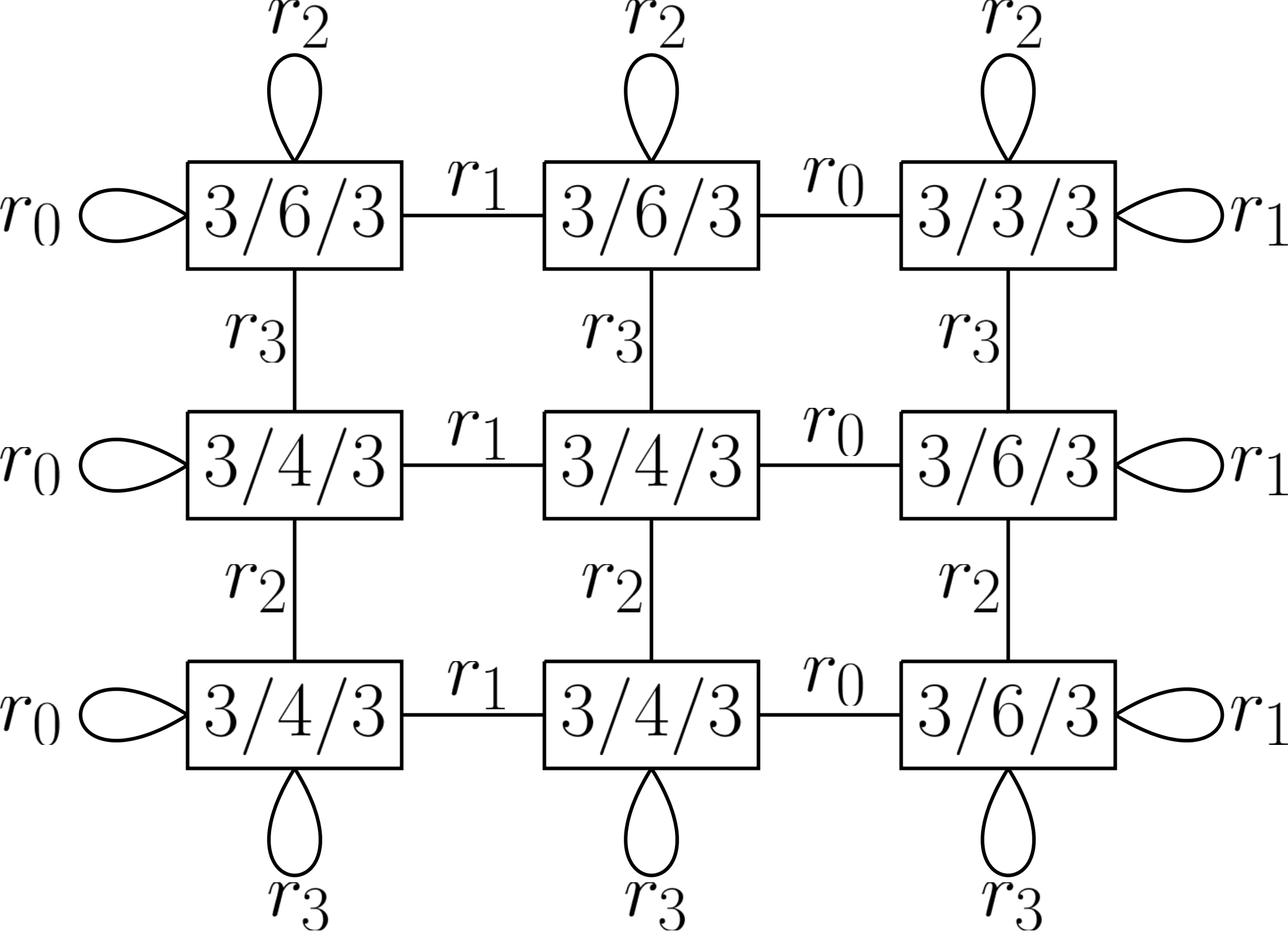}
        \vspace{.2 cm}
        \caption*{The duality isomorphism is given by reflecting about the line $y=x$.}
    \end{subfigure}
    \caption{Two self-dual fault-tolerant cluster states  which optimize locality of the decoder graph or the graph state, respectively, and their final covers. The first represents the triamond cluster state studied in \cite{nickerson2018measurement}.  The second is new, and was generated by our search.  It corresponds to a fault-tolerant degree-$3$ graph state.}
\label{cubic_vs_diamond_2}
\end{figure}

Mirroring the $2$D case, these two parameters experience a tradeoff. For example, any spanning surface in a $3$D tiling must itself obey the zero-curvature of a $2$D tiling.  As described in Section \ref{simple}, we expect the performance of fault-tolerant cluster states to vary based on the relative strength of different noise sources.  In Figures \ref{cubic_vs_diamond_1} and \ref{cubic_vs_diamond_2}, we depict four highly symmetric self-dual fault-tolerant cluster states spanning this tradeoff.  While the first three are well-known, the last one generates an optimally local (regular) graph state, and was found from our search.  In particular, the underlying graph state requires only degree-$3$ connectivity.  

The cubic, diamond, and triamond \cite{nickerson2018measurement} lattices are carried by special self-dual tilings.  Up to a maximal group of symmetries, they only have one type of volume, face, edge, and vertex, although only the cubic tiling has a single type of flag.  It has been conjectured that, among a subclass of tilings called natural, these are the only self-dual tilings with this property \cite{friedrichs2003three}.\footnote{However, natural tilings play no special role for our purposes, and non-natural tilings with only one type of volume, face, edge, and vertex have been constructed \cite{friedrichs2003semi}.  In fact, this construction also yields a degree-$3$ graph state, and can be realized as a gluing of $3$-cells in the tetrahedral-octahedral honeycomb.}  However, the degree-$3$ self-dual fault-tolerant cluster state described in Figure \ref{cubic_vs_diamond_1} is also highly symmetric: both the graph state and the decoder graph are regular.  In fact, up to a maximal group of symmetries, it has only two types of volumes, faces, edges, and vertices.\footnote{This can be determined by removing one of the $d+1$ reflection generators, and then counting distinct isomorphism classes of the connected components of the new $(d-1)$-dimensional symbol.}  Although the decoder graph is relatively dense, the graph state locality can make it robust to particular noise processes.

\begin{figure}[htb!]
\centering
\hspace{-.1cm}
\begin{subfigure}[b]{0.24\textwidth}
        \centering
        \includegraphics[width=\linewidth]{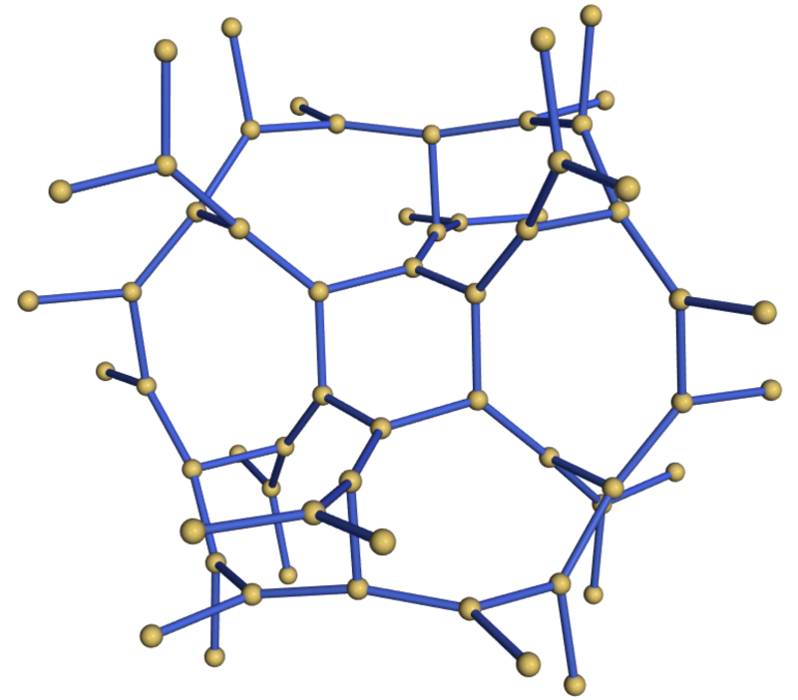}
    \end{subfigure}
    \hspace{.25cm}
    \begin{subfigure}[b]{0.21\textwidth}
        \centering
        \includegraphics[width=\linewidth]{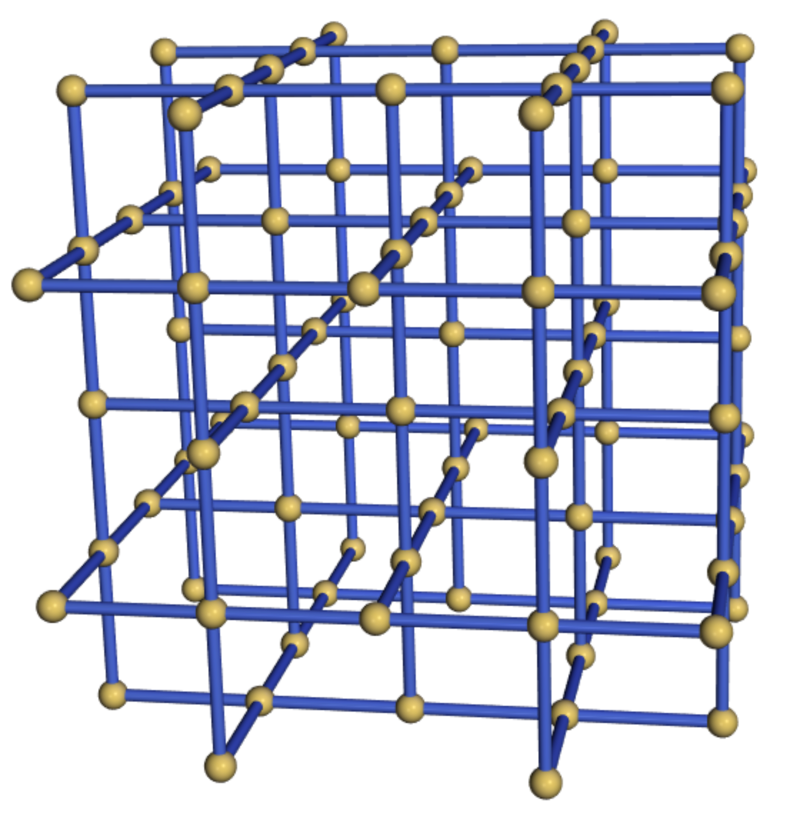}
    \end{subfigure}
    \hspace{.25cm}
    \begin{subfigure}[b]{0.21\textwidth}
        \centering
        \includegraphics[width=\linewidth]{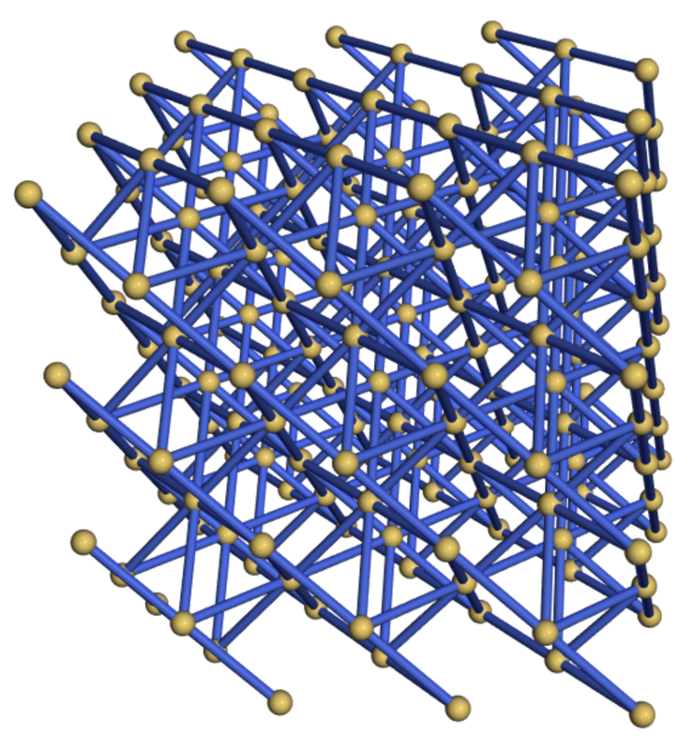}
    \end{subfigure}
    \hspace{.25cm}
    \begin{subfigure}[b]{0.22\textwidth}
        \centering
        \includegraphics[width=\linewidth]{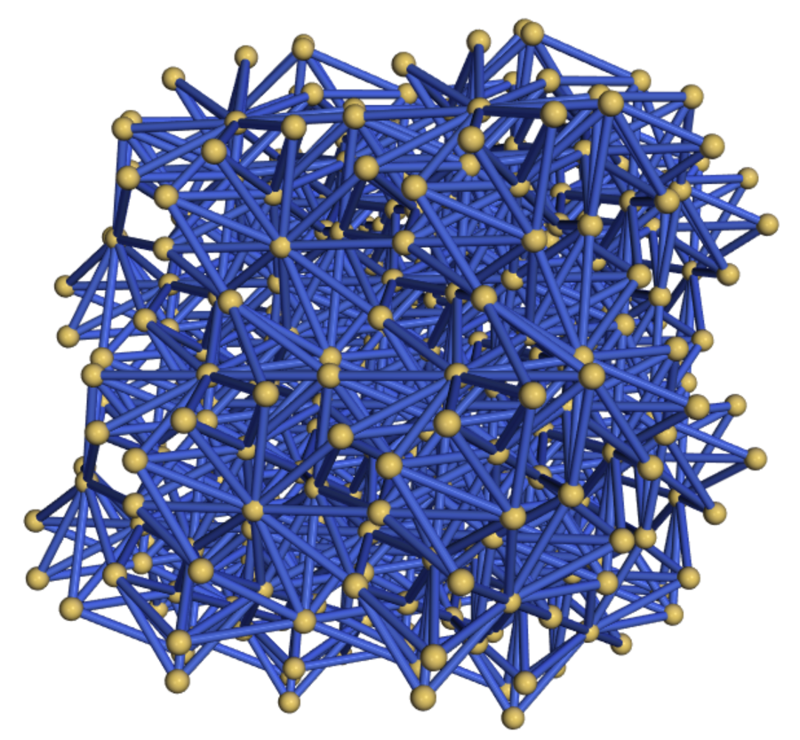}
    \end{subfigure}
    
    \hspace{-.1cm}
    \begin{subfigure}[b]{0.24\textwidth}
        \centering
        \includegraphics[width=\linewidth]{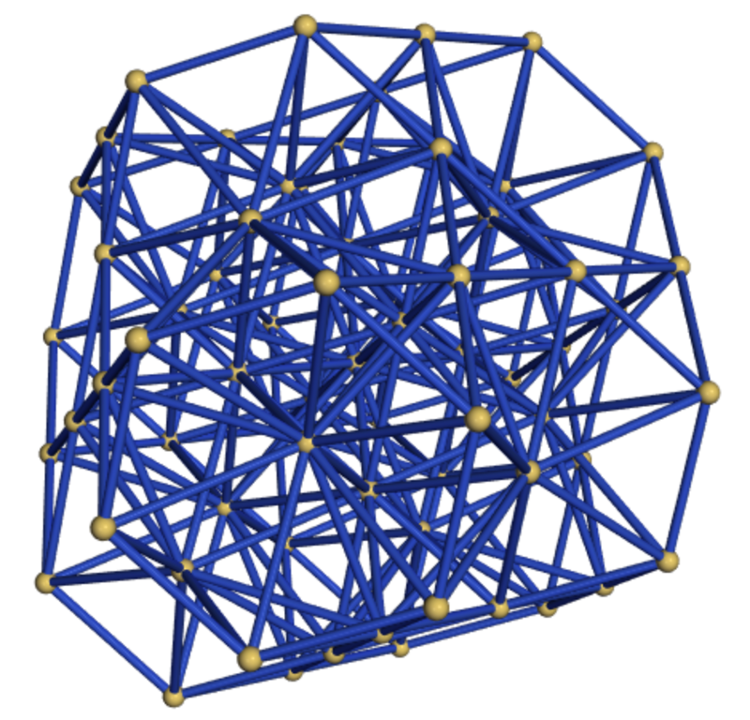}
        \caption{$3$-regular graph state, $12$-regular decoder graph.}
    \end{subfigure}
    \hspace{.25cm}
    \begin{subfigure}[b]{0.21\textwidth}
        \centering
        \includegraphics[width=\linewidth]{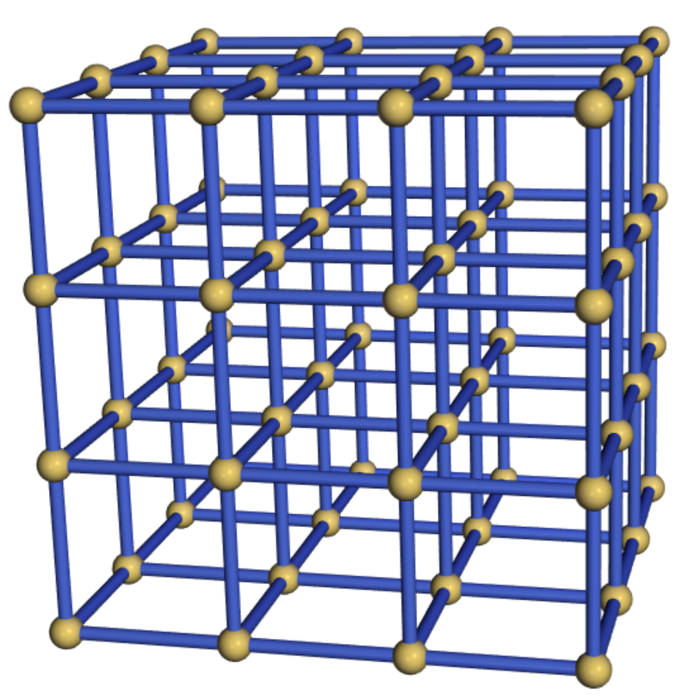}
        \caption{$4$-regular graph state, $6$-regular decoder graph.}
    \end{subfigure}
    \hspace{.25cm}
    \begin{subfigure}[b]{0.21\textwidth}
        \centering
        \includegraphics[width=\linewidth]{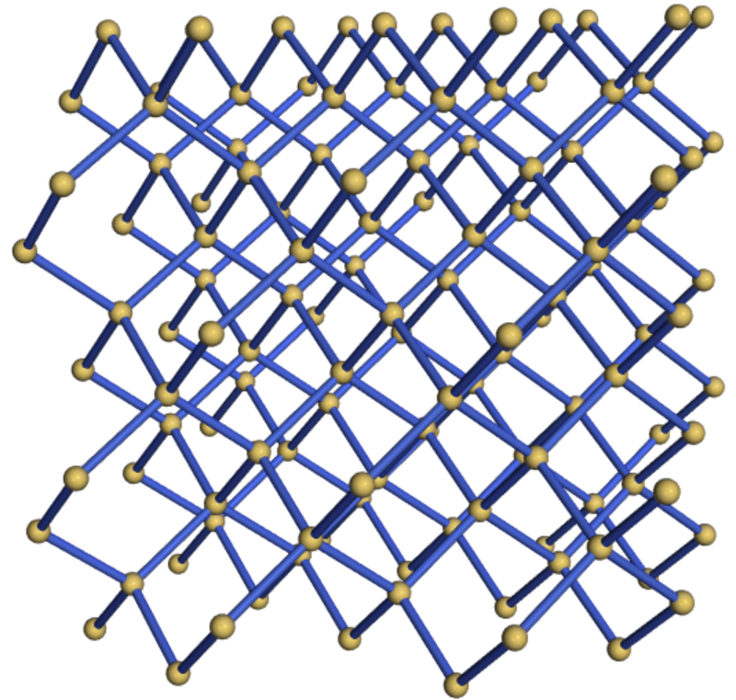}
        \caption{$6$-regular graph state, $4$-regular decoder graph.}
    \end{subfigure}
    \hspace{.25cm}
    \begin{subfigure}[b]{0.22\textwidth}
        \centering
        \includegraphics[width=\linewidth]{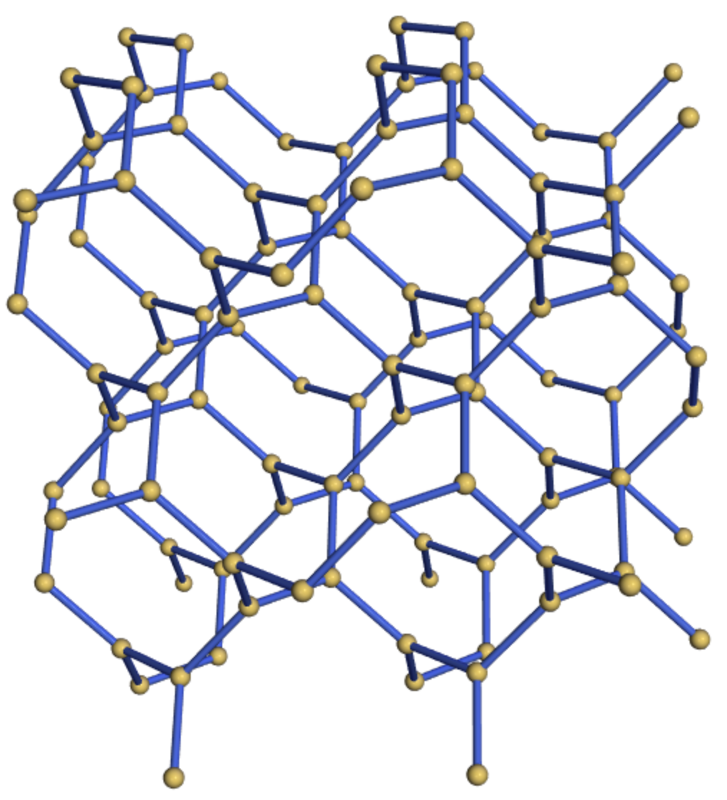}
        \caption{$10$-regular graph state, $3$-regular decoder graph.}
    \end{subfigure}
    \caption{Visualizing the tradeoff between the degree of a graph state (top) and the degree of its corresponding decoder graph (bottom).  We comment that locality often plays a role in mitigating errors that may not appear in a gate error model \cite{brown2019handling}, with recent work constructing a $2$D quantum code on a degree-$3$ graph for this purpose \cite{chamberland2020topological, chamberland2020triangular}.}
\label{degree_tradeoff}
\end{figure}

\begin{figure}[htb!]
\centering
\begin{subfigure}[b]{0.35\textwidth}
        \centering
        \includegraphics[width=\linewidth]{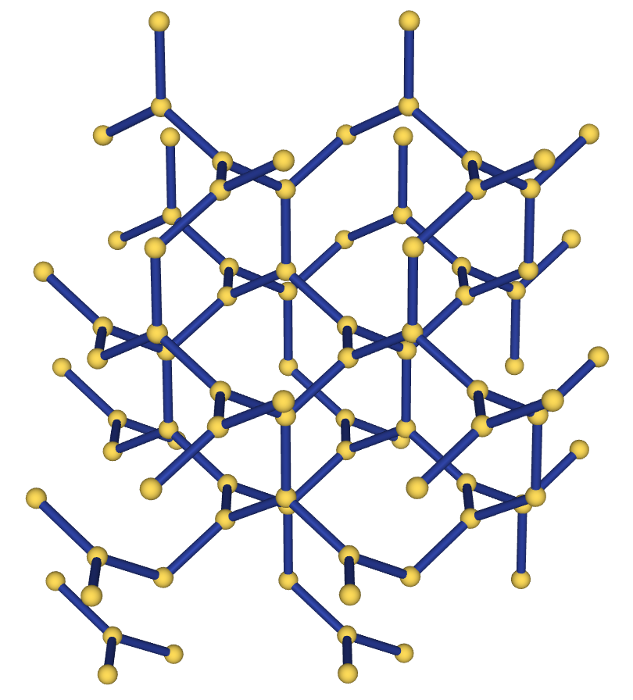}
    \end{subfigure}
    \hspace{1cm}
    \begin{subfigure}[b]{0.35\textwidth}
        \centering
        \includegraphics[width=\linewidth]{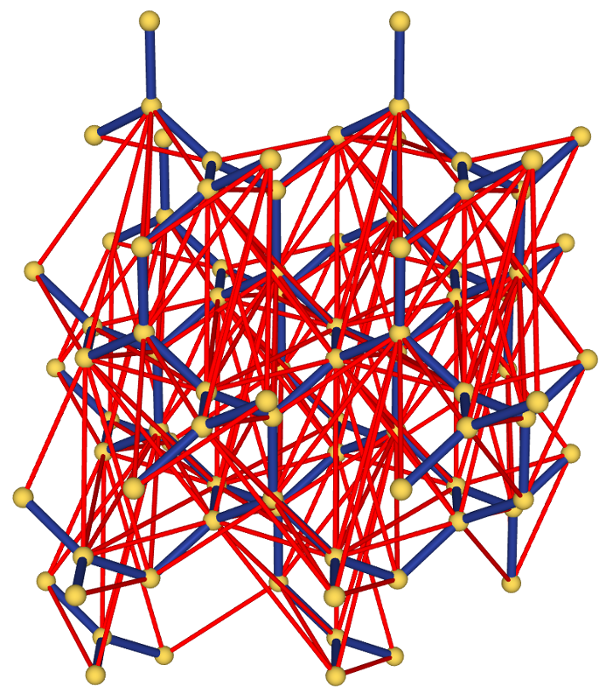}
    \end{subfigure}
\caption{The decoder graph of the triamond lattice both with (right) and without (left) correlated errors.  Red bonds correspond to $X$-type errors during the construction of the state.  One must carefully balance the robustness of error-correction against the cost of construction (figures rendered in VESTA \cite{momma2011vesta}).}
\label{decoder_graphs}
\end{figure}

\subsubsection{Families of degree-3, -4, and -5 self-dual fault-tolerant cluster states}\label{numerical_search}

Next, we describe how simple heuristics, combined with the prescription in Section \ref{constructions}, can yield a wealth of robust self-dual fault-tolerant cluster state candidates.  Although one should ideally choose a cluster state that is optimal for a particular noise model, we focus on only $3$-, $4$-, and $5$-regular graph state families for simplicity.  Very broadly, we would expect local families to perform better in the presence of stronger correlated errors, and denser families to perform better in the presence of stronger independent errors.  In particular, this search yielded the aforementioned optimally local example, among many others.

To begin the search, we must write down a family of valid, regular, self-dual symbols corresponding to $n$-regular graph states.  This can be tricky given the many conditions they must satisfy.  However, here is one prescription for doing so.
\begin{itemize}
\itemsep0em
    \item[($i$)] For a desired $n$, choose any $k|n$ or $k = 2n$.  
    \item[($ii$)] Form a $k \times k$ matrix of nodes defining the vertex set of $S$.
    \item[($iii$)] Indexing the nodes from $(x,y)=(1,1)$ in the bottom-left corner to $(k,k)$ in the top-right corner, insert the following edges.
    \begin{itemize}
    \itemsep0em
        \item[($a$)] For every $i \neq k$, connect $(i,j)$ and $(i+1,j)$ with an $r_0$-edge if $i$ is even, or with an $r_1$-edge if $i$ is odd.
        \item[($b$)] For every $j \neq k$, connect $(i,j)$ and $(i,j+1)$ with an $r_3$-edge if $j$ is even, or with an $r_2$-edge if $j$ is odd.
        \item[($c$)] If $k$ is odd, complete the symbol with self-loops along the boundary.  If $k=2n$, complete the symbol with periodic boundaries.  If $k \neq 2n$ is even, complete the symbol with either self-loops along the boundary or with periodic boundaries.
    \end{itemize}
    \item[($iv$)] Define $m_{01}(s) = m_{23}(s) = n$ for all $s \in S$.
    \item[($v$)] For every $\langle r_1,r_2 \rangle$-orbit lying either below or intersecting the diagonal, and for all $s$ in the orbit, choose $m_{12}(s) \in \{2,4,6,8,12\}$ if the orbit size is greater than one, else choose $m_{12}(s) \in \{1,2,3,4,6\}$. This limitation is imposed by the crystallographic restrictions on rotational symmetries.
    \item[$(vi)$] Define the remaining $m_{12}(s_{ij}) = m_{12}(s_{ji})$.
\end{itemize}
One example with $k=n=3$ appears in Figure \ref{cubic_vs_diamond_1}. These are valid regular symbols, as for $|i-j|>1$, each $\langle r_i, r_j \rangle$-orbit forms a cycle of length $1,2,$ or $4$.  If the choice of the $m_{12}(s)$ yields a tiling of $3$-space, then the underlying graph state will be an $(m_{01},m_{23})$-biregular graph. Thus, if this algorithm succeeds, it yields an $n$-regular graph state.  By construction, the symbol is self-dual, with reflection about $y=x$ inducing the duality isomorphism.  From this definition, $S$ and its associated $\mathcal{R}$-action are fixed.  The only free parameters are the $m_{12}(s)$, which lie in restricted sets, forming $5^{\binom{\lceil \frac{k}{2}\rceil + 1}{2}}$ potential configurations to check for flatness.\footnote{Small changes to this prescription, like mixed boundary conditions or exchanging $r_0\leftrightarrow r_1$ and $r_1 \leftrightarrow r_2$ also yield valid symbols.  We make specific choices to limit our search.}  

\begin{figure}[htb!]
\centering
    \begin{subfigure}[b]{0.45\textwidth}
        \centering
        \includegraphics[width=\linewidth]{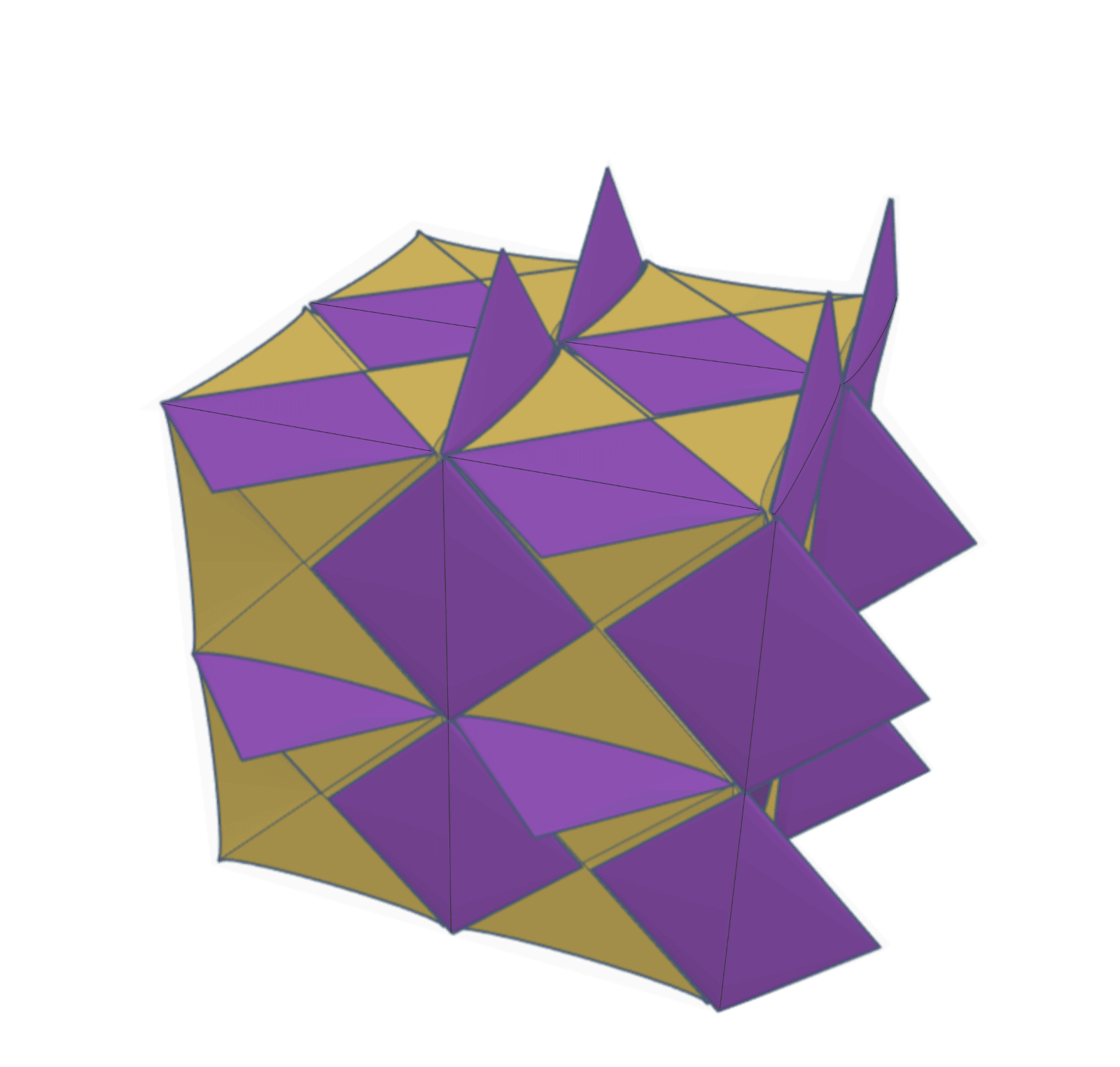}
        \vspace{.01cm}
        \caption{The self-dual tiling corresponding to $m_{12}$-vector $(6$ $4$ $4$ $2)$.  It consists of interwoven (purple) $4$-sided cones and $24$-sided (yellow) cube-like tiles (drawn in Tinkercad \cite{tinkercad}). In particular, the purple objects are four glued triangles with a thin interior.}
    \end{subfigure}%
    \hspace{.5cm}
    \begin{subfigure}[b]{0.4\textwidth}
        \centering
        \includegraphics[width=\linewidth]{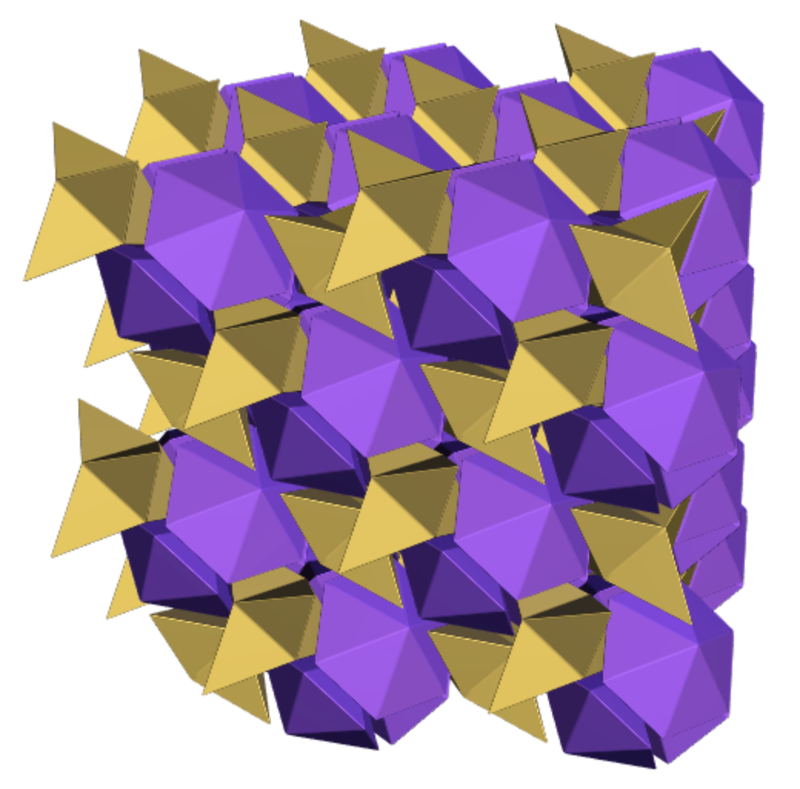}
        \caption{The self-dual tiling corresponding to $m_{12}$-vector $(4$ $6$ $6$ $3)$.  It consists of interwoven (purple) $12$-sided hexagonal bipyramids and $12$-sided star-convex polyhedra.}
    \end{subfigure}
    
    \caption{The two self-dual tilings found by searching $(n,k) = (3,3)$ according to Section \ref{numerical_search}.  To obtain the corresponding graph states, qubits are placed on faces and edges, with an edge between them if they are adjacent (an edge within a face) in the tiling.  For example, the purple objects on the left have ten associated qubits: four face qubits and six edge qubits.}
\label{33s}
\end{figure}

Gavrog is an open-source software built explicitly for analyzing tilings \cite{Gavrog, delgado2003identification, delgado2003barycentric, delgado2005crystal, delgado2005equilibrium}.  Using its internal methods (and more generally GAP \cite{GAP4}), which apply an optimized version of the construction in Section \ref{constructions}, we search over these symbols to identify self-dual fault-tolerant cluster states with $n$-regular underlying graph states.  Among the $\gg \hspace{-0.08cm}10^9$ candidates we searched over, the vast majority generated by $k=10$, we found $475$ unique symbols realizing self-dual cluster states. The search results are summarized in Table \ref{main_table}, which are complete for all but $(n,k) = (5,10)$.  Rendering these can be difficult, depending on the complexity of the symmetry group.  However, to fully translate from algebraic representations to concrete tilings, we graph the two examples described by $(n,k) = (3,3)$ in Figure \ref{33s}.

\begin{table}[htb!]
\resizebox{\linewidth}{!}{
\begin{tabular}{|c|c|c|c|c|}
\hline
\textbf{(\textit{n,k})} & \textbf{Periodic?} & \textbf{$\#$(Candidates)} & \textbf{$\#$(Cluster States)} & \textbf{Example $(m_{12})$-vectors}                                                                                                                          \\ \hhline{|=|=|=|=|=|}
$(3,1)$          & N/A               & $5$                     & $0$                         & $\varnothing$                                                                                                                                                 \\ \hline
$(3,3)$          & No                 & $125$                   & $2$                         & $(6$ $4$ $4$ $2); (4$ $6$ $6$ $3)$                                                                                                                                  \\ \hline
$(3,6)$          & Yes                & $15625$                 & $21$                        & $(2$ $4$ $4$ $4$ $6$ $6$ $4$ $6$ $6); (6$ $4$ $8$ $4$ $2$ $4$ $8$ $4$ $4); \ldots$                                                \\ \hline
$(4,1)$          & N/A                & $5$                     & $1$                         & $(3)$\footnote{The cubic tiling.}                                                                                                                                                         \\ \hline
$(4,2)$          & Yes                & $5$                     & $0$                         & $\varnothing$                                                                                                                                                 \\ \hline
$(4,2)$          & No                 & $5$                     & $0$                         & $\varnothing$                                                                                                                                                 \\ \hline
$(4,4)$          & Yes                & $125$                   & $0$                         & $\varnothing$                                                                                                                                                 \\ \hline
$(4,4)$          & No                 & $125$                   & $1$                         & $(2$ $2$ $2$ $3)$\footnote{With $r_0 \leftrightarrow r_1$ and $r_2 \leftrightarrow r_3$, this realizes the double-edge cubic cluster state in \cite{nickerson2018measurement}.}                                                                                                                                                 \\ \hline
$(4,8)$          & Yes                & $9765625$               & $376$                       & \begin{tabular}[c]{@{}c@{}}(2 2 2 2 2 2 4 4 2 4 6 6 2 4 6 6); \\ (6 2 6 2 2 4 6 4 6 6 4 2 2 4 2 4), \ldots\end{tabular}                      \\ \hline
$(5,1)$          & N/A                & $5$                     & $0$                         & $\varnothing$                                                                                                                                                 \\ \hline
$(5,5)$          & No                 & $15625$                 & $4$                         & $(6$ $2$ $4$ $2$ $2$ $2$ $4$ $2$ $2);  (4$ $2$ $6$ $2$ $2$ $2$ $6$ $2$ $3);$ \ldots                                                         \\ \hline
$(5,10)$         & Yes                & $\approx 3\times10^{10}$  & $70$                        & \begin{tabular}[c]{@{}c@{}}(2 2 2 2 2 2 2 4 4 4 2 4 2 4 4 2 4 4 4 2 2 4 4 2 4);
 \\ (4 2 2 4 2 2 4 4 4 2 2 4 4 2 4 4 4 2 2 4 2 2 4 4 4); \ldots\end{tabular} \\ \hline
\end{tabular}}
\caption{A brute-force search for self-dual cluster states with $n$-regular underlying graph states.  The nodes of the $k \times k$ matrix are ordered alphabetically from left-to-right, and then from bottom-to-top.  The components of the $m_{12}$-vectors represent $m_{12}$ for each orbit $\langle r_1,r_2\rangle\cdot s$ with respect to this ordering, determining the symbol.  When possible, a pair of random examples are chosen from each search.} \label{main_table}
\end{table}

In summary, we can search for fault-tolerant cluster states with desired properties by defining a heuristic on graphs representing symbols that satisfy those properties. Then, one can convert this to a computational problem on groups, and identify candidates whose geometric realization gives the desired cluster state.  Finally, one can construct this cluster state from its graphical representation, and study robustness of the resulting state.  

Although we stop at degree-$5$, we could continue this search to higher degree graph states.  However, at higher degrees, one might also look for more symmetric tilings using other heuristics.  Alternatively, one could search for low degree decoder graphs, rather than low degree graph states, although this constraint is not as straightforward to impose on a symbol.\footnote{For smaller symbols, one can determine vertex degrees by solving systems of equations involving the Euler characteristic.  The maximum $m_{12}$ of the symbol also provides a simple lower-bound on the minimum degree.}

\section{Benchmarking crystalline cluster states} \label{cs_numerics}

Crystals are of fundamental importance in chemistry, and so there are also an abundance of crystals carried by self-dual tilings already in the literature.  In addition to building fault-tolerant cluster states from symbols, we can search over existing databases for promising candidates.  One example is the RCSR database, which has indexed $38$ self-dual tilings \cite{o2008reticular}.  Each lattice is labeled with a three-letter identifier, which we adopt to refer to each tiling.\footnote{From a practical perspective, we focus on benchmarking lattices with smaller unit cells (see Appendix \ref{others}), as there can be a large gap in the computational expense of different lattice simulations.}

In \cite{nickerson2018measurement}, the cubic (pcu), diamond (dia), double-edged cubic, and triamond (srs) lattices were benchmarked with a union-find decoder \cite{delfosse2017almost} in weighted phenomenological and weighted erasure error models \cite{barrett2010fault}.  Although the union-find decoder is highly efficient, it is often suboptimal when compared with the less efficient minimum-weight perfect matching (MWPM) decoder \cite{edmonds2009paths, kolmogorov2009blossom, huang2020fault}.  In this section, we extend this analysis in three ways.  First, we benchmark additional examples from the aforementioned set of self-dual fault-tolerant cluster states, along with the optimally local example constructed in Section \ref{local}.  In total, we benchmarked three degree-$3$ lattices drawn from our search; the one included was the top-performer among the three.  Second, we perform these simulations in a circuit-level noise model that captures the tradeoff between robust error-correction and cluster state preparation.  Our simulations span four different relative parameter regimes in order to better characterize the effects of this tradeoff. Finally, we use the high-performance MWPM decoder to obtain a closer estimate of the maximum thresholds achievable using this approach.

In fact, we can say more using a well-established connection between thresholds for quantum error-correction and statistical mechanical models.  For each of these crystal lattices with cell complex $\mathcal{C}$, we can associate a $\mathbb{Z}_2$-lattice gauge theory with quenched disorder mirroring the random-plaquette $\mathbb{Z}_2$-gauge model on the cubic lattice \cite{Dennis:2002, wang2003confinement, ohno2004phase}. The corresponding Hamiltonian is given by assigning spins to edges and defining $$H = -\sum\limits_{f \in C_2}\tau_f \prod\limits_{e \in f} S_e$$ where $\tau_f \in\{\pm 1\}$ is a quenched random variable and the local $\mathbb{Z}_2$-symmetry is obtained by flipping the spins on all edges incident to a vertex.  The optimal phenomenological threshold for each lattice then corresponds to a phase transition of its corresponding model along the Nishimori line \cite{Dennis:2002, nishimori1986geometry, kubica2018three, chubb2018statistical}.  

The MWPM decoder selects, for each syndrome, an error class with the most likely individual error.  While this is generally suboptimal, it can be viewed from the statistical mechanics perspective as the decoder that chooses a recovery string minimizing the free energy at zero temperature.  Correspondingly, up to the relative weighting of edges, our dephasing thresholds can be interpreted as estimates for zero temperature phase transitions for lattice gauge theories with disorder defined on a variety of crystal lattices, whose behavior may be of independent interest. The circuit-level thresholds can also be interpreted similarly according to extensions of this model to correlated errors \cite{chubb2018statistical}.  However, we emphasize that these should only be regarded as rough estimates, as degeneracy among minimum-weight strings for a syndrome can cause such Monte Carlo estimates to be imprecise \cite{stace2010error}.

To simulate our noise model on a particular lattice, we first embed the lattice on a $3$-torus of linear-size $L$.  We choose a translation-invariant ordering of CZ gates in each face, and then form an augmented $1$-skeleton with edges corresponding to $Z$-error gate failures on the primal qubit support and $X$-error gate failures on the dual qubit support (see Appendix \ref{others}).  For this initial comparative benchmarking, we make this choice arbitrarily. On the one hand, this gate ordering is likely sub-optimal.  On the other, we do not benchmark the dual lattice, for which the gate ordering may be worse.  The effects of this choice will grow with $p_X$, and we leave further optimization to future work.  

For each edge $e$, let $z_e$ ($x_e$) be the number of $Z$-type ($X$-type) gate failures that could excite $e$.  Then the probability of exciting edge $e$ with a $Z$-type gate failure is given by $$P_{e,Z} \coloneqq \sum\limits_{k\text{ odd}}^{z_e} \binom{z_e}{k} p_Z^k(1-p_Z)^{(z_e-k)} = \frac{1}{2}(1-(1-2p_Z)^{z_e}),$$ and the probability of exciting that edge due to an $X$-type gate failure $P_{e,X}$ is given analogously.  The total probability of exciting an edge $e$ is then given by $$P_e\coloneqq \sum\limits_{\vect{v} \in \{0,1\}^3: |\vect{v}|\text{ odd}} p_m^{v_1}(1-p_m)^{(1-v_1)}P_{e,Z}^{v_2}(1-P_{e,Z})^{(1-v_2)}P_{e,X}^{v_3}(1-P_{e,X})^{(1-v_3)}.$$
In each trial, we excite edges according to $P_e$ and form the resulting syndrome.  To choose a recovery string, we run MWPM \cite{edmonds2009paths,kolmogorov2009blossom,leculier1999,leculier2002} with edges weighted according to $-\ln(P_e)$ \cite{stace2009thresholds, wang2011surface, raussendorf2007fault, raussendorf2007topological}.  We declare failure if we have introduced a homologically nontrivial cycle after recovery. For each data point we ran $10^6$ trials, adding up to approximately $40,000$ computational hours in total, and observed good agreement with previous benchmarks \cite{wang2003confinement, barrett2010fault}.  

We reemphasize that this a non-standard error model which is very severe.  In particular, the threshold for the $3$D cluster state in a standard depolarizing model is $\approx 0.75\%$ \cite{raussendorf2007fault}.  Consequently, these numbers should be used only comparatively: while our error model allows us to more efficiently simulate these lattices, we leave benchmarking in standard models to future work.

We simulate seven different lattices, referring to them by their three-letter RCSR identifiers.\footnote{We omit the double-edge cubic lattice, which can be viewed as a concatenation of the $3$D cluster state with a phase-flip repetition code, and was studied in \cite{stephens2013high, rudolph2017optimistic, nickerson2018measurement}.} We denote the degree-$3$ cluster state by \emph{bst} for bipyramid-star with triangular faces, and because it is best with respect to locality. Our simulation sizes span $L=4$ through $L=12$, depending on the lattice.  Although we do observe some finite size effects, it is important to note that several of the lattices have many vertices in their fundamental cell, so that their distance is much greater than $L$.  The additional three lattices were chosen from the RCSR database to validate our hypotheses on the effects of noise; we expect there are many more robust candidates.  Renderings of these new lattices, along with a short summary of the self-dual tilings available in the database, are given in Appendix \ref{others}.  In addition, we run four noise models in which $p_Z, p_X,$ and $p_m$ are weighted differently.  The resulting threshold estimates are summarized in Table \ref{thresholds}, and the raw data is available in Appendix \ref{data}.

\begin{table}[htb!]
\hspace{0.85cm}
\begin{tabular}{|c|c|c|c|c|c|c|c|}
\hline
                      & \textbf{bst} & \textbf{pcu} & \textbf{cdq} & \textbf{hms} & \textbf{dia} & \textbf{ctn} & \textbf{srs} \\ \hhline{|=|=|=|=|=|=|=|=|} $p_Z$ (Union-Find \cite{nickerson2018measurement})&N/A& $0.65\%$& N/A& N/A &$0.87\%$ &N/A & $0.95\%$ \\ \hline
$p_Z$                 & $0.35\%$     & $0.76\%$     & $0.83\%$     & $0.94\%$     & $1.01\%$     & $1.02\%$     & $1.16\%$     \\ \hline
$p_Z = 10p_X = 10p_m$ & $0.32\%$     & $0.66\%$     & $0.71\%$     & $0.80\%$     & $0.81\%$     & $0.79\%$     & $0.75\%$     \\ \hline
$p_Z = p_X = p_m$     & $0.18\%$     & $0.32\%$     & $0.32\%$     & $0.33\%$     & $0.31\%$     & $0.31\%$     & $0.19\%$     \\ \hline
$p_X = 10p_Z = 10p_m$ & $0.47\%$     & $0.65\%$     & $0.58\%$     & $0.56\%$     & $0.48\%$     & $0.51\%$     & $0.25\%$     \\ \hline
\end{tabular}
\caption{Threshold estimates for seven different lattices in four different noise regimes.  The graph state (decoder graph) degree increases (decreases) from left to right. The relative impact of $Z$-type ($X$-type) errors decreases (increases) from top to bottom. We include the corresponding union-find thresholds from \cite{nickerson2018measurement} for comparison.  The total error-rate $p$ is defined as the largest individual error-rate among $p_Z, p_X,$ and $p_m$.  The lower thresholds reflect the harsh noise model due to our definition of $p_X,p_Z$ and the the double sampling of errors for each two-qubit gate described in Section \ref{simple}.  By comparison, the $3$D cluster state has a standard depolarizing threshold of $\approx 0.75\%$ \cite{raussendorf2007fault}.}\label{thresholds}
\end{table}

Supporting the discussion in Section \ref{simple}, the performance of each cluster state generally reflects the tradeoff between the degree of the graph state and of the decoder graph with respect to the relative strength of $X$-errors and $Z$-errors.  One can see the peak performer moving from right to left as correlated errors become more prevalent, and so we would expect the degree-$3$ cluster state to be a top performer in an entirely correlated error model.  However, as cluster states are constructed entirely from CZ gates, we believe that the noise regime in which $p_Z$ dominates but $p_X$ and $p_m$ remain significant warrants the most interest \cite{PhysRevA.78.052331,aliferis2009fault, stephens2013high}.  Indeed, this reflects the growing effort to take advantage of biased models \cite{Tuckett:2018,tuckett2018tailoring,tuckett2019fault, xu2018high, puri2019bias, li20192d, PhysRevA.92.062309,PhysRevA.87.032310}.  However, even when noise is symmetric, there are several candidates competitive with the surface code, and it appears that some lattices are generally more robust than others.

Unsurprisingly, the MWPM decoder performs better than the union-find decoder.  While this improvement is noticeable across all three compared lattices, the difference in performance only ranges from $\approx0.1\%-0.2\%$.  Mirroring the results in \cite{delfosse2017almost} on the cubic lattice, we interpret this as positive evidence that the union-find decoder may be a promising alternative to MWPM, given its essentially linear-time efficiency. However, it could be the case that this performance gap increases in the presence of correlated errors, for which the matching decoder can be tailored.

One outlier for symmetric and $Z$-biased noise is the degree-$3$ cluster state.  Although the trend suggests that it may be a top-performer in a fully $X$-type error model, for more realistic models its thresholds are lower.  Generally, achieving fault-tolerance with degree-$3$ connectivity or $3$-local interactions comes with difficult assumptions, or reductions in threshold.  However, the sacrifice is often worth the cost, as locality is more valuable than it might appear at the level of gate errors \cite{chamberland2020topological, chamberland2020triangular, bravyi2013subsystem}. Consequently, we believe that the performance of the degree-$3$ cluster state, whose threshold is only a factor of $2$ away from the top-performing surface code in our simplified error model, is very encouraging.

\section{Conclusions}\label{final_conclusions}

Although there are many similarities between them, fault-tolerant MBQC has certain freedoms that CBQC with error-correcting codes lacks.  We can take advantage of these freedoms to build more robust fault-tolerance protocols beyond those based on error-correcting codes.  At the circuit level, this freedom manifests as passing logical information between qubits which are regularly measured and reinitialized, without a distinction between data and ancilla.  This stands in stark contrast to the more restrictive CBCQ fault-tolerance model, where logical information is held statically within the code qubits until a final measurement.

Using combinatorial tiling theory, we gave a prescription for building fault-tolerant cluster states.  Different types of errors have different effects, and one must balance locality of both the decoder graph and the graph state according to different relative error strengths and architectural considerations. We also expect many of these cluster states to be particularly resilient to measurement errors, as their relative error strength does not scale with the construction cost of the underlying graph state. Finally, we analyzed several examples of fault-tolerant cluster states in a variety of noise models emulating circuit-level errors, and benchmarked them with the high-performing MWPM decoder.  These results draw closer to optimal thresholds for topological MBQC, and relate to statistical mechanical models defined on different crystal structures.

There are several immediate extensions worth exploring.  It would be interesting to consider more varied noise models; for example, we expect local graph state constructions to perform well in the presence of correlated noise.  Additionally, this framework is naturally robust to leakage, as no qubit is long-lived.  In comparison, thresholds of quantum error-correcting codes require additional teleportation gadgets which lower thresholds \cite{suchara2015leakage}.  Further understanding these cluster states' resilience to hybridized loss and Pauli models would also be of interest \cite{nickerson2018measurement, varnava2007loss}.  While we have tested only a few examples to take a snapshot of their performance, a more thorough investigation of promising tilings and their robustness is warranted. 

There are also several broader avenues worth exploring.  Following \cite{nickerson2018measurement}, we have restricted ourselves to geometrically defined cluster states.  While this immediately extends to a universal fault-tolerance scheme, it is not required by the underlying homological problem of preserving distinguished primal and dual symmetries.  It would be interesting to study non-foliated codes that are not geometrically defined, although it may be difficult to build abstract sparse chain complexes with the desired properties.

To this end, in the prescription for building fault-tolerant cluster states from simplicial structures on orbifolds, we discarded those whose symmetries were not compatible with a Euclidean metric.  However, for computational devices with all-to-all connectivity, there is no need to restrict ourselves to periodic tilings of Euclidean $3$-space.  Candidates with a $\mathbb{Z}^3$ subgroup in their fundamental group are particularly interesting, as this subgroup serves the role of growing the cluster state in the Euclidean case. One could additionally look at generalized notions of periodicity which have less restrictive symmetries.

There has also been significant work aimed at constructing high rate codes in hyperbolic geometries \cite{breuckmann2017hyperbolic,breuckmann2016constructions}.  Often, the encoding rate can be described as a function of the combinatorics of an associated regular tiling \cite{breuckmann2020single}.  It would be interesting to try to algorithmically construct non-regular self-dual tilings with higher rates, although recognizing symbols corresponding to tilings with negative curvature would likely prove more difficult.

Finally, there is no fundamental barrier to extending this approach to build fault-tolerant cluster states from tilings of higher dimensional spaces.  For example, a fault-tolerant cluster state built from a $5$D cubic lattice will inherit string-like excitations as a foliation of the $4$D toric code.  This comes with a number of favorable properties, including high phenomenological thresholds and local decoders, stemming from the $4$D toric code's thermodynamic stability \cite{duivenvoorden2018renormalization, breuckmann2016local, kubica2018cellular}.  However, it also requires higher degree graph states due to the $4$D toric code's weight-$6$ stabilizers, which are damaging at the circuit-level \cite{duivenvoorden2018renormalization}.  Analogous to dimension-$3$, could we use a similar prescription in higher dimensions to construct low-degree fault-tolerant cluster states that exhibit string-like excitations?  We could also take the opposite approach, and try minimizing the degree of the decoder graph by constructing a higher-dimensional analogue of the triamond lattice, similar to the hypercubic lattice.  Understanding this tradeoff in higher dimensions, which allows for richer symmetry, would be of significant interest.

\section{Acknowledgments}
The authors are indebted to Olaf Delgado-Friedrichs for various personal communications about tilings and his open-source software package, Gavrog.  They are also grateful to Austin Daniel, Akimasa Miyake, Naomi Nickerson, and anonymous reviewers for their helpful comments.  They thank Dripto Debroy, Shilin Huang, and Muyuan Li for their assistance in running numerical simulations.  The computational resources for simulations were provided by the Duke Computing Cluster.  This research was supported in part by the NSF STAQ project (1717523), the ODNI/IARPA LogiQ program through ARO grant (W911NF-16-1-0082), ARO MURI (W911NF-16-1-0349 and W911NF-18-1-0218), and EPiQC -- an NSF Expedition in Computing (1730104).

\bibliographystyle{unsrtnat}
\bibliography{bibliography}

\begin{thebibliography}{112}
\providecommand{\natexlab}[1]{#1}
\providecommand{\url}[1]{\texttt{#1}}
\expandafter\ifx\csname urlstyle\endcsname\relax
  \providecommand{\doi}[1]{doi: #1}\else
  \providecommand{\doi}{doi: \begingroup \urlstyle{rm}\Url}\fi

\bibitem[Aliferis et~al.(2006)Aliferis, Gottesman, and Preskill]{Aliferis:2006}
Panos Aliferis, Daniel Gottesman, and John Preskill.
\newblock Quantum accuracy threshold for concatenated distance-3 codes.
\newblock \emph{Quantum Information \& Computation}, 6\penalty0 (2):\penalty0
  97--165, 2006.

\bibitem[Knill et~al.(1996)Knill, Laflamme, and Zurek]{Knill:1996b}
Emanuel Knill, Raymond Laflamme, and W~Zurek.
\newblock Threshold accuracy for quantum computation.
\newblock \emph{arXiv preprint quant-ph/9610011}, 1996.

\bibitem[Aharonov and Ben-Or(1997)]{Aharonov:1997}
Dorit Aharonov and Michael Ben-Or.
\newblock Fault-tolerant quantum computation with constant error.
\newblock In \emph{Proceedings of the Twenty-ninth Annual ACM Symposium on
  Theory of Computing}, pages 176--188. ACM, 1997.
\newblock \doi{https://doi.org/10.1145/258533.258579}.

\bibitem[Terhal and Burkard(2005)]{terhal2005fault}
Barbara~M Terhal and Guido Burkard.
\newblock Fault-tolerant quantum computation for local non-markovian noise.
\newblock \emph{Physical Review A}, 71\penalty0 (1):\penalty0 012336, 2005.
\newblock \doi{https://doi.org/10.1103/physreva.71.012336}.

\bibitem[Aliferis and Terhal(2007)]{aliferis2005fault}
Panos Aliferis and Barbara~M Terhal.
\newblock Fault-tolerant quantum computation for local leakage faults.
\newblock \emph{Quantum Information \& Computation}, 7\penalty0 (4):\penalty0
  139--156, 2007.

\bibitem[Knill(2005)]{knill2005quantum}
Emanuel Knill.
\newblock Quantum computing with realistically noisy devices.
\newblock \emph{Nature}, 434\penalty0 (7029):\penalty0 39, 2005.
\newblock \doi{https://doi.org/10.1038/nature03350}.

\bibitem[Dennis et~al.(2002)Dennis, Kitaev, Landahl, and Preskill]{Dennis:2002}
Eric Dennis, Alexei Kitaev, Andrew Landahl, and John Preskill.
\newblock Topological quantum memory.
\newblock \emph{Journal of Mathematical Physics}, 43\penalty0 (9):\penalty0
  4452--4505, 2002.
\newblock \doi{https://doi.org/10.1063/1.1499754}.

\bibitem[Raussendorf and Harrington(2007)]{raussendorf2007fault}
Robert Raussendorf and Jim Harrington.
\newblock Fault-tolerant quantum computation with high threshold in two
  dimensions.
\newblock \emph{Physical Review Letters}, 98\penalty0 (19):\penalty0 190504,
  2007.
\newblock \doi{https://doi.org/10.1103/physrevlett.98.190504}.

\bibitem[Wang et~al.(2003)Wang, Harrington, and Preskill]{wang2003confinement}
Chenyang Wang, Jim Harrington, and John Preskill.
\newblock Confinement-higgs transition in a disordered gauge theory and the
  accuracy threshold for quantum memory.
\newblock \emph{Annals of Physics}, 303\penalty0 (1):\penalty0 31--58, 2003.
\newblock \doi{https://doi.org/10.1016/s0003-4916(02)00019-2}.

\bibitem[Ohno et~al.(2004)Ohno, Arakawa, Ichinose, and Matsui]{ohno2004phase}
Takuya Ohno, Gaku Arakawa, Ikuo Ichinose, and Tetsuo Matsui.
\newblock Phase structure of the random-plaquette z2 gauge model: accuracy
  threshold for a toric quantum memory.
\newblock \emph{Nuclear Physics B}, 697\penalty0 (3):\penalty0 462--480, 2004.
\newblock \doi{https://doi.org/10.1016/j.nuclphysb.2004.07.003}.

\bibitem[Wang et~al.(2010)Wang, Fowler, Stephens, and Hollenberg]{Wang:2009}
DS~Wang, AG~Fowler, AM~Stephens, and LCL Hollenberg.
\newblock Threshold error rates for the toric and planar codes.
\newblock \emph{Quantum Information \& Computation}, 10\penalty0 (5):\penalty0
  456--469, 2010.

\bibitem[Fowler et~al.(2009)Fowler, Stephens, and Groszkowski]{fowler2009high}
Austin~G Fowler, Ashley~M Stephens, and Peter Groszkowski.
\newblock High-threshold universal quantum computation on the surface code.
\newblock \emph{Physical Review A}, 80\penalty0 (5):\penalty0 052312, 2009.
\newblock \doi{https://doi.org/10.1103/physreva.80.052312}.

\bibitem[Fowler et~al.(2012)Fowler, Whiteside, and
  Hollenberg]{fowler2012towards}
Austin~G Fowler, Adam~C Whiteside, and Lloyd~CL Hollenberg.
\newblock Towards practical classical processing for the surface code.
\newblock \emph{Physical Review Letters}, 108\penalty0 (18):\penalty0 180501,
  2012.
\newblock \doi{https://doi.org/10.1103/physrevlett.108.180501}.

\bibitem[Suchara et~al.(2015{\natexlab{a}})Suchara, Cross, and
  Gambetta]{Suchara:2014}
Martin Suchara, Andrew~W Cross, and Jay~M Gambetta.
\newblock Leakage suppression in the toric code.
\newblock \emph{Quantum Information \& Computation}, 15\penalty0
  (11-12):\penalty0 997--1016, 2015{\natexlab{a}}.
\newblock \doi{https://doi.org/10.1109/isit.2015.7282629}.

\bibitem[Wang et~al.(2011)Wang, Fowler, and Hollenberg]{wang2011surface}
David~S Wang, Austin~G Fowler, and Lloyd~CL Hollenberg.
\newblock Surface code quantum computing with error rates over 1\%.
\newblock \emph{Physical Review A}, 83\penalty0 (2):\penalty0 020302, 2011.
\newblock \doi{https://doi.org/10.1103/physreva.83.020302}.

\bibitem[Stephens et~al.(2013)Stephens, Munro, and Nemoto]{stephens2013high}
Ashley~M Stephens, William~J Munro, and Kae Nemoto.
\newblock High-threshold topological quantum error correction against biased
  noise.
\newblock \emph{Physical Review A}, 88\penalty0 (6):\penalty0 060301, 2013.

\bibitem[Shor(1996)]{shor1996fault}
Peter~W Shor.
\newblock Fault-tolerant quantum computation.
\newblock In \emph{Foundations of Computer Science, 1996. Proceedings., 37th
  Annual Symposium on}, pages 56--65. IEEE, 1996.
\newblock \doi{https://doi.org/10.1109/sfcs.1996.548464}.

\bibitem[Shor(1995)]{Shor:1995b}
Peter~W Shor.
\newblock Scheme for reducing decoherence in quantum computer memory.
\newblock \emph{Physical Review A}, 52\penalty0 (4):\penalty0 R2493, 1995.
\newblock \doi{https://doi.org/10.1103/physreva.52.r2493}.

\bibitem[Calderbank and Shor(1996)]{calderbank1996good}
A~Robert Calderbank and Peter~W Shor.
\newblock Good quantum error-correcting codes exist.
\newblock \emph{Physical Review A}, 54\penalty0 (2):\penalty0 1098, 1996.
\newblock \doi{https://doi.org/10.1103/PhysRevA.54.1098}.

\bibitem[Steane(1996)]{steane1996error}
Andrew~M Steane.
\newblock Error correcting codes in quantum theory.
\newblock \emph{Physical Review Letters}, 77\penalty0 (5):\penalty0 793, 1996.
\newblock \doi{https://doi.org/10.1103/physrevlett.77.793}.

\bibitem[Steane(1997)]{steane1997active}
Andrew~M Steane.
\newblock Active stabilization, quantum computation, and quantum state
  synthesis.
\newblock \emph{Physical Review Letters}, 78\penalty0 (11):\penalty0 2252,
  1997.
\newblock \doi{https://doi.org/10.1103/physrevlett.78.2252}.

\bibitem[Gottesman(1997)]{Gottesman:1997}
Daniel Gottesman.
\newblock Stabilizer codes and quantum error correction.
\newblock \emph{arXiv preprint quant-ph/9705052}, 1997.

\bibitem[Raussendorf et~al.(2003)Raussendorf, Browne, and
  Briegel]{raussendorf2003measurement}
Robert Raussendorf, Daniel~E Browne, and Hans~J Briegel.
\newblock Measurement-based quantum computation on cluster states.
\newblock \emph{Physical Review A}, 68\penalty0 (2):\penalty0 022312, 2003.
\newblock \doi{https://doi.org/10.1103/physreva.68.022312}.

\bibitem[Briegel et~al.(2009)Briegel, Browne, D{\"u}r, Raussendorf, and Van~den
  Nest]{briegel2009measurement}
Hans~J Briegel, David~E Browne, Wolfgang D{\"u}r, Robert Raussendorf, and
  Maarten Van~den Nest.
\newblock Measurement-based quantum computation.
\newblock \emph{Nature Physics}, 5\penalty0 (1):\penalty0 19, 2009.
\newblock \doi{https://doi.org/10.1038/nphys1157}.

\bibitem[Raussendorf et~al.(2007)Raussendorf, Harrington, and
  Goyal]{raussendorf2007topological}
Robert Raussendorf, Jim Harrington, and Kovid Goyal.
\newblock Topological fault-tolerance in cluster state quantum computation.
\newblock \emph{New Journal of Physics}, 9\penalty0 (6):\penalty0 199, 2007.
\newblock \doi{https://doi.org/10.1088/1367-2630/9/6/199}.

\bibitem[Raussendorf et~al.(2006)Raussendorf, Harrington, and
  Goyal]{raussendorf2006fault}
Robert Raussendorf, Jim Harrington, and Kovid Goyal.
\newblock A fault-tolerant one-way quantum computer.
\newblock \emph{Annals of Physics}, 321\penalty0 (9):\penalty0 2242--2270,
  2006.
\newblock \doi{https://doi.org/10.1016/j.aop.2006.01.012}.

\bibitem[Raussendorf et~al.(2005)Raussendorf, Bravyi, and
  Harrington]{raussendorf2005long}
Robert Raussendorf, Sergey Bravyi, and Jim Harrington.
\newblock Long-range quantum entanglement in noisy cluster states.
\newblock \emph{Physical Review A}, 71\penalty0 (6):\penalty0 062313, 2005.
\newblock \doi{https://doi.org/10.1103/physreva.71.062313}.

\bibitem[Bolt et~al.(2016)Bolt, Duclos-Cianci, Poulin, and
  Stace]{bolt2016foliated}
A~Bolt, G~Duclos-Cianci, D~Poulin, and TM~Stace.
\newblock Foliated quantum error-correcting codes.
\newblock \emph{Physical Review Letters}, 117\penalty0 (7):\penalty0 070501,
  2016.
\newblock \doi{https://doi.org/10.1103/physrevlett.117.070501}.

\bibitem[Bolt et~al.(2018)Bolt, Poulin, and Stace]{bolt2018decoding}
A~Bolt, D~Poulin, and TM~Stace.
\newblock Decoding schemes for foliated sparse quantum error-correcting codes.
\newblock \emph{Physical Review A}, 98\penalty0 (6):\penalty0 062302, 2018.
\newblock \doi{https://doi.org/10.1103/physreva.98.062302}.

\bibitem[Brown and Roberts(2018)]{brown2018universal}
Benjamin Brown and Sam Roberts.
\newblock Universal fault-tolerant measurement-based quantum computation.
\newblock \emph{arXiv preprint arXiv:1811.11780}, 2018.

\bibitem[Nickerson and Bomb{\'\i}n(2018)]{nickerson2018measurement}
Naomi Nickerson and H{\'e}ctor Bomb{\'\i}n.
\newblock Measurement based fault tolerance beyond foliation.
\newblock \emph{arXiv preprint arXiv:1810.09621}, 2018.

\bibitem[Nguyen et~al.(1994)Nguyen, Pezaris, Pratt, and Ward]{nguyen1994three}
John Nguyen, John Pezaris, Gill Pratt, and Steve Ward.
\newblock Three-dimensional network topologies.
\newblock In \emph{International Workshop on Parallel Computer Routing and
  Communication}, pages 101--115. Springer, 1994.
\newblock \doi{https://doi.org/10.1007/3-540-58429-3_31}.

\bibitem[Parhami and Kwai(2001)]{parhami2001unified}
Behrooz Parhami and Ding-Ming Kwai.
\newblock A unified formulation of honeycomb and diamond networks.
\newblock \emph{IEEE Transactions on Parallel and Distributed Systems},
  12\penalty0 (1):\penalty0 74--80, 2001.
\newblock \doi{https://doi.org/10.1109/71.899940}.

\bibitem[Barrett and Stace(2010)]{barrett2010fault}
Sean~D Barrett and Thomas~M Stace.
\newblock Fault tolerant quantum computation with very high threshold for loss
  errors.
\newblock \emph{Physical Review Letters}, 105\penalty0 (20):\penalty0 200502,
  2010.
\newblock \doi{https://doi.org/10.1103/physrevlett.105.200502}.

\bibitem[Dress(1987)]{dress1987presentations}
Andreas~WM Dress.
\newblock Presentations of discrete groups, acting on simply connected
  manifolds, in terms of parametrized systems of coxeter matrices—a
  systematic approach.
\newblock \emph{Advances in Mathematics}, 63\penalty0 (2):\penalty0 196--212,
  1987.
\newblock \doi{https://doi.org/10.1016/0001-8708(87)90053-3}.

\bibitem[Delgado-Friedrichs(2001)]{delgado2001recognition}
Olaf Delgado-Friedrichs.
\newblock Recognition of flat orbifolds and the classification of tilings in
  {R3}.
\newblock \emph{Discrete \& Computational Geometry}, 26\penalty0 (4):\penalty0
  549--571, 2001.
\newblock \doi{https://doi.org/10.1007/s00454-001-0022-2}.

\bibitem[Delfosse and Nickerson(2017)]{delfosse2017almost}
Nicolas Delfosse and Naomi~H Nickerson.
\newblock Almost-linear time decoding algorithm for topological codes.
\newblock \emph{arXiv preprint arXiv:1709.06218}, 2017.

\bibitem[Huang et~al.(2020)Huang, Newman, and Brown]{huang2020fault}
Shilin Huang, Michael Newman, and Kenneth~R Brown.
\newblock Fault-tolerant weighted union-find decoding on the toric code.
\newblock \emph{arXiv preprint arXiv:2004.04693}, 2020.

\bibitem[Chubb and Flammia(2018)]{chubb2018statistical}
Christopher~T Chubb and Steven~T Flammia.
\newblock Statistical mechanical models for quantum codes with correlated
  noise.
\newblock \emph{arXiv preprint arXiv:1809.10704}, 2018.

\bibitem[Bomb{\'\i}n and Martin-Delgado(2007)]{bombin2007homological}
Hector Bomb{\'\i}n and Miguel~A Martin-Delgado.
\newblock Homological error correction: Classical and quantum codes.
\newblock \emph{Journal of Mathematical Physics}, 48\penalty0 (5):\penalty0
  052105, 2007.
\newblock \doi{https://doi.org/10.1063/1.2731356}.

\bibitem[Kitaev(2003)]{kitaev2003fault}
A~Yu Kitaev.
\newblock Fault-tolerant quantum computation by anyons.
\newblock \emph{Annals of Physics}, 303\penalty0 (1):\penalty0 2--30, 2003.
\newblock \doi{https://doi.org/10.1016/s0003-4916(02)00018-0}.

\bibitem[Freedman et~al.(2002)Freedman, Meyer, and Luo]{freedman2002z2}
Michael~H Freedman, David~A Meyer, and Feng Luo.
\newblock Z2-systolic freedom and quantum codes.
\newblock In \emph{Mathematics of Quantum Computation}, pages 287--320. CRC
  Press, 2002.
\newblock \doi{https://doi.org/10.1201/9781420035377.ch12}.

\bibitem[Hatcher(2000)]{Hatcher:478079}
Allen Hatcher.
\newblock \emph{{Algebraic Topology}}.
\newblock Cambridge Univ. Press, Cambridge, 2000.
\newblock URL \url{https://cds.cern.ch/record/478079}.

\bibitem[Bravyi and Hastings(2014)]{bravyi2014homological}
Sergey Bravyi and Matthew~B Hastings.
\newblock Homological product codes.
\newblock In \emph{Proceedings of the Forty-Sixth Annual ACM Symposium on
  Theory of Computing}, pages 273--282. ACM, 2014.

\bibitem[Hastings(2016)]{hastings2016quantum}
Matthew~B Hastings.
\newblock Quantum codes from high-dimensional manifolds.
\newblock \emph{arXiv preprint arXiv:1608.05089}, 2016.

\bibitem[Londe and Leverrier(2019)]{londe2017golden}
Vivien Londe and Anthony Leverrier.
\newblock Golden codes: quantum {LDPC} codes built from regular tessellations
  of hyperbolic 4-manifolds.
\newblock \emph{Quantum Information \& Computation}, 19\penalty0
  (5-6):\penalty0 361--391, 2019.

\bibitem[Audoux and Couvreur(2019)]{audoux2015tensor}
Benjamin Audoux and Alain Couvreur.
\newblock On tensor products of {CSS} codes.
\newblock \emph{Annales de l’Institut Henri Poincar{\'e} D}, 2019.
\newblock \doi{https://doi.org/10.4171/aihpd/71}.

\bibitem[Campbell(2019)]{campbell2019theory}
Earl Campbell.
\newblock A theory of single-shot error correction for adversarial noise.
\newblock \emph{Quantum Science and Technology}, 4:\penalty0 025006, 2019.
\newblock \doi{https://doi.org/10.1088/2058-9565/aafc8f}.

\bibitem[Bomb{\'\i}n(2015)]{bombin2015single}
H{\'e}ctor Bomb{\'\i}n.
\newblock Single-shot fault-tolerant quantum error correction.
\newblock \emph{Physical Review X}, 5\penalty0 (3):\penalty0 031043, 2015.
\newblock \doi{https://doi.org/10.1103/physrevx.5.031043}.

\bibitem[Jochym-O'Connor(2019)]{jochym2019fault}
Tomas Jochym-O'Connor.
\newblock Fault-tolerant gates via homological product codes.
\newblock \emph{Quantum}, 3:\penalty0 Art--No, 2019.
\newblock \doi{https://doi.org/10.22331/q-2019-02-04-120}.

\bibitem[Krishna and Poulin(2019)]{krishna2019fault}
Anirudh Krishna and David Poulin.
\newblock Fault-tolerant gates on hypergraph product codes.
\newblock \emph{arXiv preprint arXiv:1909.07424}, 2019.

\bibitem[Bomb{\'\i}n et~al.(2012)Bomb{\'\i}n, Andrist, Ohzeki, Katzgraber, and
  Mart{\'\i}n-Delgado]{bombin2012strong}
H{\'e}ctor Bomb{\'\i}n, Ruben~S Andrist, Masayuki Ohzeki, Helmut~G Katzgraber,
  and Miguel~A Mart{\'\i}n-Delgado.
\newblock Strong resilience of topological codes to depolarization.
\newblock \emph{Physical Review X}, 2\penalty0 (2):\penalty0 021004, 2012.
\newblock \doi{https://doi.org/10.1103/physrevx.2.021004}.

\bibitem[Li et~al.(2019)Li, Miller, Newman, Wu, and Brown]{li20192d}
Muyuan Li, Daniel Miller, Michael Newman, Yukai Wu, and Kenneth~R Brown.
\newblock {2D} compass codes.
\newblock \emph{Physical Review X}, 9\penalty0 (2):\penalty0 021041, 2019.
\newblock \doi{https://doi.org/10.1103/physrevx.9.021041}.

\bibitem[Kubica and Yoshida(2018)]{kubica2018ungauging}
Aleksander Kubica and Beni Yoshida.
\newblock Ungauging quantum error-correcting codes.
\newblock \emph{arXiv preprint arXiv:1805.01836}, 2018.

\bibitem[Daniel et~al.(2020)Daniel, Alexander, and
  Miyake]{daniel2019computational}
Austin~K Daniel, Rafael~N Alexander, and Akimasa Miyake.
\newblock Computational universality of symmetry-protected topologically
  ordered cluster phases on 2d archimedean lattices.
\newblock \emph{Quantum}, 4:\penalty0 228, 2020.
\newblock \doi{https://doi.org/10.22331/q-2020-02-10-228}.

\bibitem[Aliferis et~al.(2009)Aliferis, Brito, DiVincenzo, Preskill, Steffen,
  and Terhal]{aliferis2009fault}
Panos Aliferis, Frederico Brito, David~P DiVincenzo, John Preskill, Matthias
  Steffen, and Barbara~M Terhal.
\newblock Fault-tolerant computing with biased-noise superconducting qubits: a
  case study.
\newblock \emph{New Journal of Physics}, 11\penalty0 (1):\penalty0 013061,
  2009.
\newblock \doi{https://doi.org/10.1088/1367-2630/11/1/013061}.

\bibitem[Aliferis and Preskill(2008)]{PhysRevA.78.052331}
Panos Aliferis and John Preskill.
\newblock Fault-tolerant quantum computation against biased noise.
\newblock \emph{Physical Review A}, 78\penalty0 (5):\penalty0 052331, 2008.
\newblock \doi{https://doi.org/10.1103/physreva.78.052331}.

\bibitem[Trout et~al.(2018)Trout, Li, Guti{\'e}rrez, Wu, Wang, Duan, and
  Brown]{trout2018simulating}
Colin~J Trout, Muyuan Li, Mauricio Guti{\'e}rrez, Yukai Wu, Sheng-Tao Wang,
  Luming Duan, and Kenneth~R Brown.
\newblock Simulating the performance of a distance-3 surface code in a linear
  ion trap.
\newblock \emph{New Journal of Physics}, 20\penalty0 (4):\penalty0 043038,
  2018.
\newblock \doi{https://doi.org/10.1088/1367-2630/aab341}.

\bibitem[Breuckmann et~al.(2017{\natexlab{a}})Breuckmann, Vuillot, Campbell,
  Krishna, and Terhal]{breuckmann2017hyperbolic}
Nikolas~P Breuckmann, Christophe Vuillot, Earl Campbell, Anirudh Krishna, and
  Barbara~M Terhal.
\newblock Hyperbolic and semi-hyperbolic surface codes for quantum storage.
\newblock \emph{Quantum Science and Technology}, 2\penalty0 (3):\penalty0
  035007, 2017{\natexlab{a}}.
\newblock \doi{https://doi.org/10.1088/2058-9565/aa7d3b}.

\bibitem[Breuckmann and Terhal(2016)]{breuckmann2016constructions}
Nikolas~P Breuckmann and Barbara~M Terhal.
\newblock Constructions and noise threshold of hyperbolic surface codes.
\newblock \emph{IEEE transactions on Information Theory}, 62\penalty0
  (6):\penalty0 3731--3744, 2016.
\newblock \doi{https://doi.org/10.1109/tit.2016.2555700}.

\bibitem[Conrad et~al.(2018)Conrad, Chamberland, Breuckmann, and
  Terhal]{conrad2018small}
Jonathan Conrad, Christopher Chamberland, Nikolas~P Breuckmann, and Barbara~M
  Terhal.
\newblock The small stellated dodecahedron code and friends.
\newblock \emph{Philosophical Transactions of the Royal Society A:
  Mathematical, Physical and Engineering Sciences}, 376\penalty0
  (2123):\penalty0 20170323, 2018.
\newblock \doi{https://doi.org/10.1098/rsta.2017.0323}.

\bibitem[Dress(1985)]{dress1985regular}
Andreas~WM Dress.
\newblock Regular polytopes and equivariant tessellations from a combinatorial
  point of view.
\newblock In \emph{Algebraic Topology G{\"o}ttingen 1984}, pages 56--72.
  Springer, 1985.
\newblock \doi{https://doi.org/10.1007/bfb0074423}.

\bibitem[Dress and Huson(1987)]{dress1987tilings}
Andreas~WM Dress and Daniel Huson.
\newblock On tilings of the plane.
\newblock \emph{Geometriae Dedicata}, 24\penalty0 (3):\penalty0 295--310, 1987.
\newblock \doi{https://doi.org/10.1007/bf00181602}.

\bibitem[Gr{\"u}nbaum and Shephard(1987)]{grunbaum1987tilings}
Branko Gr{\"u}nbaum and Geoffrey~Colin Shephard.
\newblock \emph{Tilings and Patterns}.
\newblock Freeman, 1987.
\newblock \doi{https://doi.org/10.2307/2323457}.

\bibitem[Huson(1993)]{huson1993generation}
Daniel~H Huson.
\newblock The generation and classification of tile-k-transitive tilings of the
  euclidean plane, the sphere and the hyperbolic plane.
\newblock \emph{Geometriae Dedicata}, 47\penalty0 (3):\penalty0 269--296, 1993.
\newblock \doi{https://doi.org/10.1007/bf01263661}.

\bibitem[Balke and Huson(1996)]{balke1996two}
Ludwig Balke and Daniel~H Huson.
\newblock Two-dimensional groups, orbifolds and tilings.
\newblock \emph{Geometriae Dedicata}, 60\penalty0 (1):\penalty0 89--106, 1996.
\newblock \doi{https://doi.org/10.1007/bf00150869}.

\bibitem[Delgado-Friedrichs et~al.(2003{\natexlab{a}})Delgado-Friedrichs,
  O'Keeffe, and Yaghi]{friedrichs2003three}
Olaf Delgado-Friedrichs, Michael O'Keeffe, and Omar~M. Yaghi.
\newblock Three-periodic nets and tilings: regular and quasiregular nets.
\newblock \emph{Acta Crystallographica Section A}, 59\penalty0 (1):\penalty0
  22--27, 2003{\natexlab{a}}.

\bibitem[Delgado-Friedrichs et~al.(1999)Delgado-Friedrichs, Dress, Huson,
  Klinowski, and Mackay]{friedrichs1999systematic}
Olaf Delgado-Friedrichs, Andreas~WM Dress, Daniel~H Huson, Jacek Klinowski, and
  Alan~L Mackay.
\newblock Systematic enumeration of crystalline networks.
\newblock \emph{Nature}, 400\penalty0 (6745):\penalty0 644, 1999.
\newblock \doi{https://doi.org/10.1038/23210}.

\bibitem[Friedrichs and Huson(1997)]{friedrichs1997orbifold}
Olaf~Delgado Friedrichs and Daniel~H Huson.
\newblock Orbifold triangulations and crystallographic groups.
\newblock \emph{Periodica Mathematica Hungarica}, 34\penalty0 (1-2):\penalty0
  29--55, 1997.

\bibitem[Thurston et~al.(1997)Thurston, Kerckhoff, Floyd, and
  Milnor]{thurston1997geometry}
William~P Thurston, Steve Kerckhoff, WJ~Floyd, and John~Willard Milnor.
\newblock \emph{The Geometry and Topology of Three-Manifolds}.
\newblock Unpublished, 1997.

\bibitem[Hyde et~al.(2006)Hyde, Delgado-Friedrichs, Ramsden, and
  Robins]{hyde2006towards}
ST~Hyde, O~Delgado-Friedrichs, SJ~Ramsden, and Vanessa Robins.
\newblock Towards enumeration of crystalline frameworks: the {2D} hyperbolic
  approach.
\newblock \emph{Solid State Sciences}, 8\penalty0 (7):\penalty0 740--752, 2006.
\newblock \doi{https://doi.org/10.1016/j.solidstatesciences.2006.04.001}.

\bibitem[Blatov et~al.(2010)Blatov, O'Keeffe, and Proserpio]{blatov2010vertex}
VA~Blatov, M~O'Keeffe, and DM~Proserpio.
\newblock Vertex-, face-, point-, {S}chl{\"a}fli-, and {D}elaney-symbols in
  nets, polyhedra and tilings: recommended terminology.
\newblock \emph{CrystEngComm}, 12\penalty0 (1):\penalty0 44--48, 2010.
\newblock \doi{https://doi.org/10.1039/b910671e}.

\bibitem[Balke and Valverde(1996)]{balke1996chamber}
Ludwig Balke and Alba Valverde.
\newblock Chamber systems, coloured graphs and orbifolds.
\newblock \emph{Contributions to Algebra and Geometry}, 37\penalty0
  (1):\penalty0 17--29, 1996.

\bibitem[Bieberbach(1911)]{bieberbach1911bewegungsgruppen}
Ludwig Bieberbach.
\newblock {\"U}ber die bewegungsgruppen der euklidischen r{\"a}ume.
\newblock \emph{Mathematische Annalen}, 70\penalty0 (3):\penalty0 297--336,
  1911.
\newblock \doi{https://doi.org/10.1007/bf01564500}.

\bibitem[Bieberbach(1912)]{bieberbach1912bewegungsgruppen}
Ludwig Bieberbach.
\newblock {\"U}ber die bewegungsgruppen der euklidischen r{\"a}ume (zweite
  abhandlung.) die gruppen mit einem endlichen fundamentalbereich.
\newblock \emph{Mathematische Annalen}, 72\penalty0 (3):\penalty0 400--412,
  1912.
\newblock \doi{https://doi.org/10.1007/bf01456724}.

\bibitem[Hemion(1992)]{hemion1992classification}
Geoffery Hemion.
\newblock \emph{The Classification of Knots and 3-Dimensional Spaces}.
\newblock Oxford University Press, 1992.

\bibitem[Delgado-Friedrichs()]{Gavrog}
Olaf Delgado-Friedrichs.
\newblock Gavrog.
\newblock \url{https://github.com/odf/gavrog}.

\bibitem[Delgado-Friedrichs et~al.(2003{\natexlab{b}})Delgado-Friedrichs,
  O'Keeffe, and Yaghi]{friedrichs2003semi}
Olaf Delgado-Friedrichs, Michael O'Keeffe, and Omar~M. Yaghi.
\newblock Three-periodic nets and tilings: semiregular nets.
\newblock \emph{Acta Crystallographica Section A}, 59\penalty0 (6):\penalty0
  515--525, 2003{\natexlab{b}}.
\newblock \doi{https://doi.org/10.1107/s0108767302018494}.

\bibitem[Brown et~al.(2019)Brown, Newman, and Brown]{brown2019handling}
Natalie~C Brown, Michael Newman, and Kenneth~R Brown.
\newblock Handling leakage with subsystem codes.
\newblock \emph{New Journal of Physics}, 21\penalty0 (1):\penalty0 073055,
  2019.
\newblock \doi{https://doi.org/10.1088/1367-2630/ab3372}.

\bibitem[Chamberland et~al.(2020{\natexlab{a}})Chamberland, Zhu, Yoder,
  Hertzberg, and Cross]{chamberland2020topological}
Christopher Chamberland, Guanyu Zhu, Theodore~J Yoder, Jared~B Hertzberg, and
  Andrew~W Cross.
\newblock Topological and subsystem codes on low-degree graphs with flag
  qubits.
\newblock \emph{Physical Review X}, 10\penalty0 (1):\penalty0 011022,
  2020{\natexlab{a}}.
\newblock \doi{https://doi.org/10.1103/physrevx.10.011022}.

\bibitem[Chamberland et~al.(2020{\natexlab{b}})Chamberland, Kubica, Yoder, and
  Zhu]{chamberland2020triangular}
Christopher Chamberland, Aleksander Kubica, Ted Yoder, and Guanyu Zhu.
\newblock Triangular color codes on trivalent graphs with flag qubits.
\newblock \emph{New Journal of Physics}, 22\penalty0 (2):\penalty0 023019,
  2020{\natexlab{b}}.
\newblock \doi{https://doi.org/10.1088/1367-2630/ab68fd}.

\bibitem[Momma and Izumi(2011)]{momma2011vesta}
Koichi Momma and Fujio Izumi.
\newblock Vesta 3 for three-dimensional visualization of crystal, volumetric
  and morphology data.
\newblock \emph{Journal of Applied Crystallography}, 44\penalty0 (6):\penalty0
  1272--1276, 2011.
\newblock \doi{https://doi.org/10.1107/s0021889811038970}.

\bibitem[Delgado-Friedrichs and O'Keeffe(2003)]{delgado2003identification}
Olaf Delgado-Friedrichs and Michael O'Keeffe.
\newblock Identification of and symmetry computation for crystal nets.
\newblock \emph{Acta Crystallographica Section A: Foundations of
  Crystallography}, 59\penalty0 (4):\penalty0 351--360, 2003.
\newblock \doi{https://doi.org/10.1107/s0108767303012017}.

\bibitem[Delgado-Friedrichs(2003)]{delgado2003barycentric}
Olaf Delgado-Friedrichs.
\newblock Barycentric drawings of periodic graphs.
\newblock In \emph{International Symposium on Graph Drawing}, pages 178--189.
  Springer, 2003.
\newblock \doi{https://doi.org/10.1007/978-3-540-24595-7_17}.

\bibitem[Delgado-Friedrichs and O'Keeffe(2005)]{delgado2005crystal}
Olaf Delgado-Friedrichs and Michael O'Keeffe.
\newblock Crystal nets as graphs: Terminology and definitions.
\newblock \emph{Journal of Solid State Chemistry}, 178\penalty0 (8):\penalty0
  2480--2485, 2005.
\newblock \doi{https://doi.org/10.1016/j.jssc.2005.06.011}.

\bibitem[Delgado-Friedrichs(2005)]{delgado2005equilibrium}
Olaf Delgado-Friedrichs.
\newblock Equilibrium placement of periodic graphs and convexity of plane
  tilings.
\newblock \emph{Discrete \& Computational Geometry}, 33\penalty0 (1):\penalty0
  67--81, 2005.
\newblock \doi{https://doi.org/10.1007/s00454-004-1147-x}.

\bibitem[GAP()]{GAP4}
GAP.
\newblock \emph{{GAP -- Groups, Algorithms, and Programming, Version 4.10.2}}.
\newblock The GAP~Group, 2019.
\newblock URL \url{https://www.gap-system.org}.

\bibitem[tin()]{tinkercad}
Tinkercad.
\newblock URL \url{http://www.tinkercad.com}.

\bibitem[O'Keeffe et~al.(2008)O'Keeffe, Peskov, Ramsden, and
  Yaghi]{o2008reticular}
Michael O'Keeffe, Maxim~A Peskov, Stuart~J Ramsden, and Omar~M Yaghi.
\newblock The reticular chemistry structure resource ({RCSR}) database of, and
  symbols for, crystal nets.
\newblock \emph{Accounts of Chemical Research}, 41\penalty0 (12):\penalty0
  1782--1789, 2008.
\newblock \doi{https://doi.org/10.1021/ar800124u}.

\bibitem[Edmonds(2009)]{edmonds2009paths}
Jack Edmonds.
\newblock Paths, trees, and flowers.
\newblock In \emph{Classic Papers in Combinatorics}, pages 361--379. Springer,
  2009.
\newblock \doi{https://doi.org/10.1007/978-0-8176-4842-8_26}.

\bibitem[Kolmogorov(2009)]{kolmogorov2009blossom}
Vladimir Kolmogorov.
\newblock Blossom {V}: a new implementation of a minimum cost perfect matching
  algorithm.
\newblock \emph{Mathematical Programming Computation}, 1\penalty0 (1):\penalty0
  43--67, 2009.
\newblock \doi{https://doi.org/10.1007/s12532-009-0002-8}.

\bibitem[Nishimori(1986)]{nishimori1986geometry}
Hidetoshi Nishimori.
\newblock Geometry-induced phase transition in the$\pm${J} ising model.
\newblock \emph{Journal of the Physical Society of Japan}, 55\penalty0
  (10):\penalty0 3305--3307, 1986.
\newblock \doi{https://doi.org/10.1143/jpsj.55.3305}.

\bibitem[Kubica et~al.(2018)Kubica, Beverland, Brand{\~a}o, Preskill, and
  Svore]{kubica2018three}
Aleksander Kubica, Michael~E Beverland, Fernando Brand{\~a}o, John Preskill,
  and Krysta~M Svore.
\newblock Three-dimensional color code thresholds via statistical-mechanical
  mapping.
\newblock \emph{Physical Review Letters}, 120\penalty0 (18):\penalty0 180501,
  2018.
\newblock \doi{https://doi.org/10.1103/physrevlett.120.180501}.

\bibitem[Stace and Barrett(2010)]{stace2010error}
Thomas~M Stace and Sean~D Barrett.
\newblock Error correction and degeneracy in surface codes suffering loss.
\newblock \emph{Physical Review A}, 81\penalty0 (2):\penalty0 022317, 2010.
\newblock \doi{https://doi.org/10.1103/physreva.81.022317}.

\bibitem[L'{E}culier(1999)]{leculier1999}
Pierre L'{E}culier.
\newblock Good parameter sets for combined multiple recursive random number
  generators.
\newblock \emph{Operations Research}, 47:\penalty0 159--164, 1999.
\newblock \doi{https://doi.org/10.1287/opre.47.1.159}.

\bibitem[L'{E}culier et~al.(2002)L'{E}culier, Simard, Chen, and
  Kelton]{leculier2002}
Pierre L'{E}culier, R.~Simard, E.~J. Chen, and W.~D. Kelton.
\newblock An objected-oriented random-number package with many long streams and
  substreams.
\newblock \emph{Operations Research}, 50:\penalty0 1073--1075, 2002.
\newblock \doi{https://doi.org/10.1287/opre.50.6.1073.358}.

\bibitem[Stace et~al.(2009)Stace, Barrett, and Doherty]{stace2009thresholds}
Thomas~M Stace, Sean~D Barrett, and Andrew~C Doherty.
\newblock Thresholds for topological codes in the presence of loss.
\newblock \emph{Physical Review Letters}, 102\penalty0 (20):\penalty0 200501,
  2009.
\newblock \doi{https://doi.org/10.1103/physrevlett.102.200501}.

\bibitem[Rudolph(2017)]{rudolph2017optimistic}
Terry Rudolph.
\newblock Why {I} am optimistic about the silicon-photonic route to quantum
  computing.
\newblock \emph{APL Photonics}, 2\penalty0 (3):\penalty0 030901, 2017.
\newblock \doi{https://doi.org/10.1063/1.4976737}.

\bibitem[Tuckett et~al.(2018)Tuckett, Bartlett, and Flammia]{Tuckett:2018}
David~K Tuckett, Stephen~D Bartlett, and Steven~T Flammia.
\newblock Ultrahigh error threshold for surface codes with biased noise.
\newblock \emph{Physical Review Letters}, 120\penalty0 (5):\penalty0 050505,
  2018.
\newblock \doi{https://doi.org/10.1103/physrevlett.120.050505}.

\bibitem[Tuckett et~al.(2019)Tuckett, Darmawan, Chubb, Bravyi, Bartlett, and
  Flammia]{tuckett2018tailoring}
David~K Tuckett, Andrew~S Darmawan, Christopher~T Chubb, Sergey Bravyi,
  Stephen~D Bartlett, and Steven~T Flammia.
\newblock Tailoring surface codes for highly biased noise.
\newblock \emph{Physical Review X}, 9\penalty0 (4):\penalty0 041031, 2019.
\newblock \doi{https://doi.org/10.1103/physrevx.9.041031}.

\bibitem[Tuckett et~al.(2020)Tuckett, Bartlett, Flammia, and
  Brown]{tuckett2019fault}
David~K Tuckett, Stephen~D Bartlett, Steven~T Flammia, and Benjamin~J Brown.
\newblock Fault-tolerant thresholds for the surface code in excess of 5\% under
  biased noise.
\newblock \emph{Physical Review Letters}, 124\penalty0 (13):\penalty0 130501,
  2020.
\newblock \doi{https://doi.org/10.1103/physrevlett.124.130501}.

\bibitem[Xu et~al.(2019)Xu, Zhao, Yuan, and Benjamin]{xu2018high}
Xiaosi Xu, Qi~Zhao, Xiao Yuan, and Simon~C Benjamin.
\newblock High-threshold code for modular hardware with asymmetric noise.
\newblock \emph{Physical Review Applied}, 12\penalty0 (6):\penalty0 064006,
  2019.
\newblock \doi{https://doi.org/10.1103/physrevapplied.12.064006}.

\bibitem[Puri et~al.(2019)Puri, St-Jean, Gross, Grimm, Frattini, Iyer, Krishna,
  Touzard, Jiang, Blais, et~al.]{puri2019bias}
Shruti Puri, Lucas St-Jean, Jonathan~A Gross, Alexander Grimm, NE~Frattini,
  Pavithran~S Iyer, Anirudh Krishna, Steven Touzard, Liang Jiang, Alexandre
  Blais, et~al.
\newblock Bias-preserving gates with stabilized cat qubits.
\newblock \emph{arXiv preprint arXiv:1905.00450}, 2019.

\bibitem[Webster et~al.(2015)Webster, Bartlett, and Poulin]{PhysRevA.92.062309}
Paul Webster, Stephen~D Bartlett, and David Poulin.
\newblock Reducing the overhead for quantum computation when noise is biased.
\newblock \emph{Physical Review A}, 92\penalty0 (6):\penalty0 062309, 2015.
\newblock \doi{https://doi.org/10.1103/physreva.92.062309}.

\bibitem[Brooks and Preskill(2013)]{PhysRevA.87.032310}
Peter Brooks and John Preskill.
\newblock Fault-tolerant quantum computation with asymmetric {B}acon-{S}hor
  codes.
\newblock \emph{Physical Review A}, 87\penalty0 (3):\penalty0 032310, 2013.
\newblock \doi{https://doi.org/10.1103/physreva.87.032310}.

\bibitem[Bravyi et~al.(2013)Bravyi, Duclos-Cianci, Poulin, and
  Suchara]{bravyi2013subsystem}
Sergey Bravyi, Guillaume Duclos-Cianci, David Poulin, and Martin Suchara.
\newblock Subsystem surface codes with three-qubit check operators.
\newblock \emph{Quantum Information \& Computation}, 13\penalty0
  (11-12):\penalty0 963--985, 2013.

\bibitem[Suchara et~al.(2015{\natexlab{b}})Suchara, Cross, and
  Gambetta]{suchara2015leakage}
Martin Suchara, Andrew~W Cross, and Jay~M Gambetta.
\newblock Leakage suppression in the toric code.
\newblock In \emph{Information Theory (ISIT), 2015 IEEE International Symposium
  on}, pages 1119--1123. IEEE, 2015{\natexlab{b}}.
\newblock \doi{https://doi.org/10.1109/isit.2015.7282629}.

\bibitem[Varnava et~al.(2007)Varnava, Browne, and Rudolph]{varnava2007loss}
Michael Varnava, Daniel~E Browne, and Terry Rudolph.
\newblock Loss tolerant linear optical quantum memory by measurement-based
  quantum computing.
\newblock \emph{New Journal of Physics}, 9\penalty0 (6):\penalty0 203, 2007.
\newblock \doi{https://doi.org/10.1088/1367-2630/9/6/203}.

\bibitem[Breuckmann and Londe(2020)]{breuckmann2020single}
Nikolas~P Breuckmann and Vivien Londe.
\newblock Single-shot decoding of linear rate {LDPC} quantum codes with high
  performance.
\newblock \emph{arXiv preprint arXiv:2001.03568}, 2020.

\bibitem[Duivenvoorden et~al.(2018)Duivenvoorden, Breuckmann, and
  Terhal]{duivenvoorden2018renormalization}
Kasper Duivenvoorden, Nikolas~P Breuckmann, and Barbara~M Terhal.
\newblock Renormalization group decoder for a four-dimensional toric code.
\newblock \emph{IEEE Transactions on Information Theory}, 65\penalty0
  (4):\penalty0 2545--2562, 2018.
\newblock \doi{https://doi.org/10.1109/tit.2018.2879937}.

\bibitem[Breuckmann et~al.(2017{\natexlab{b}})Breuckmann, Duivenvoorden,
  Michels, and Terhal]{breuckmann2016local}
Nikolas~P Breuckmann, Kasper Duivenvoorden, Dominik Michels, and Barbara~M
  Terhal.
\newblock Local decoders for the {2D} and {4D} toric code.
\newblock \emph{Quantum Information \& Computation}, 17\penalty0
  (3-4):\penalty0 181--208, 2017{\natexlab{b}}.

\bibitem[Kubica and Preskill(2019)]{kubica2018cellular}
Aleksander Kubica and John Preskill.
\newblock Cellular-automaton decoders with provable thresholds for topological
  codes.
\newblock \emph{Physical Review Letters}, 123\penalty0 (2):\penalty0 020501,
  2019.
\newblock \doi{https://doi.org/10.1103/physrevlett.123.020501}.

\end{thebibliography}

\clearpage

\appendix
\section{Additional tiling information} \label{others}
\begin{figure}[htb!]
\centering
    \includegraphics[width=\linewidth]{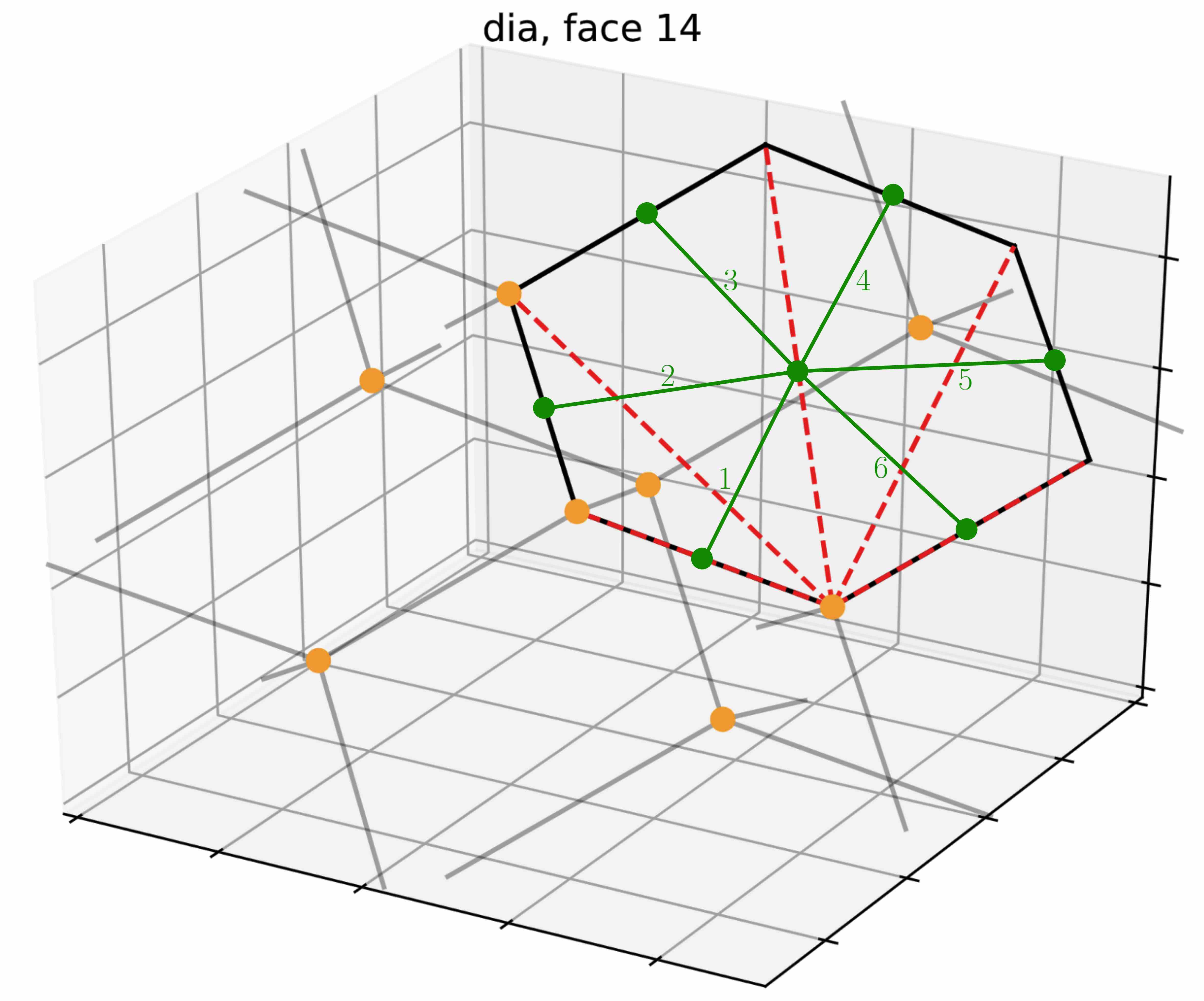}
    \caption{A guiding example for reading the gate orderings in \texttt{lib/doc} taken from the $14$th face of the decoder graph of the diamond (dia) lattice.  Here, the green is overlayed on the actual decoder graph shown in \texttt{lib/doc}.  Qubits and CZ gates (green) are placed on edges and faces.  The red dotted lines indicate correlated errors due to $X$-type errors.  The gate ordering always begins with a CZ gate supported on a qubit causing a weight one error, and ends with a CZ gate supported on a qubit causing a weight one error.  The ordering proceeds either clockwise or counter-clockwise (the choice of which, in our error model, will not have an effect); in this case, it is labeled clockwise.  The unit cell vertices and edges of the tiling are depicted in yellow and black and grey, respectively.}
\label{guide}
\end{figure}

Here, we include renderings of the tilings used in our benchmarking simulations. For completeness, we also include a summary of self-dual tilings appearing in the RCSR database.  This information can be found at \url{http://rcsr.anu.edu.au/}.  The code used to construct and perform matching on these lattices, along with additional lattice benchmarks, can be found at \url{https://github.com/leoadec/crystal-thresholds}.  In particular, unit cells and gate orderings for each lattice can be found in the \texttt{lib/doc} subdirectory.  See Figure \ref{guide} for a description.

\begin{figure}[htb!]
\centering
    \begin{subfigure}[b]{0.25\textwidth}
        \centering
        \includegraphics[width=\linewidth]{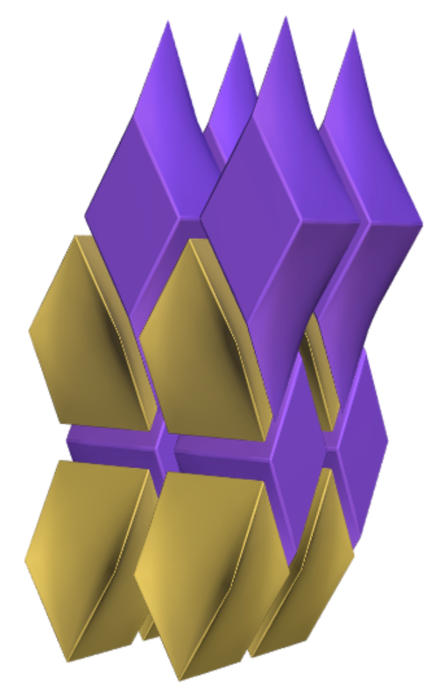}
        \caption{\textbf{cdq}}
    \end{subfigure}%
    \hspace{0.5 cm}
    \begin{subfigure}[b]{0.3\textwidth}
        \centering
        \includegraphics[width=\linewidth]{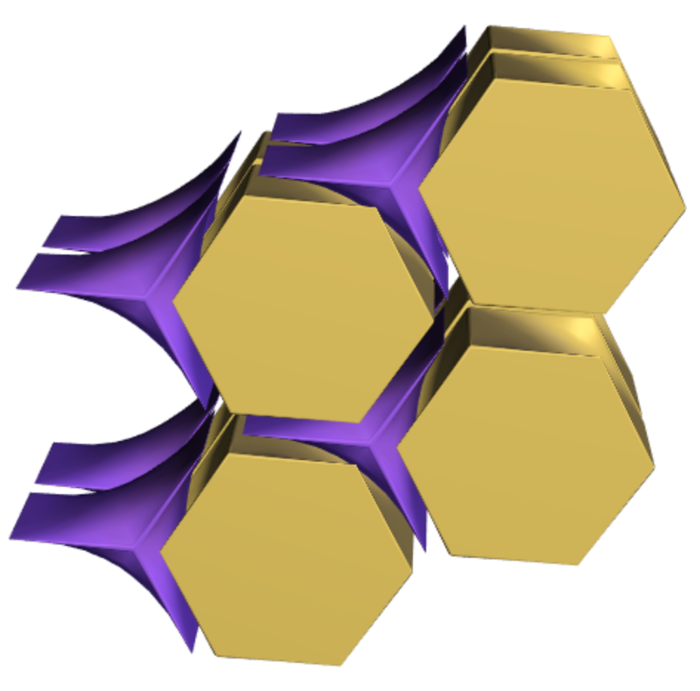}
        \caption{\textbf{hms}}
    \end{subfigure}
    \hspace{0.5 cm}
    \begin{subfigure}[b]{0.3\textwidth}
        \centering
        \includegraphics[width=\linewidth]{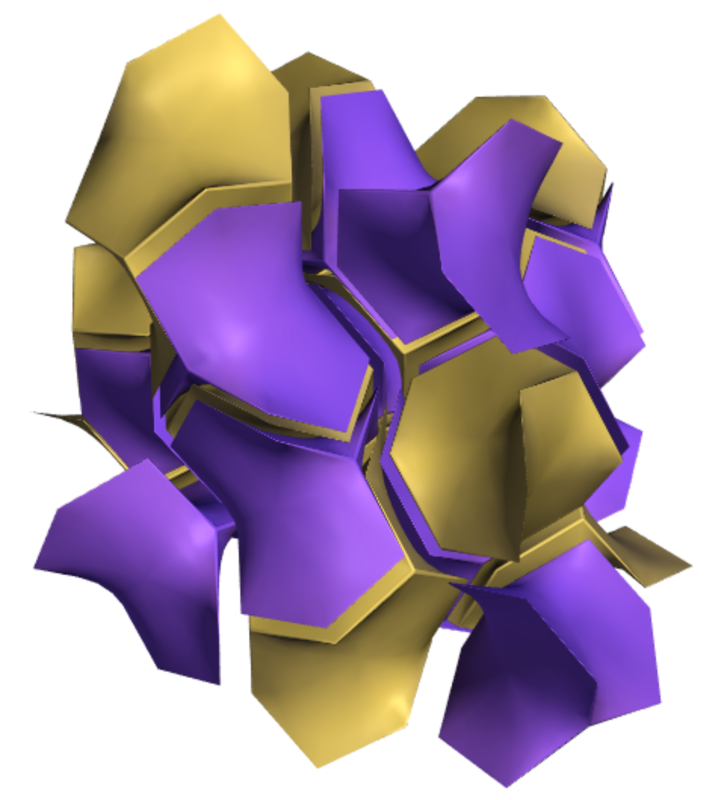}
        \caption{\textbf{ctn}}
    \end{subfigure}
    \caption{The benchmarked lattices from the RCSR database, ordered from left to right by increasing (decreasing) graph state (decoder graph) degree.  For a summary of their parameters, see Table \ref{networks}.}
\label{other_tilings}
\end{figure}

\begin{table}[htb!]
\resizebox{\columnwidth}{!}{$
\begin{tabular}{|c|c|c|c|c|c|c|c|c|c|c|c|c|c|}
\hline
\multicolumn{1}{|l|}{}             & \textbf{bbr} & \textbf{cbs} & \textbf{cdq} & \textbf{cds} & \textbf{ctn} & \textbf{dia} & \textbf{est} & \textbf{ete} & \textbf{fsf} & \textbf{ftw} & \textbf{gsi} & \textbf{hms} & \textbf{hst} \\ \hline
\textbf{Avg. Decoder Graph Degree} & 4            & 6            & 5            & 4            & 3.43         & 4            & 5            & 3         & 5            & 6            & 4            & 4            & 3.33         \\ \hline
\textbf{Avg. Graph State Degree}   & 7.5          & 4            & 4.8          & 7            & 8            & 6            & 5.375        & 10.33        & 5            & 4            & 7.5          & 6            & 9.6         \\ \hline
\end{tabular}$}

\vspace{1cm}
\resizebox{\columnwidth}{!}{$
\begin{tabular}{|c|c|c|c|c|c|c|c|c|c|c|c|c|c|}
\hline
\textbf{}                          & \textbf{lcy} & \textbf{mab} & \textbf{mcf} & \textbf{mco} & \textbf{mgc} & \textbf{pcu} & \textbf{pte} & \textbf{pyr} & \textbf{qtz} & \textbf{rtw} & \textbf{sda} & \textbf{smt} & \textbf{srs} \\ \hline
\textbf{Avg. Decoder Graph Degree} & 6            & 6            & 4            & 3.33         & 8            & 6            & 4.67         & 4            & 6            & 4.66         & 6            & 6            & 3            \\ \hline
\textbf{Avg. Graph State Degree}   & 5            & 5.33         & 6            & 8            & 4            & 4            & 5.14         & 6            & 4.66         & 5.42         & 4.33         & 4.66         & 10           \\ \hline
\end{tabular}$}

\vspace{1cm}
\resizebox{\columnwidth}{!}{$
\begin{tabular}{|c|c|c|c|c|c|c|c|c|c|c|c|c|c|}
\hline
                                   & \textbf{sto} & \textbf{svn} & \textbf{swl} & \textbf{sxd} & \textbf{tfa} & \textbf{tfc} & \textbf{ths} & \textbf{tph} & \textbf{ttv} & \textbf{unj} & \textbf{vck} & \textbf{vtx} & \textbf{bst} \\ \hline
\textbf{Avg. Decoder Graph Degree} & 3.33         & 7            & 7            & 6            & 3.33         & 3.33         & 3            & 6.86         & 4.29         & 4            & 7            & 5.33  &12      \\ \hline
\textbf{Avg. Graph State Degree}   & 8.8          & 4.29         & 3.71         & 4.33         & 8            & 8.8          & 10.66        & 4            & 4.8          & 6.5          & 3.71         & 4.5   &3       \\ \hline
\end{tabular}$}
\vspace{0.5cm}
\caption{A list of the self-dual $3$D nets in the RCSR database, along with their relevant average degrees.  We have also included the degree-$3$ bst lattice, which does not appear in the RCSR database to the best of our knowledge.} \label{networks}
\end{table}

\clearpage
\section{Numerical results} \label{data}

\subsection*{BST Lattice}

\begin{center}
\includegraphics[width=.49\textwidth]{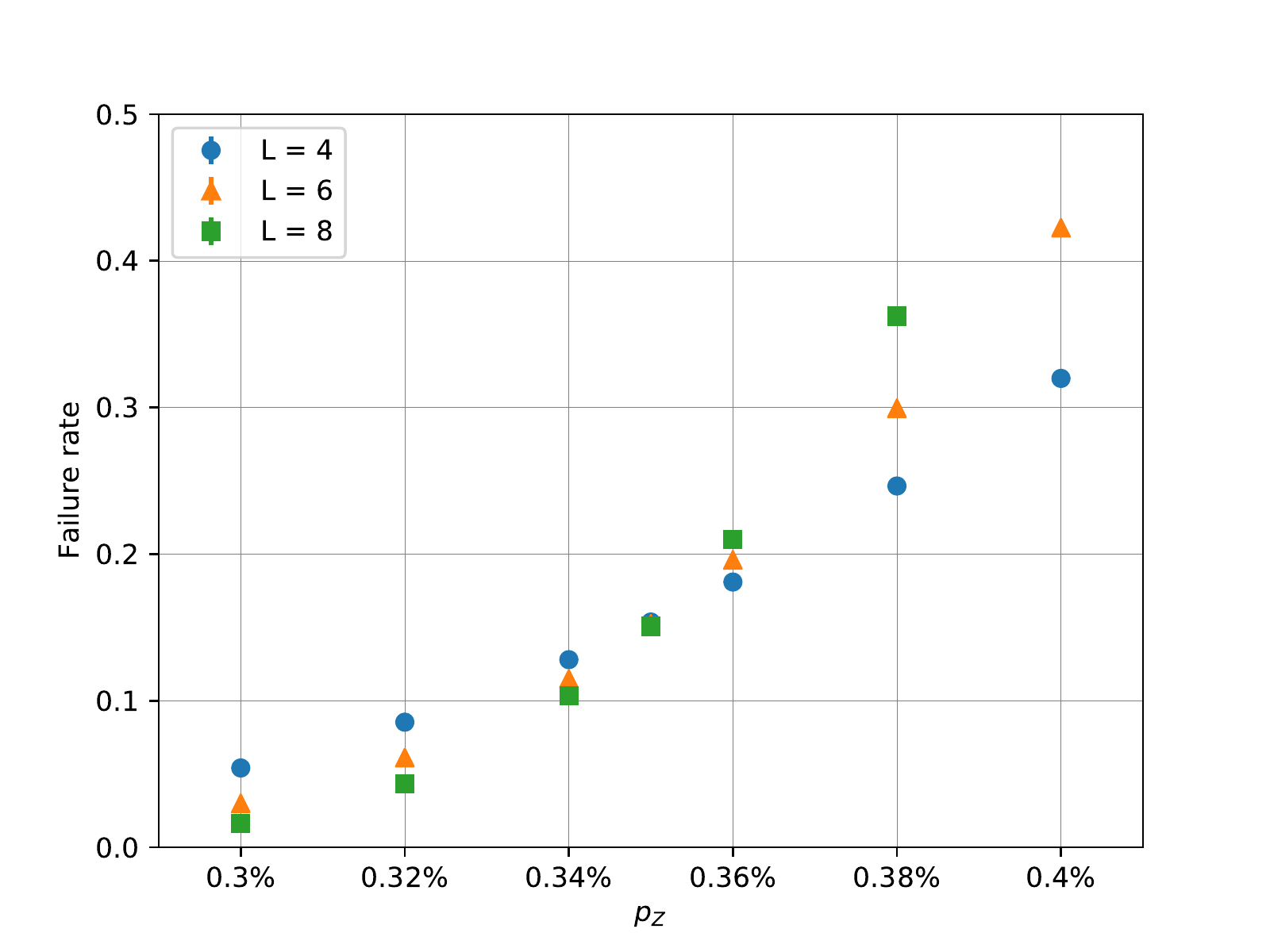}
\includegraphics[width=.49\textwidth]{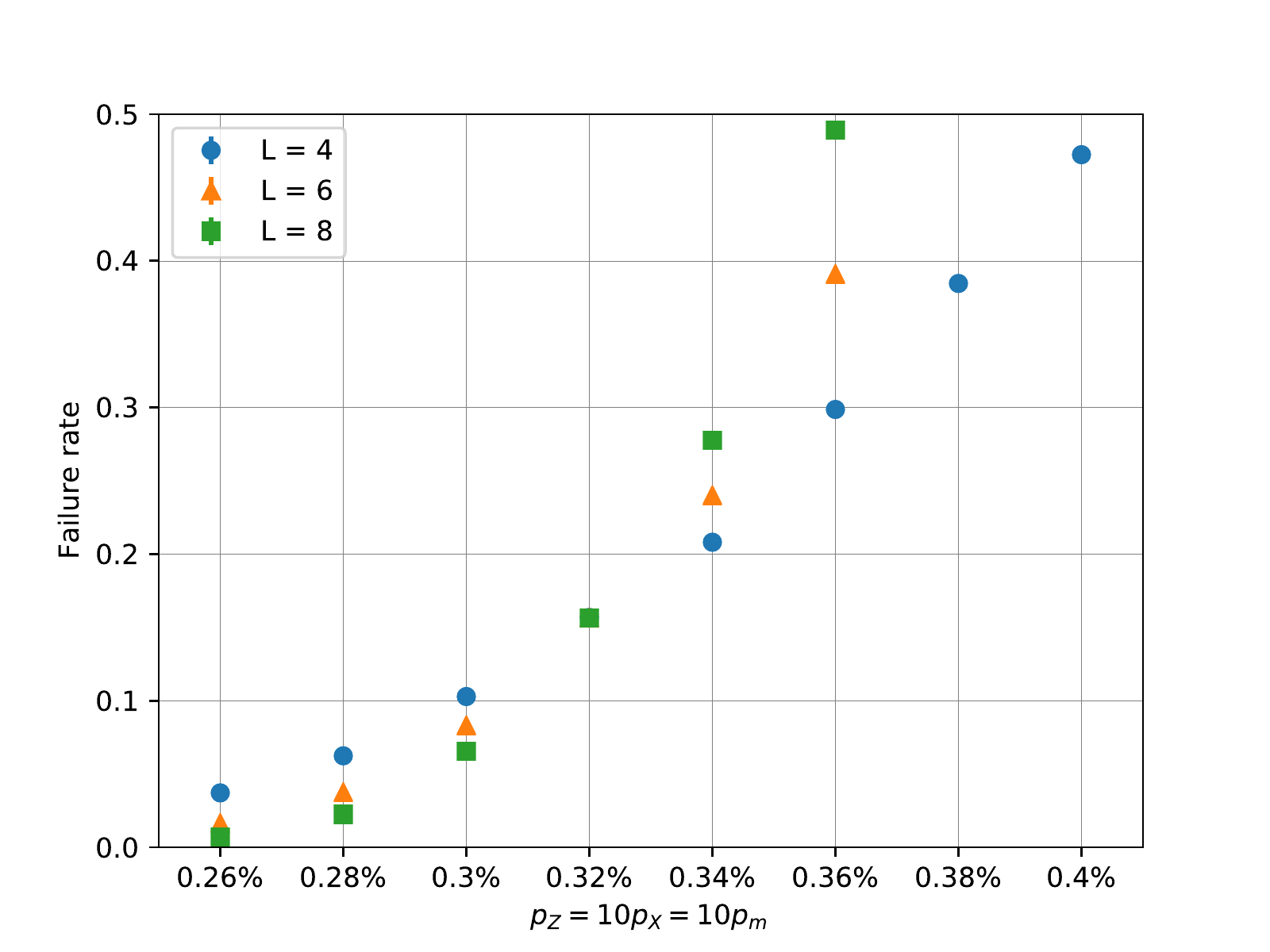}

\includegraphics[width=.49\textwidth]{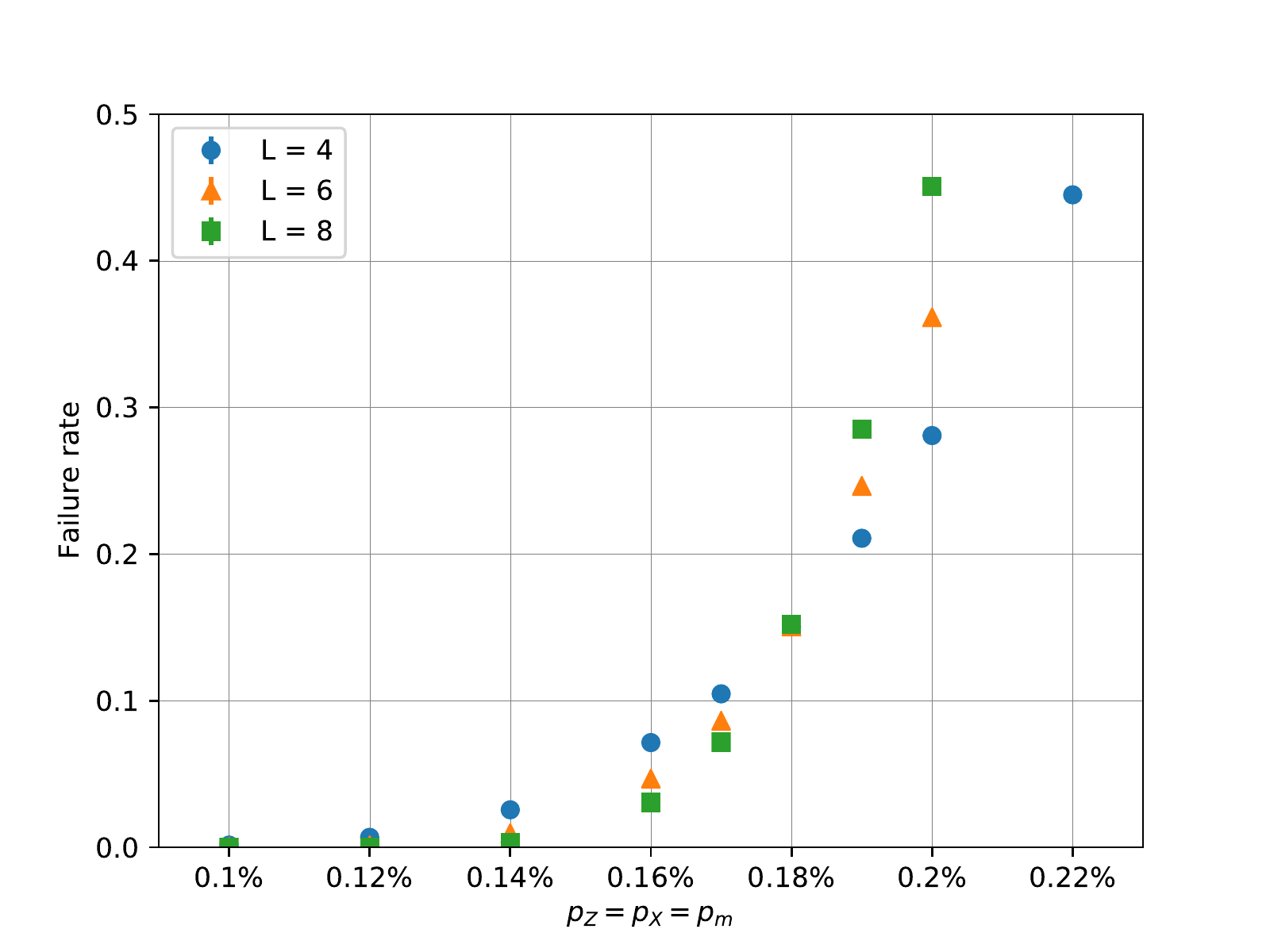}
\includegraphics[width=.49\textwidth]{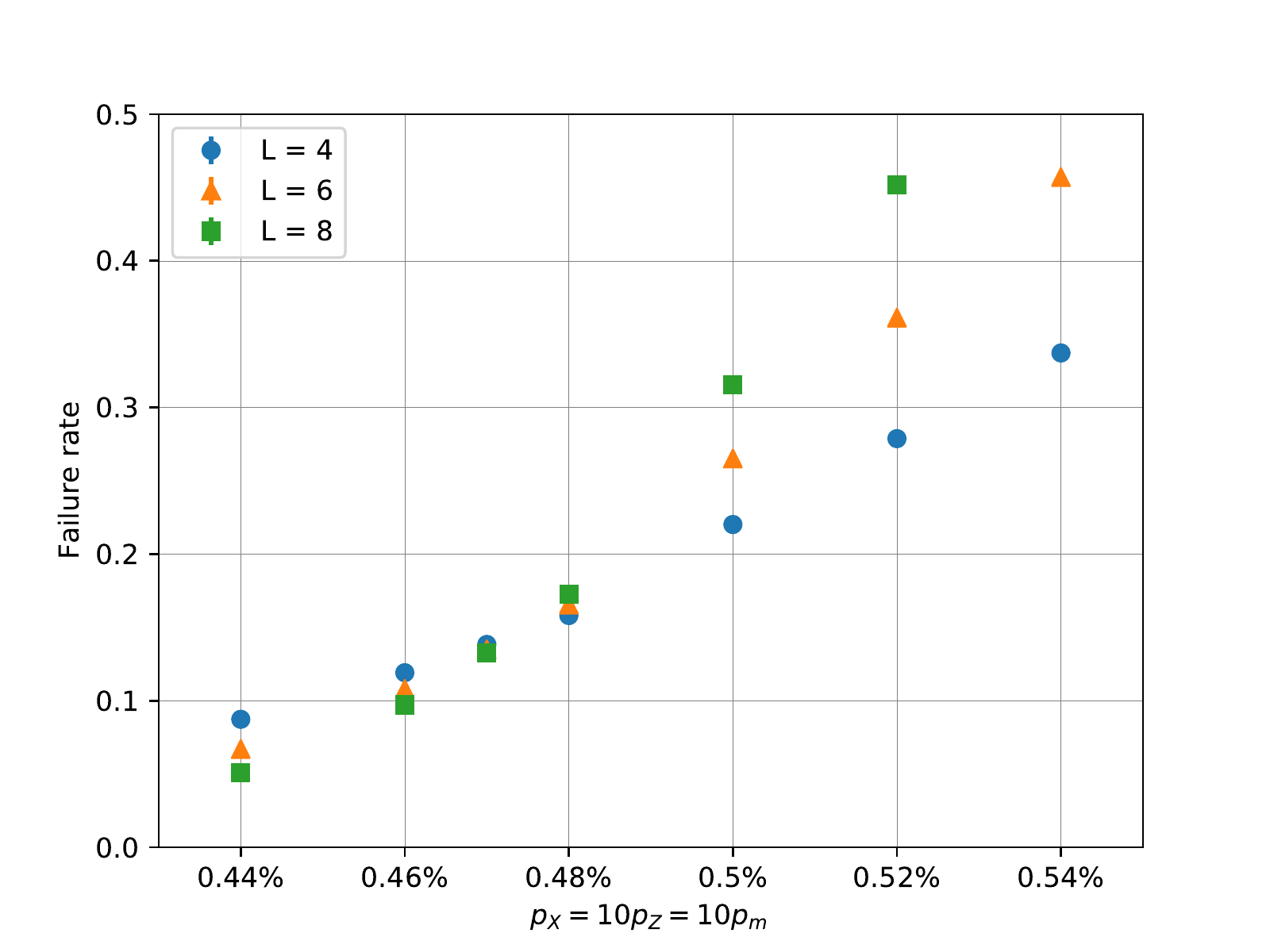}
\end{center}

\subsection*{PCU Lattice}

\begin{center}
\includegraphics[width=.49\textwidth]{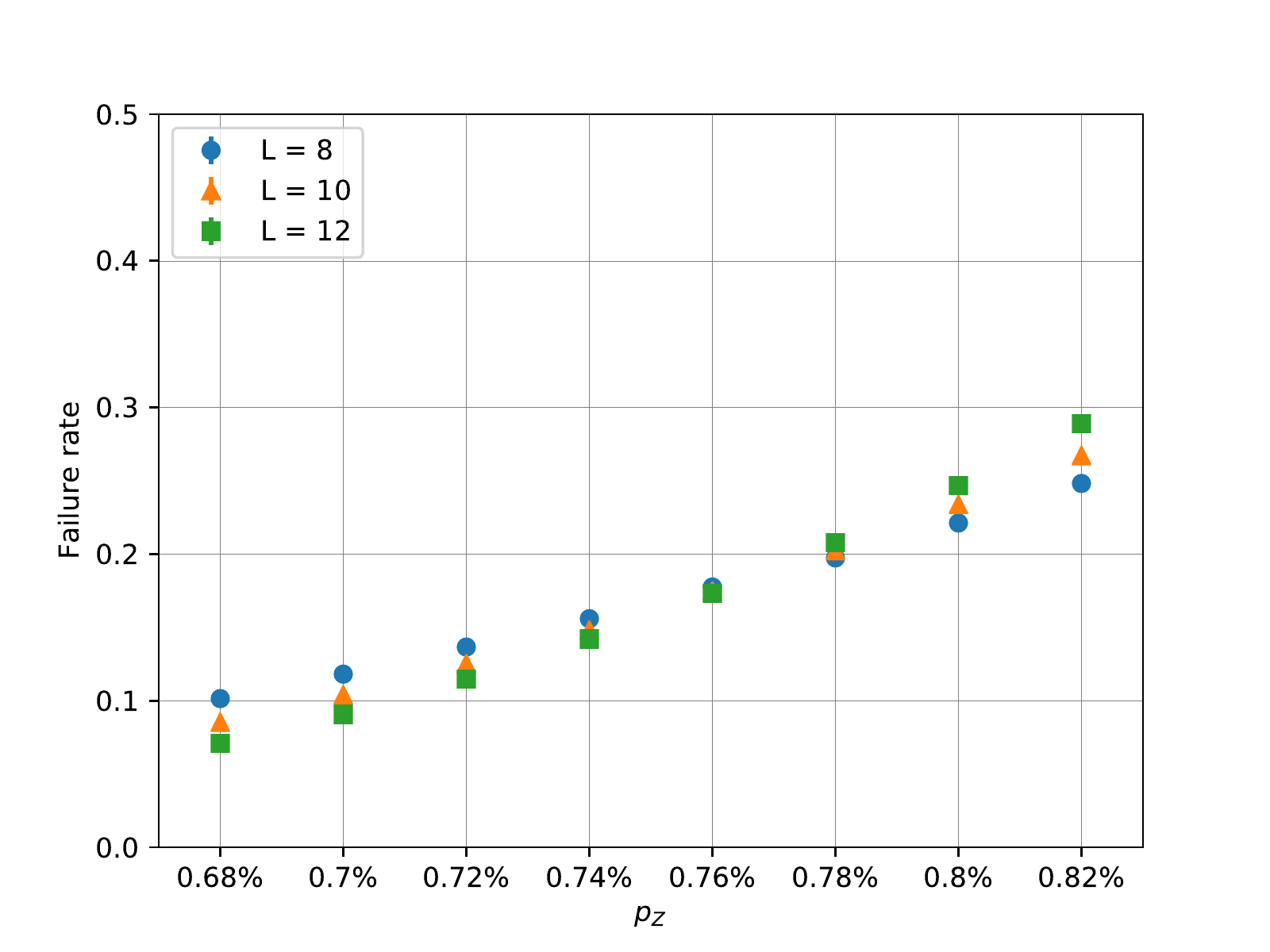}
\includegraphics[width=.49\textwidth]{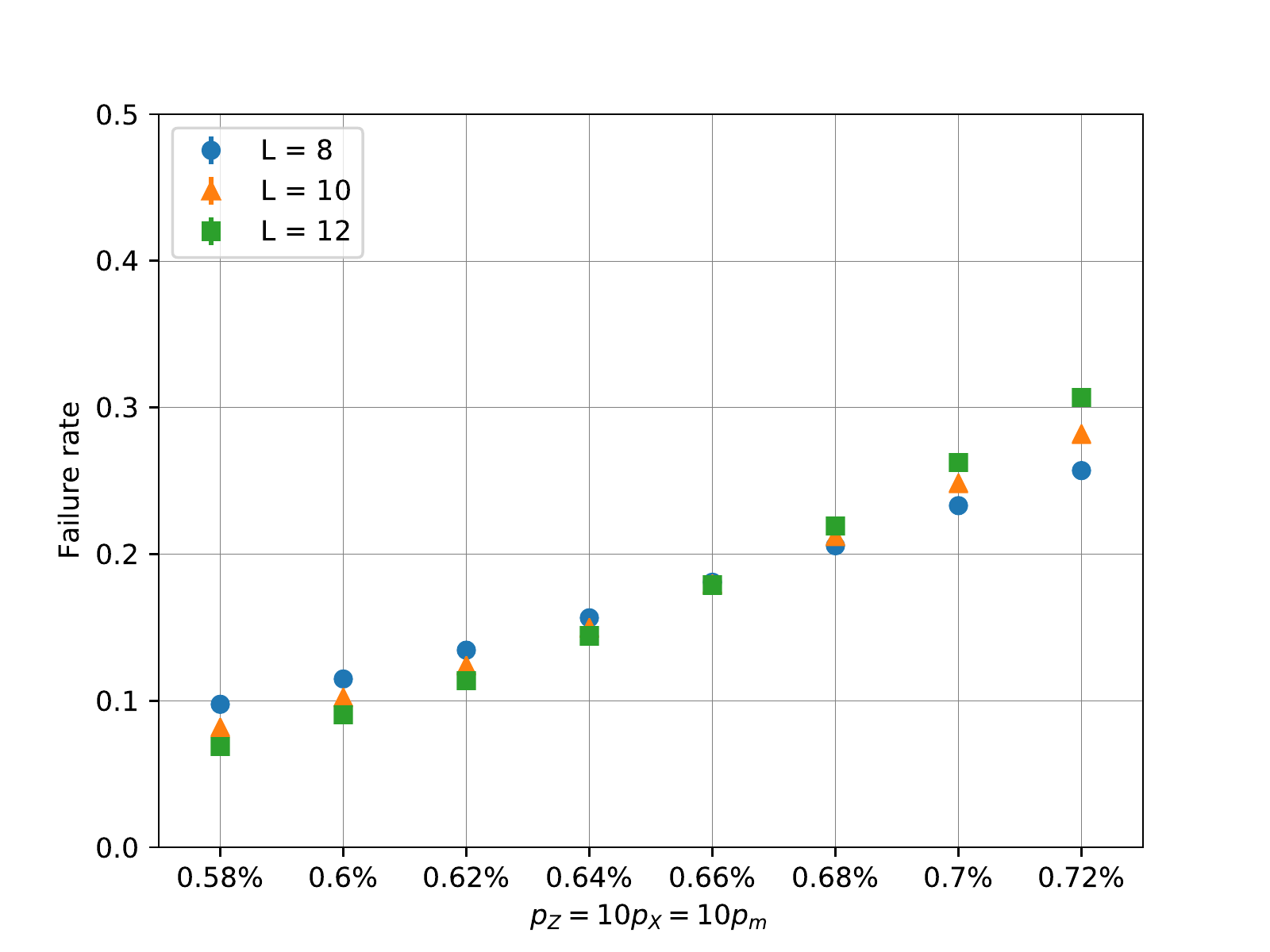}

\includegraphics[width=.49\textwidth]{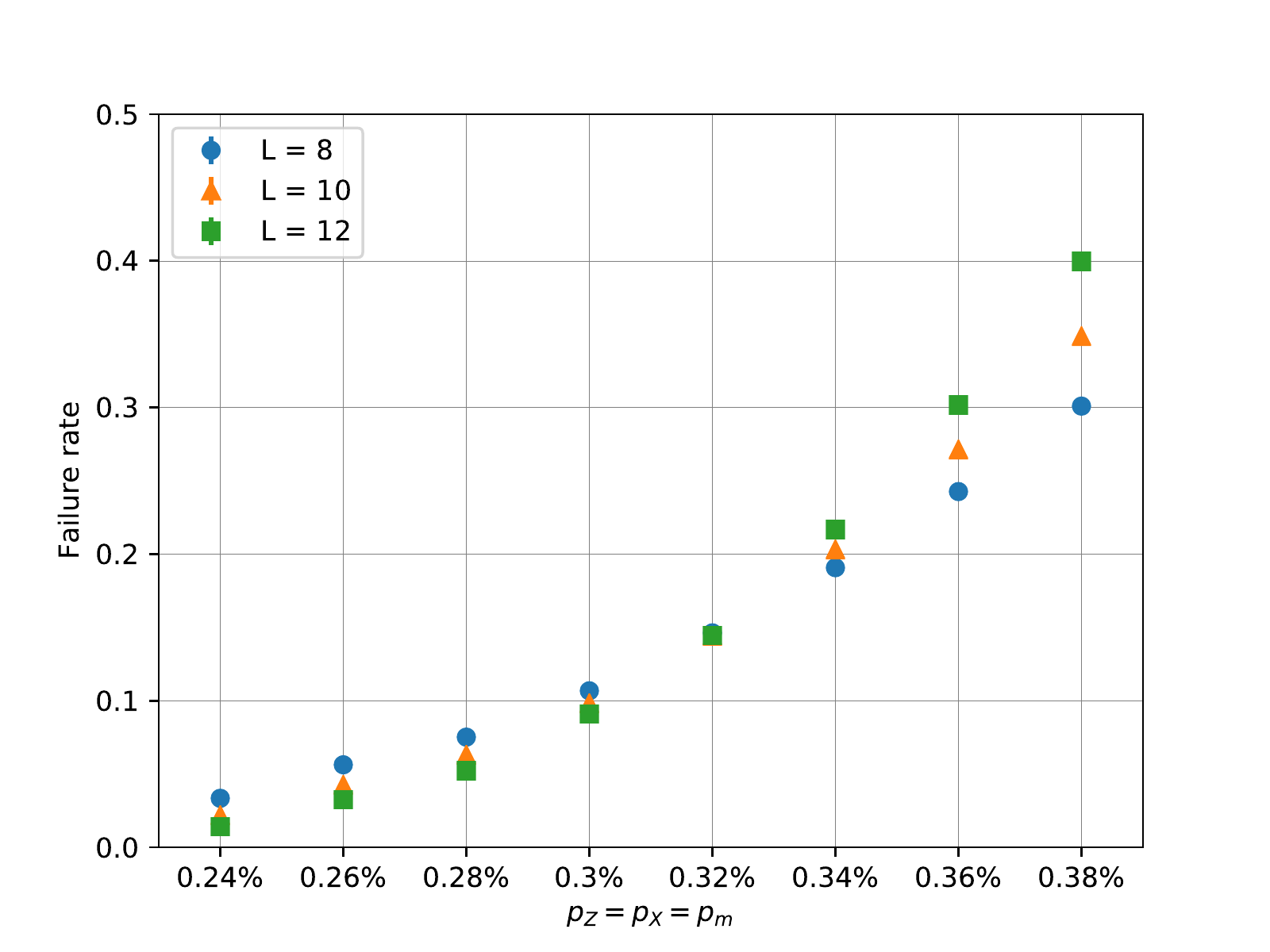}
\includegraphics[width=.49\textwidth]{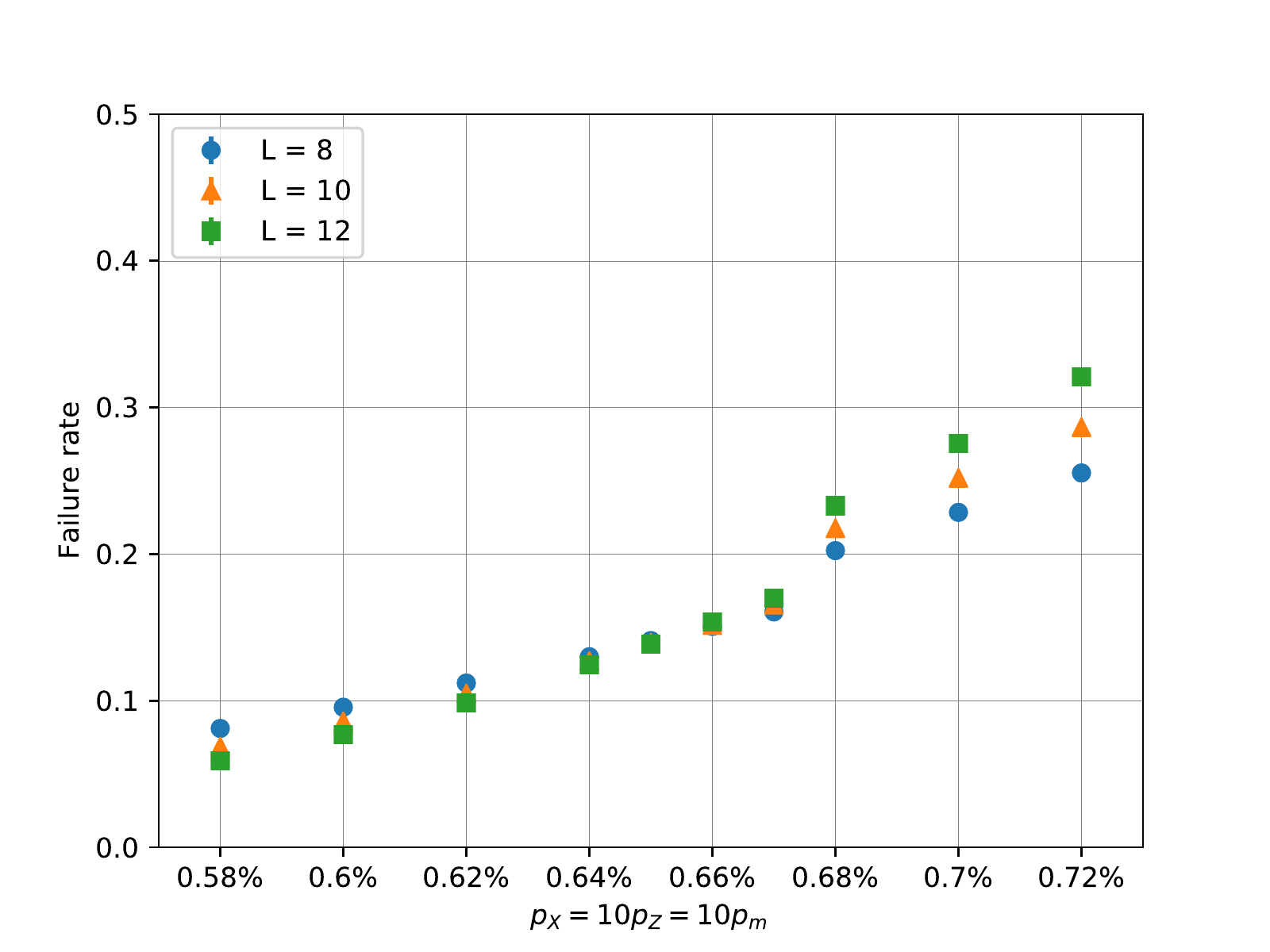}
\end{center}

\subsection*{CDQ Lattice}

\begin{center}
\includegraphics[width=.49\textwidth]{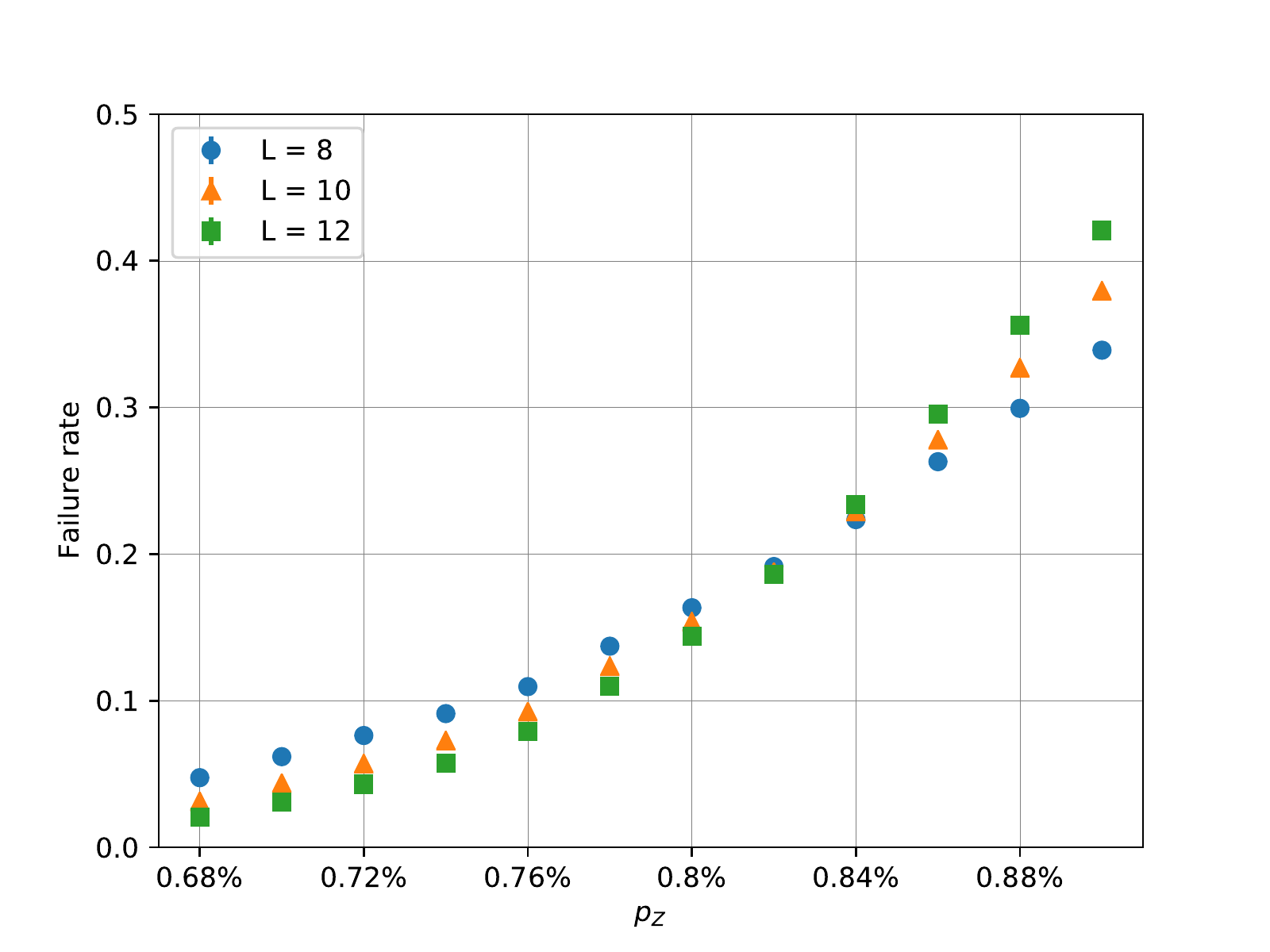}
\includegraphics[width=.49\textwidth]{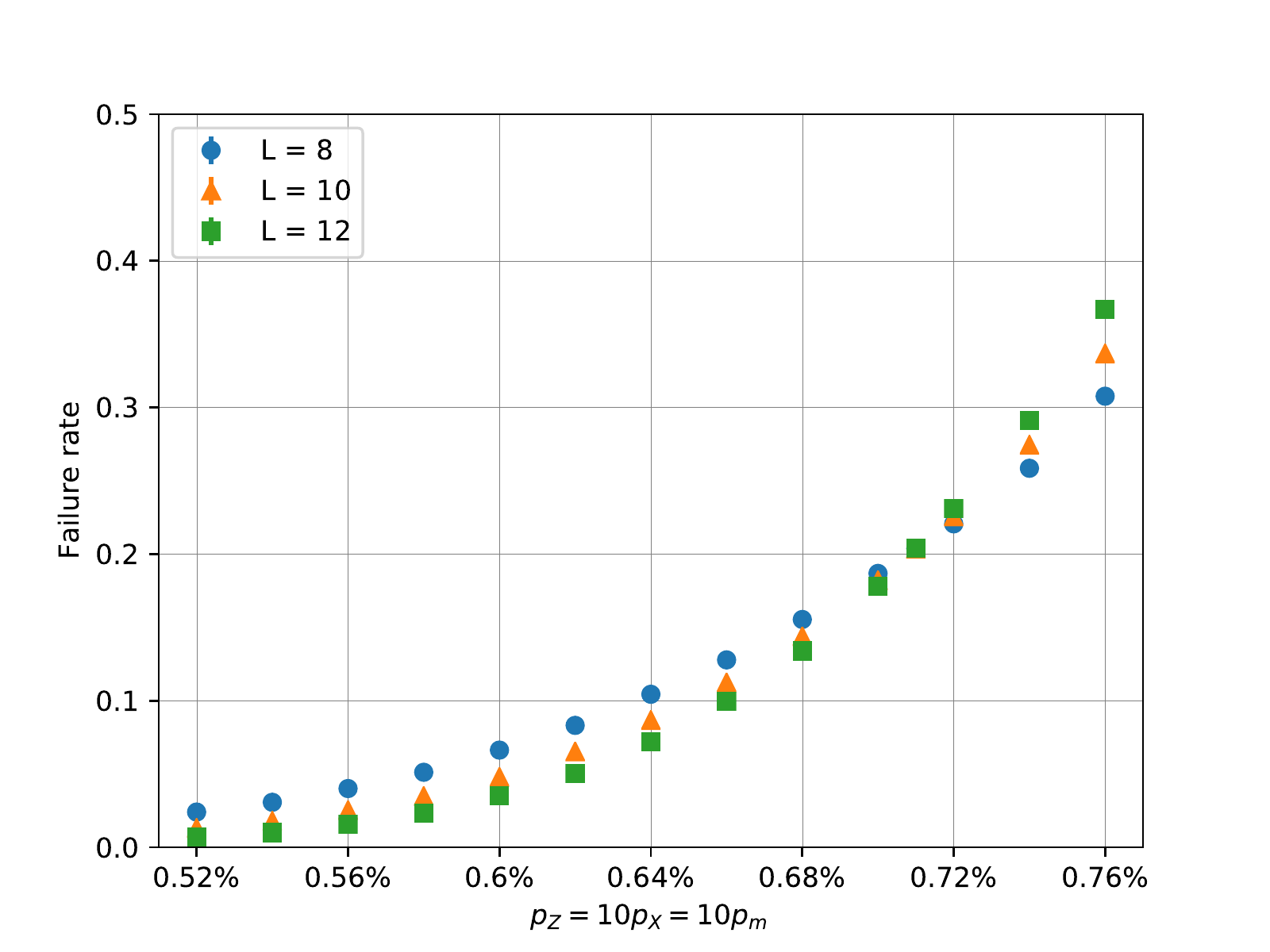}

\includegraphics[width=.49\textwidth]{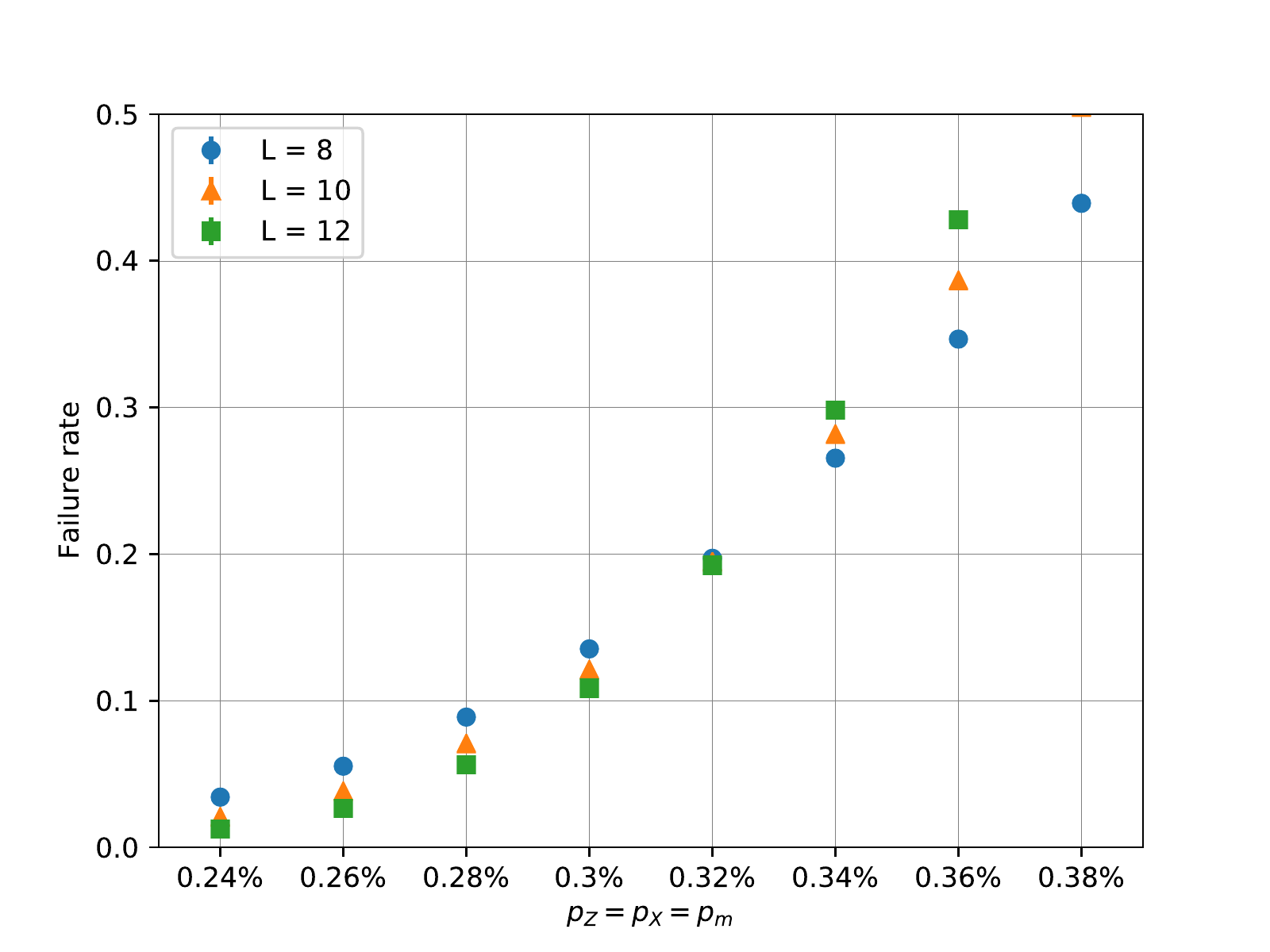}
\includegraphics[width=.49\textwidth]{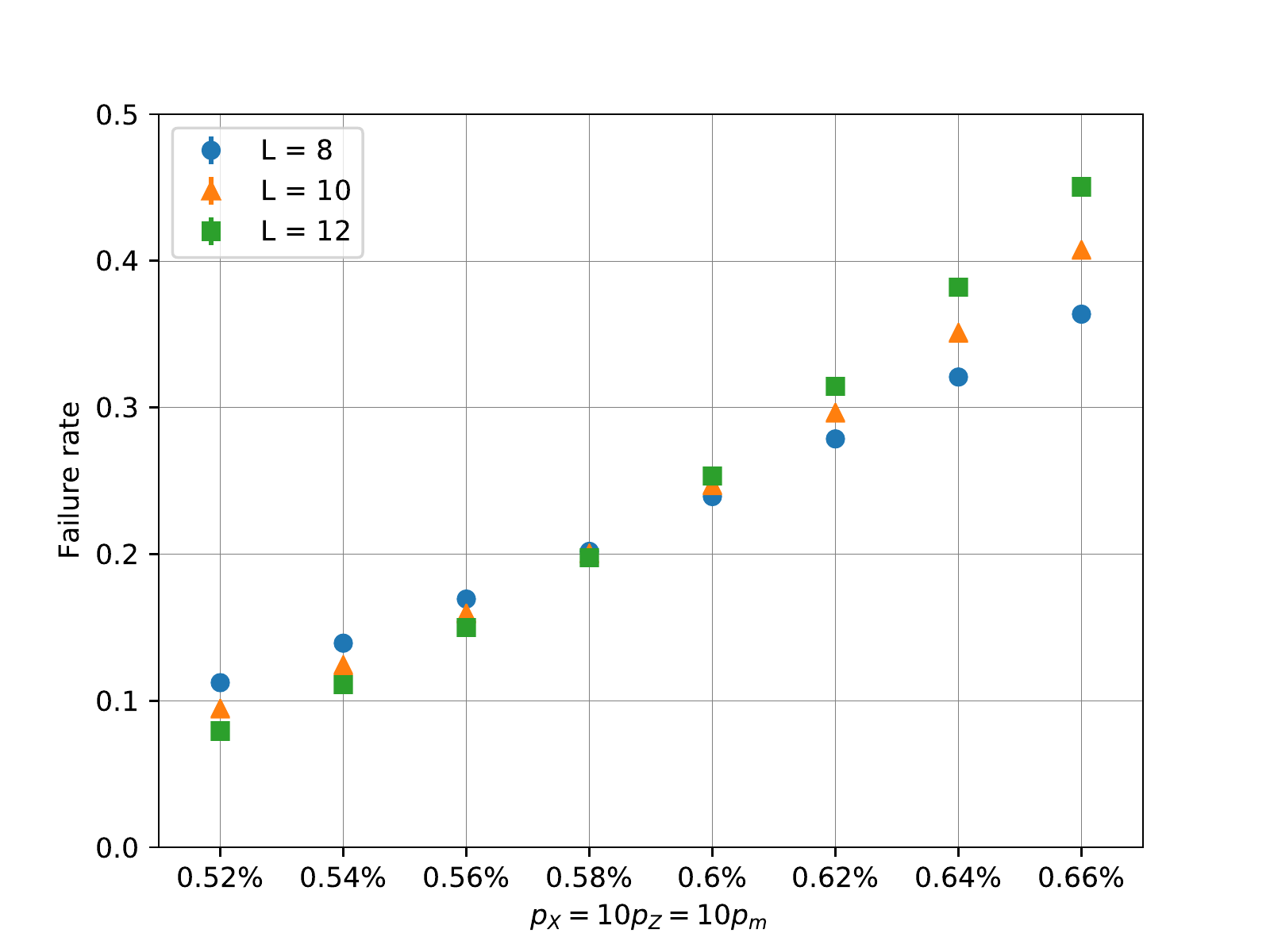}
\end{center}

\subsection*{HMS Lattice}

\begin{center}
\includegraphics[width=.49\textwidth]{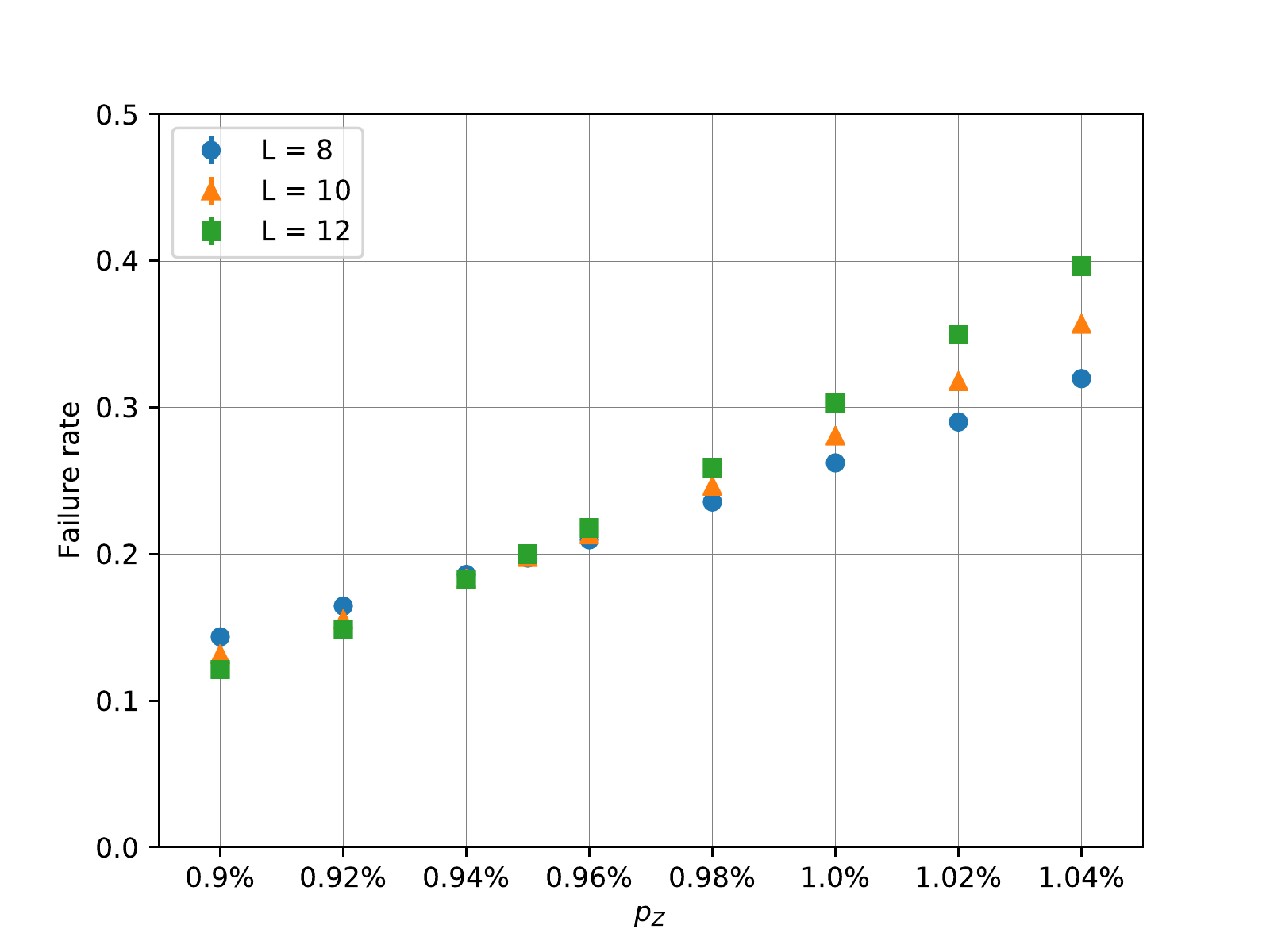}
\includegraphics[width=.49\textwidth]{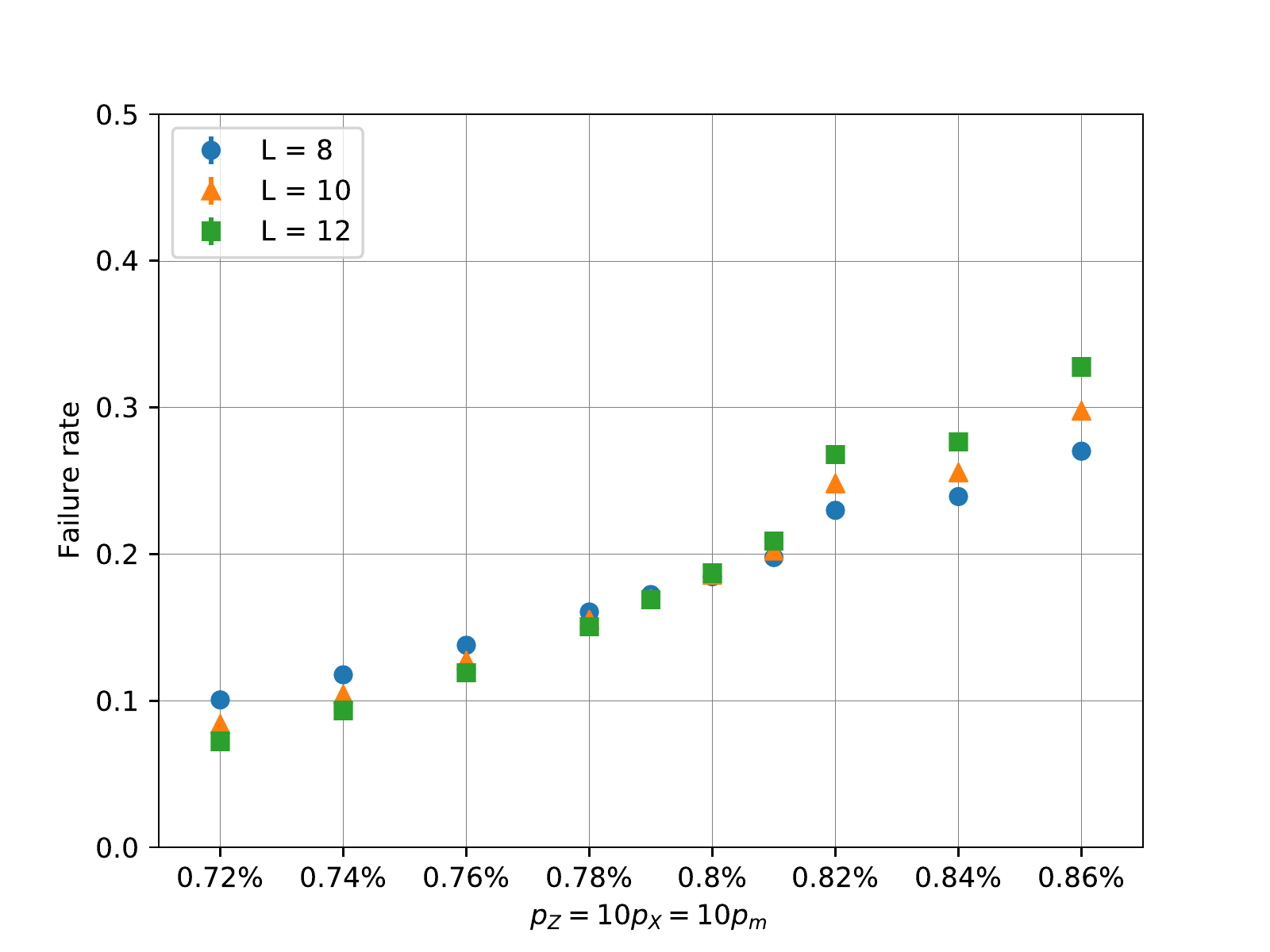}

\includegraphics[width=.49\textwidth]{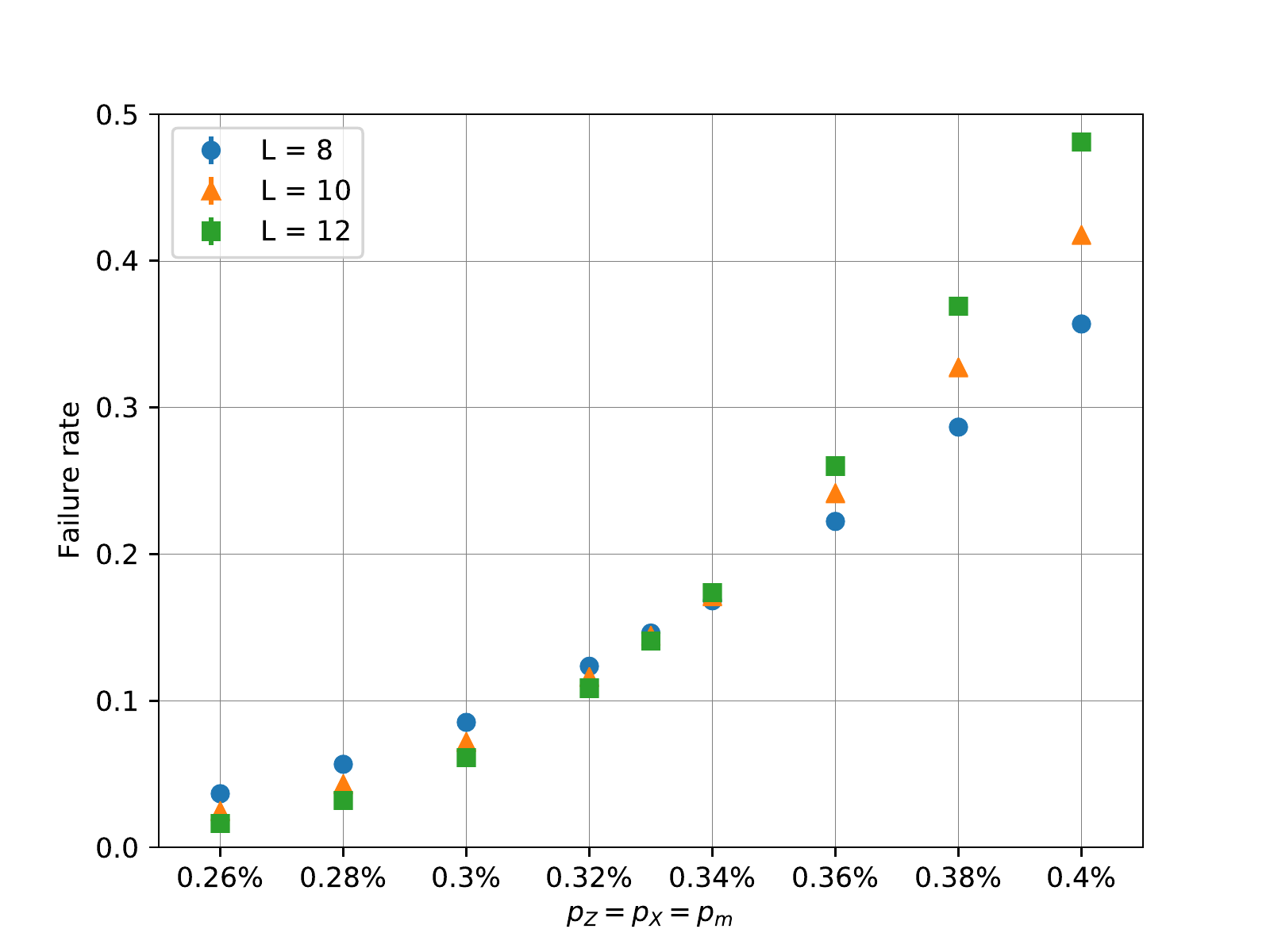}
\includegraphics[width=.49\textwidth]{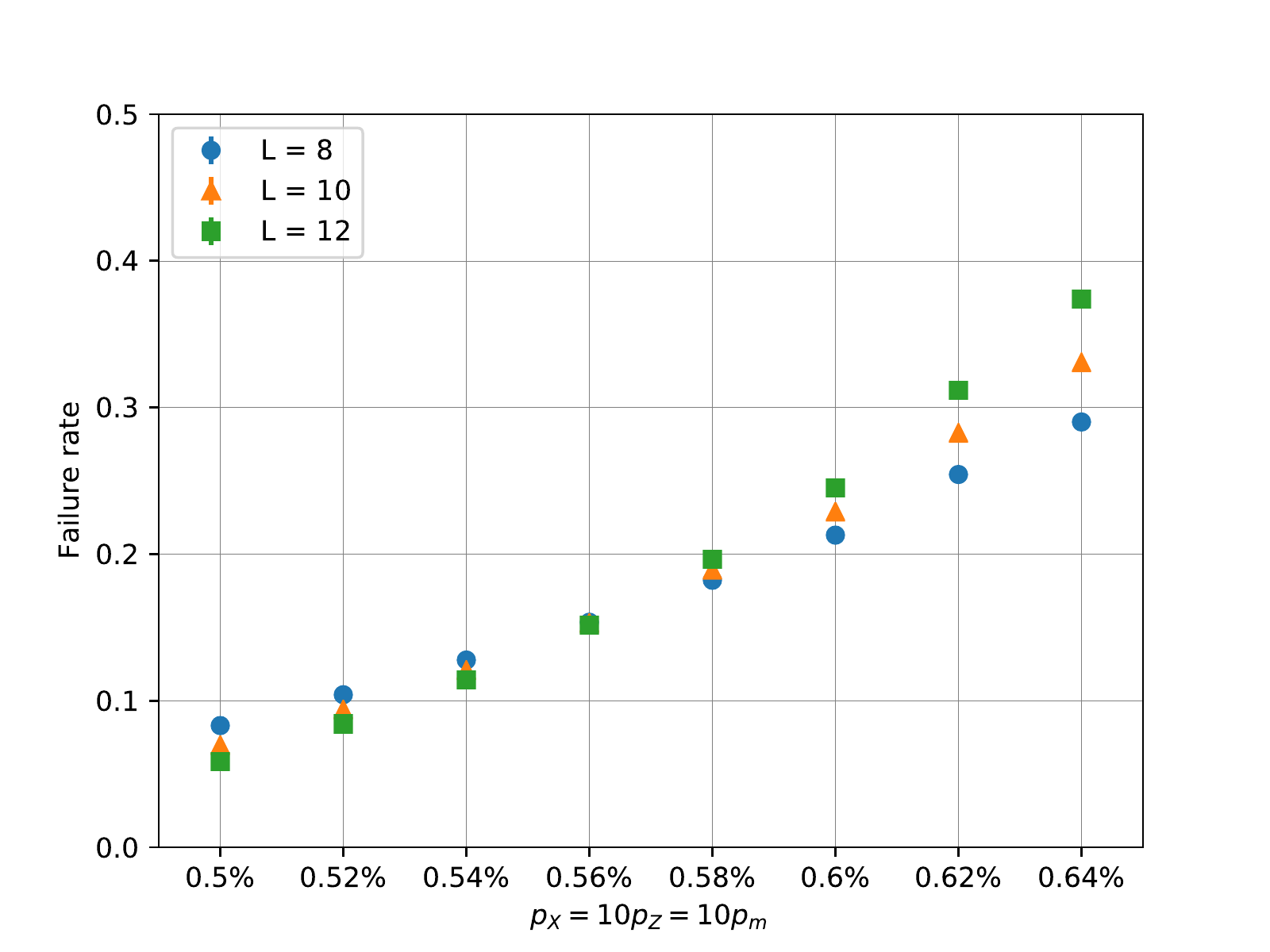}
\end{center}

\subsection*{DIA Lattice}

\begin{center}
\includegraphics[width=.49\textwidth]{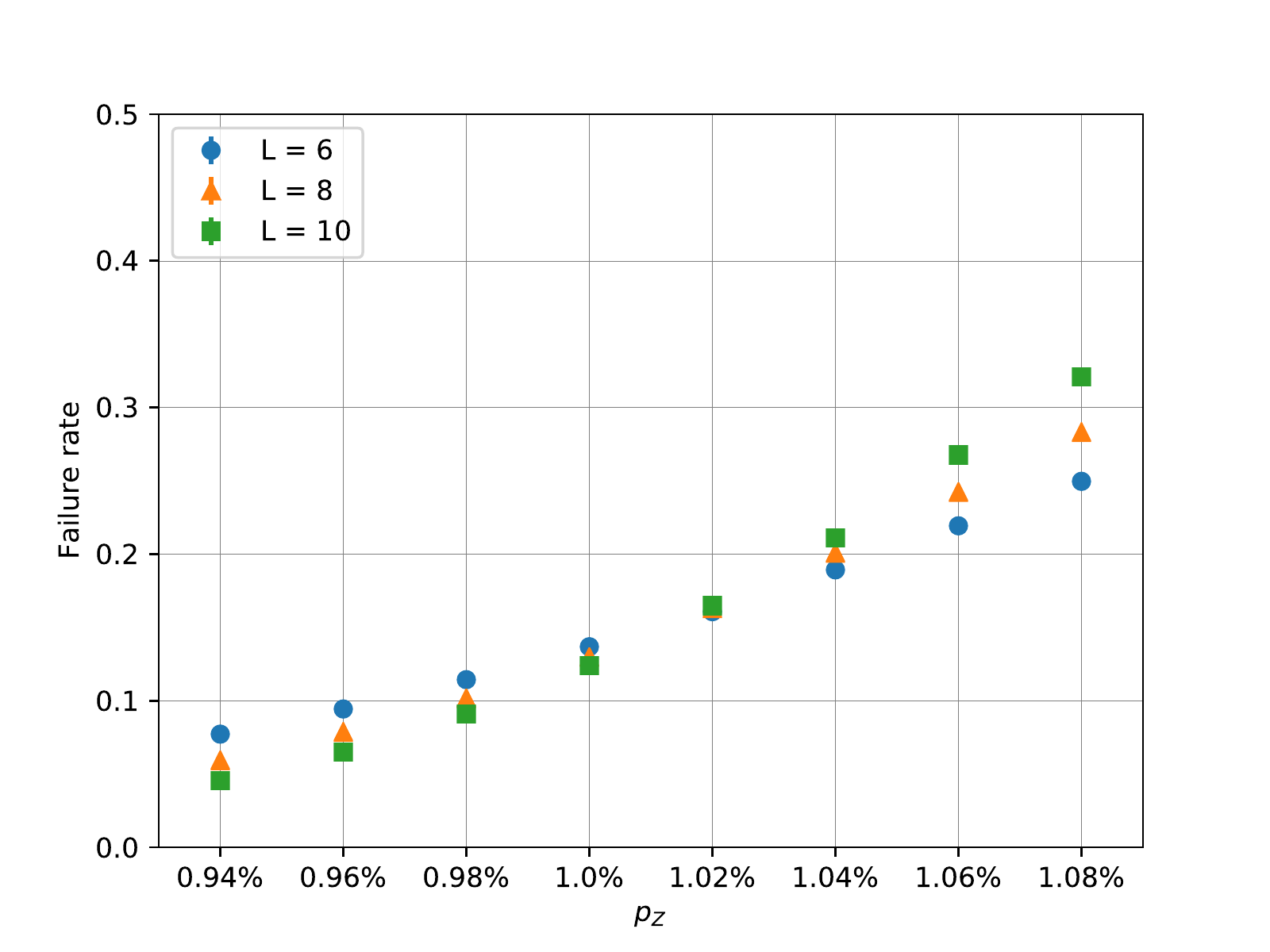}
\includegraphics[width=.49\textwidth]{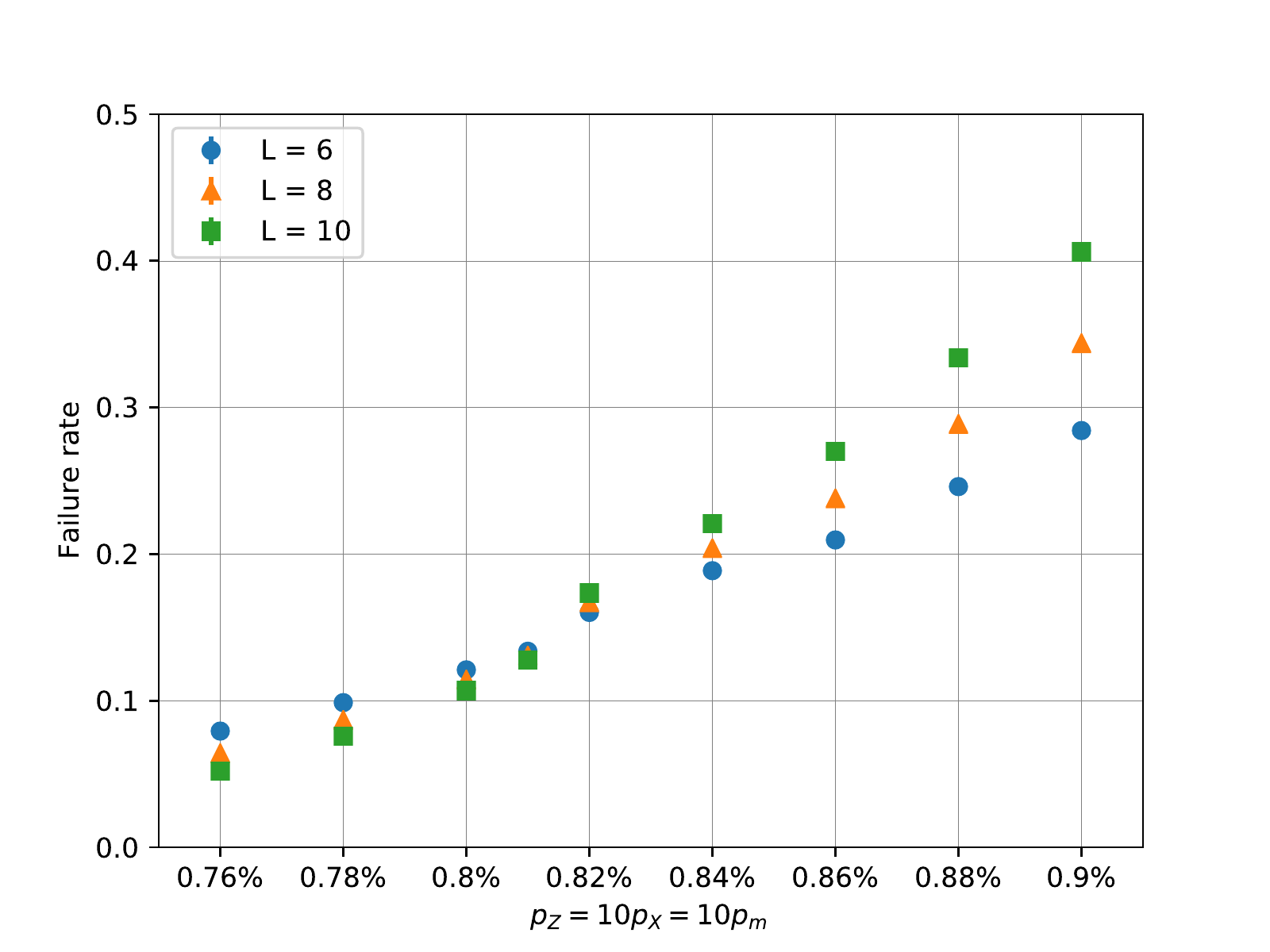}

\includegraphics[width=.49\textwidth]{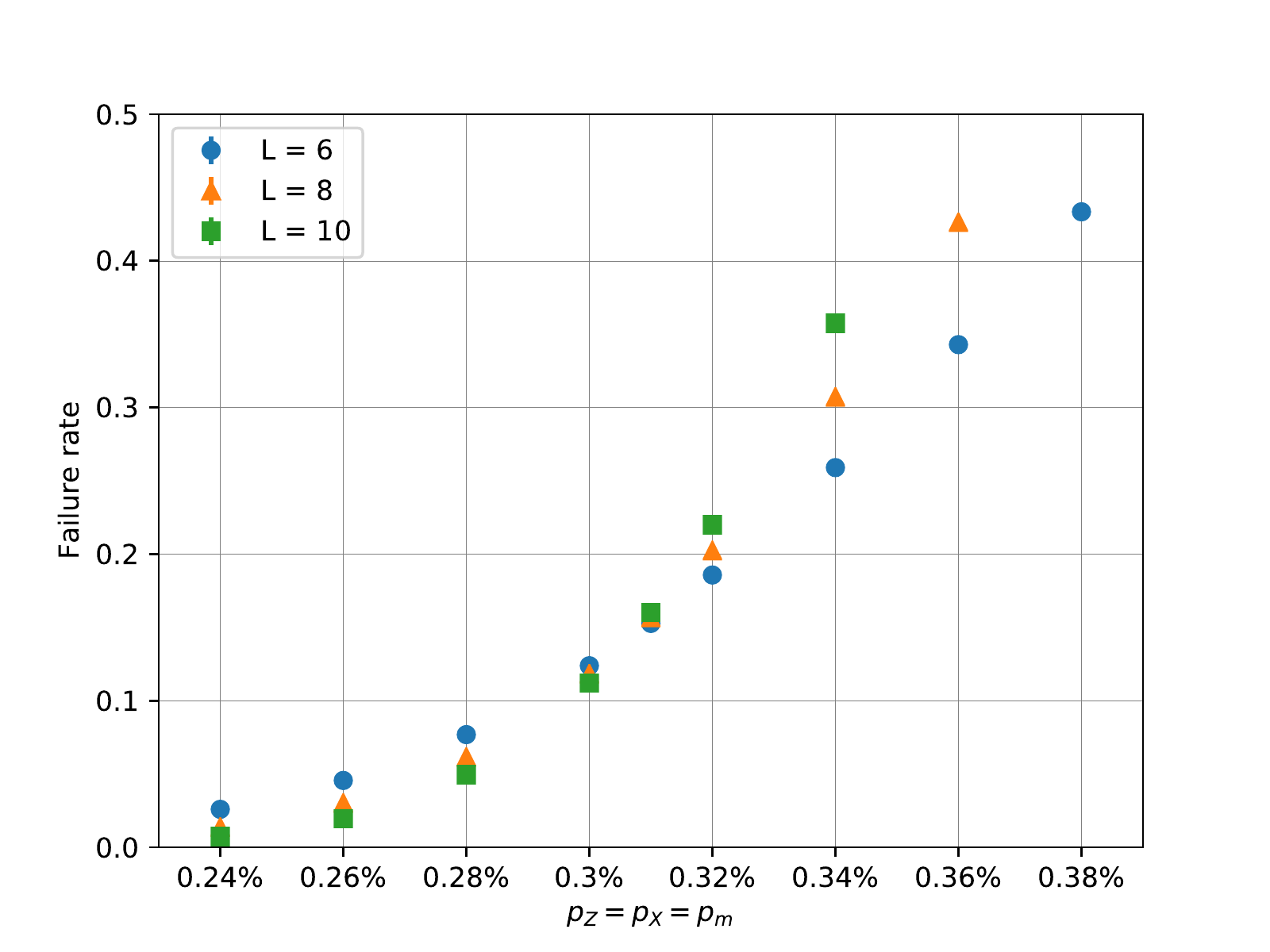}
\includegraphics[width=.49\textwidth]{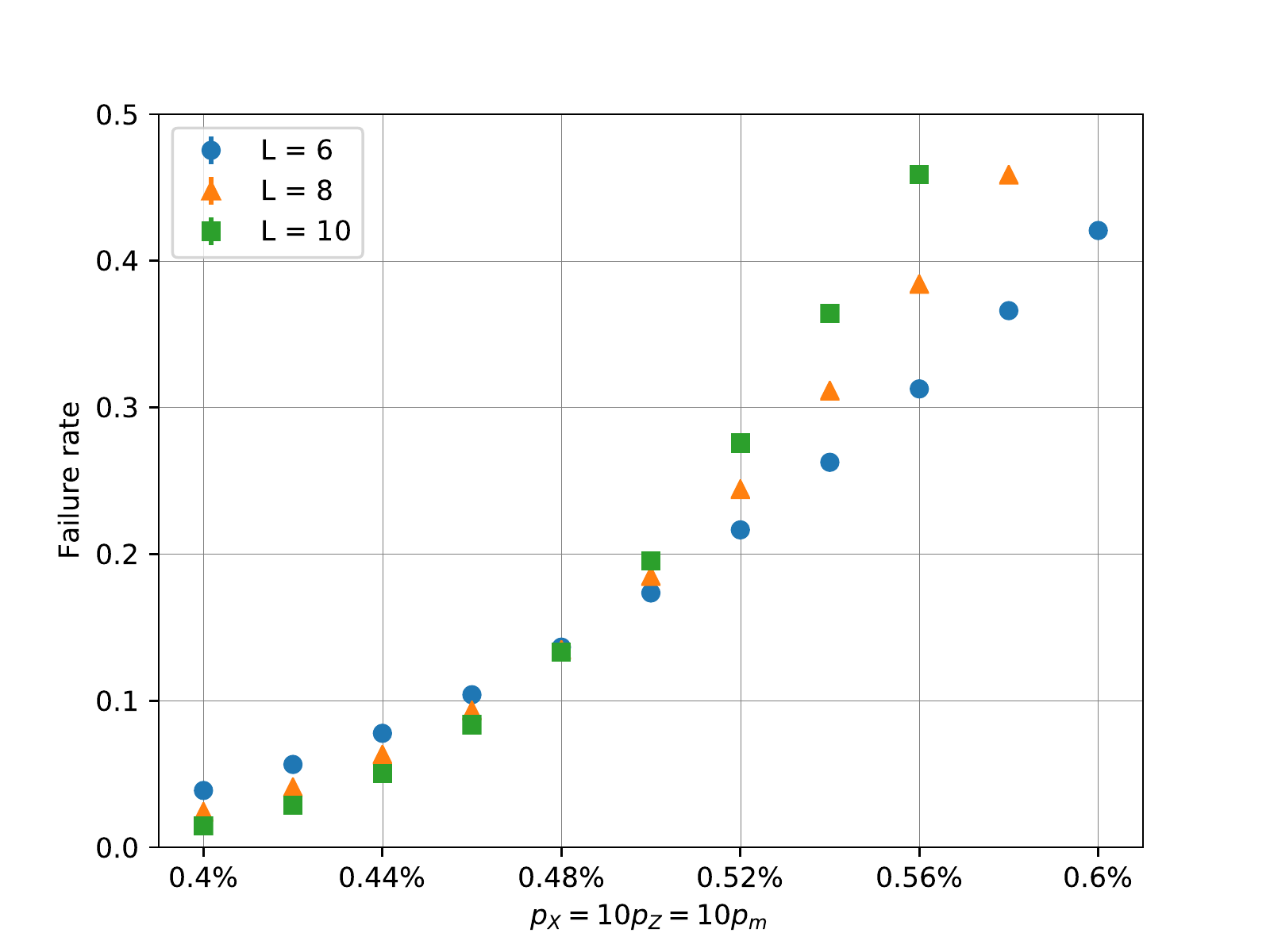}
\end{center}

\subsection*{CTN Lattice}

\begin{center}
\includegraphics[width=.49\textwidth]{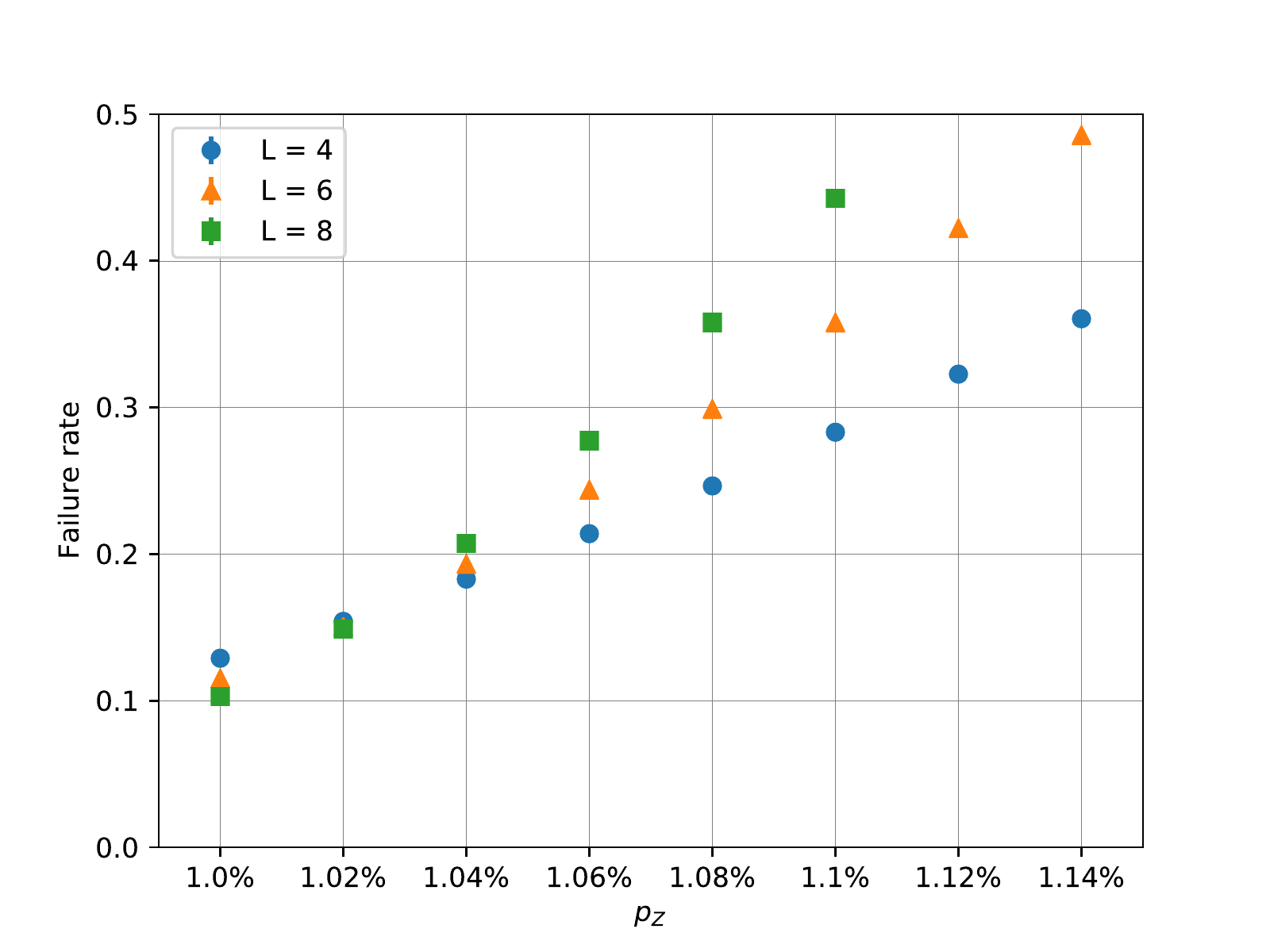}
\includegraphics[width=.49\textwidth]{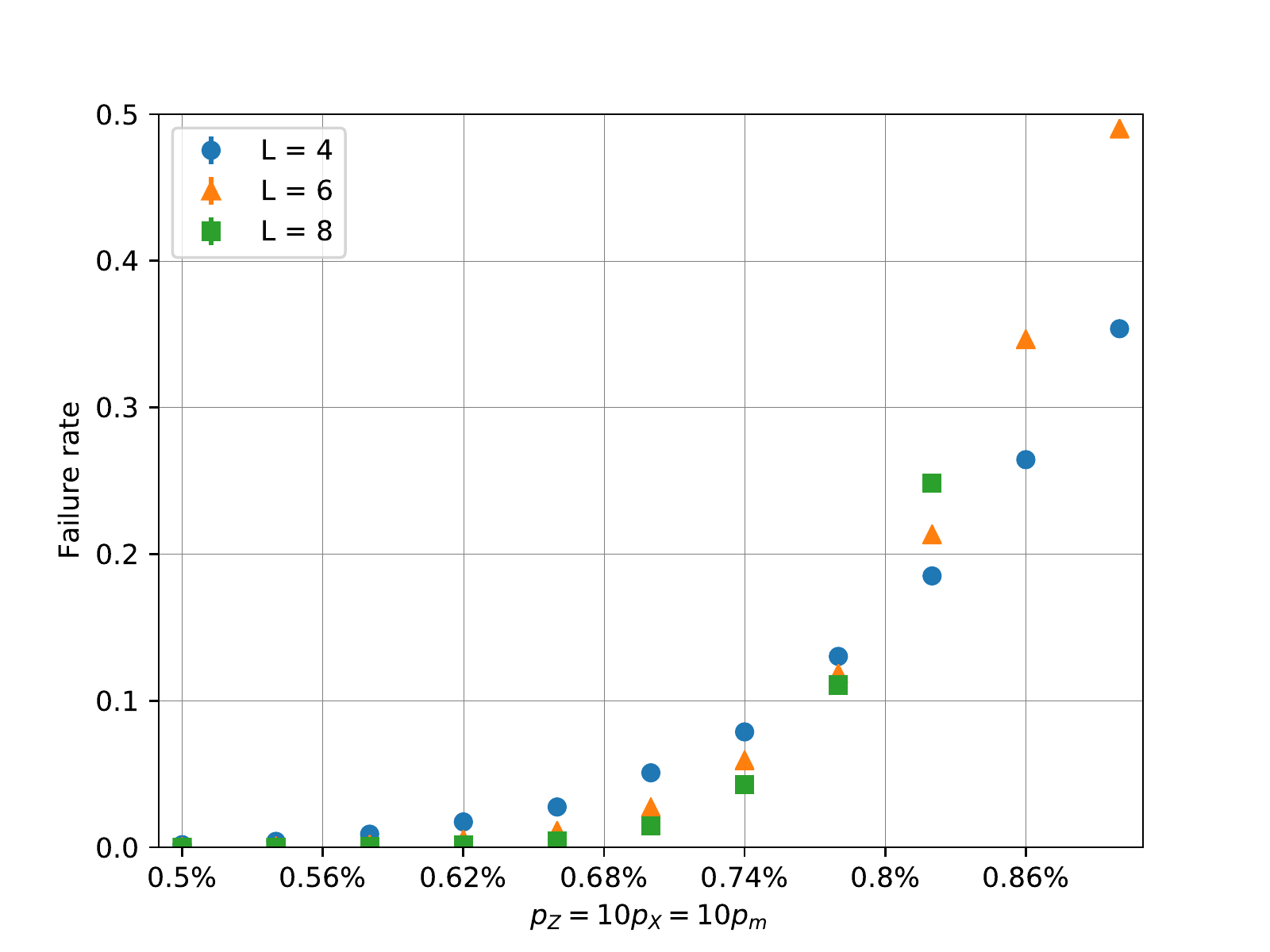}

\includegraphics[width=.49\textwidth]{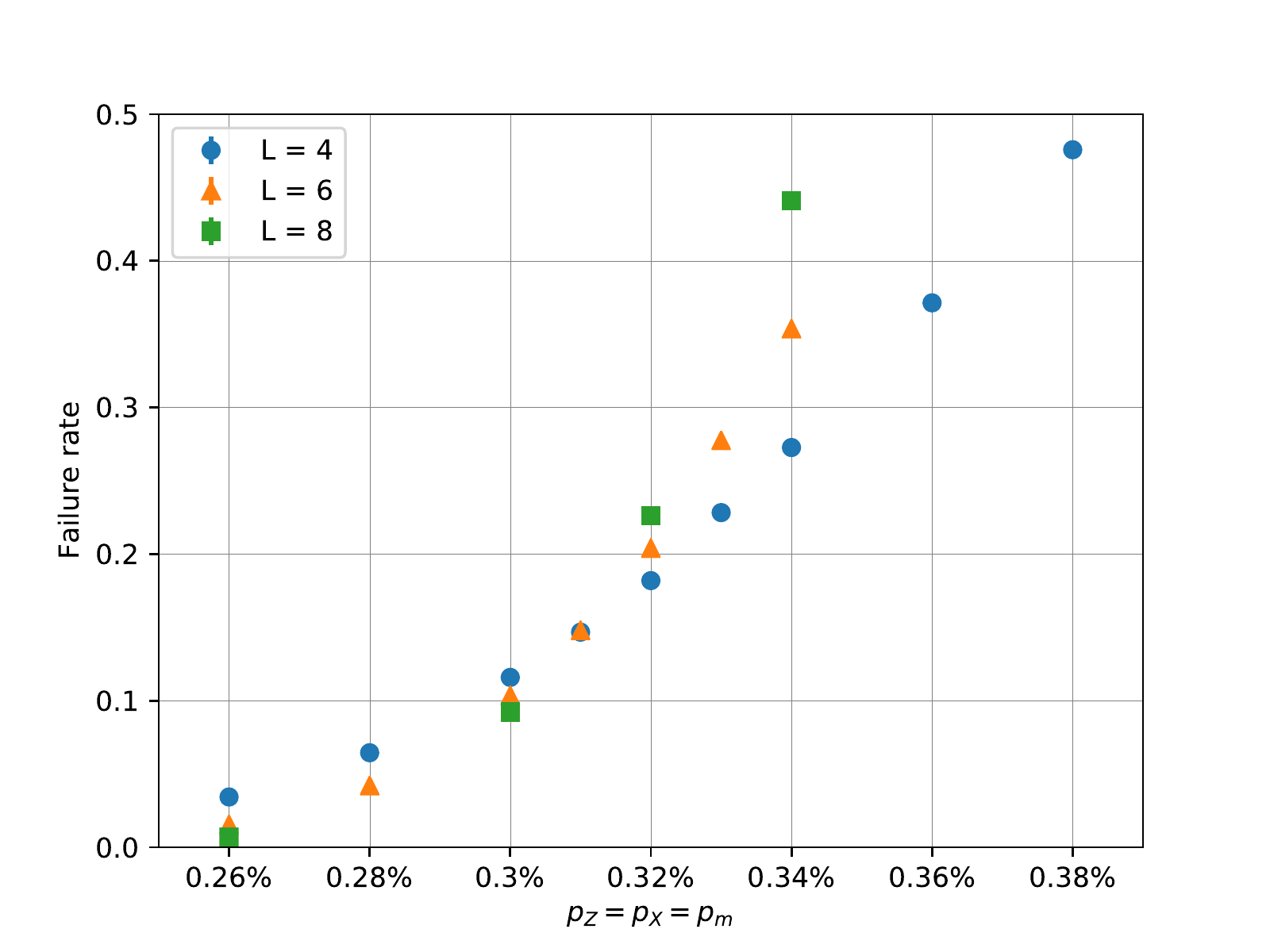}
\includegraphics[width=.49\textwidth]{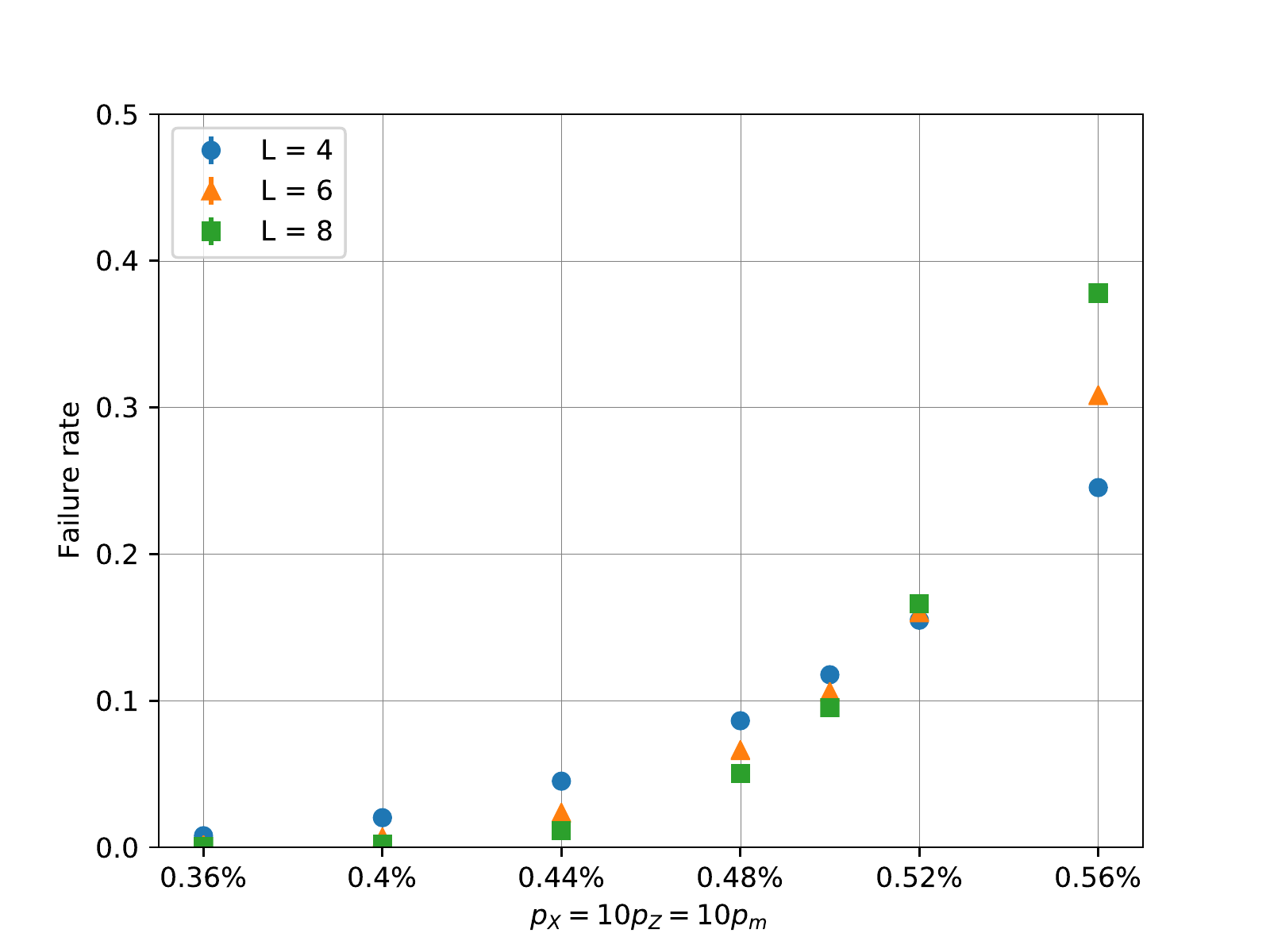}
\end{center}

\subsection*{SRS Lattice}

\begin{center}
\includegraphics[width=.49\textwidth]{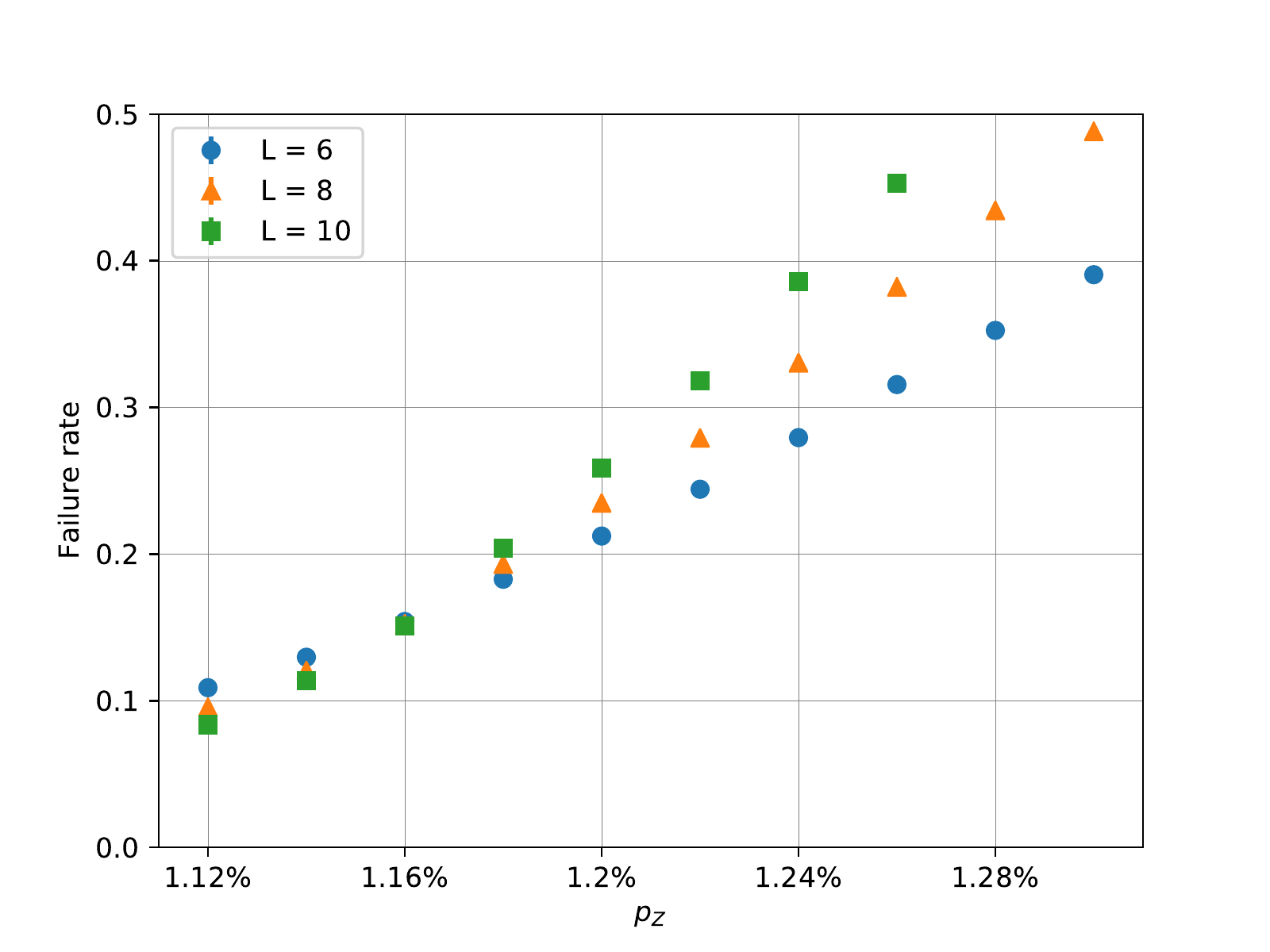}
\includegraphics[width=.49\textwidth]{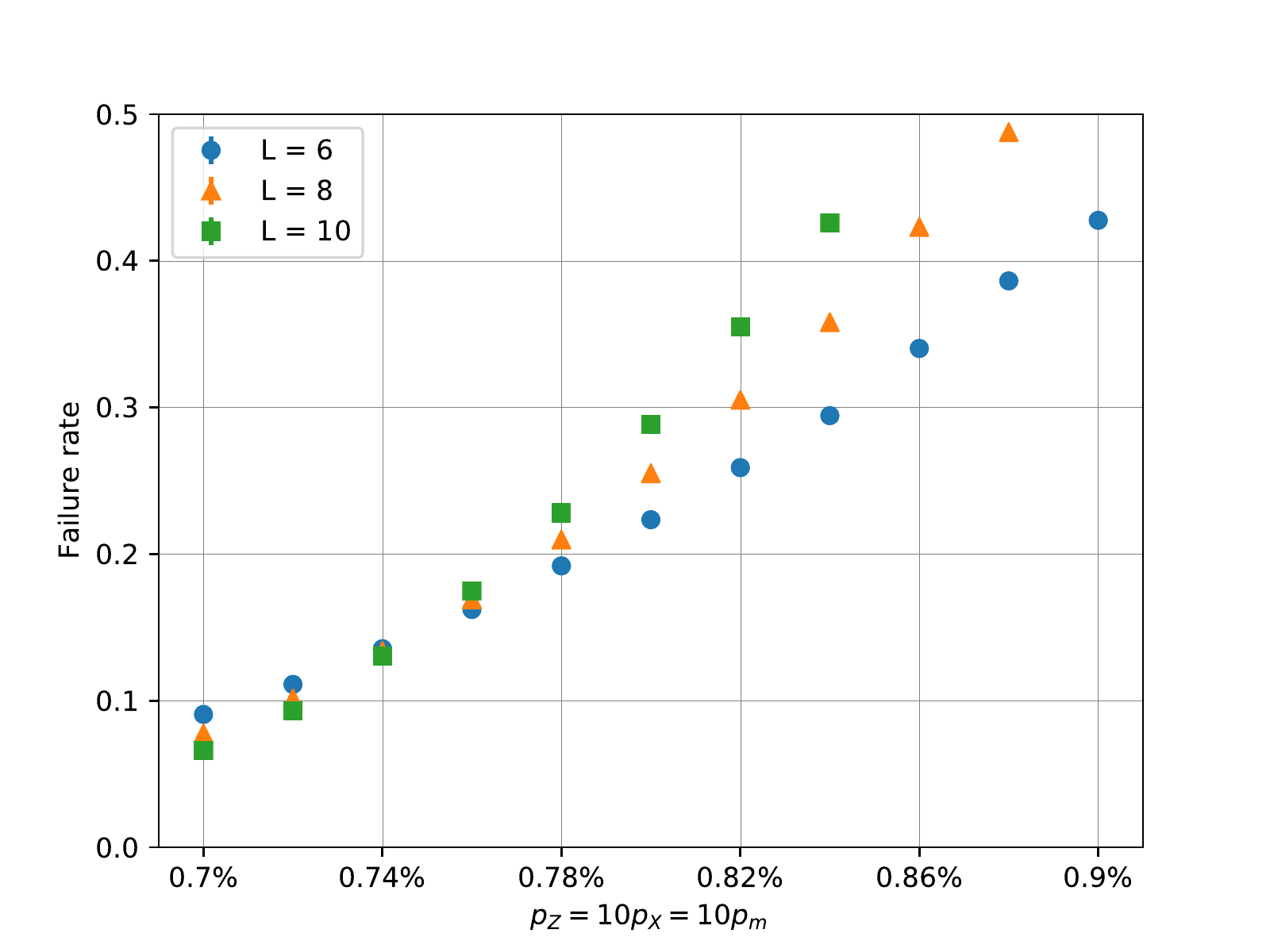}

\includegraphics[width=.49\textwidth]{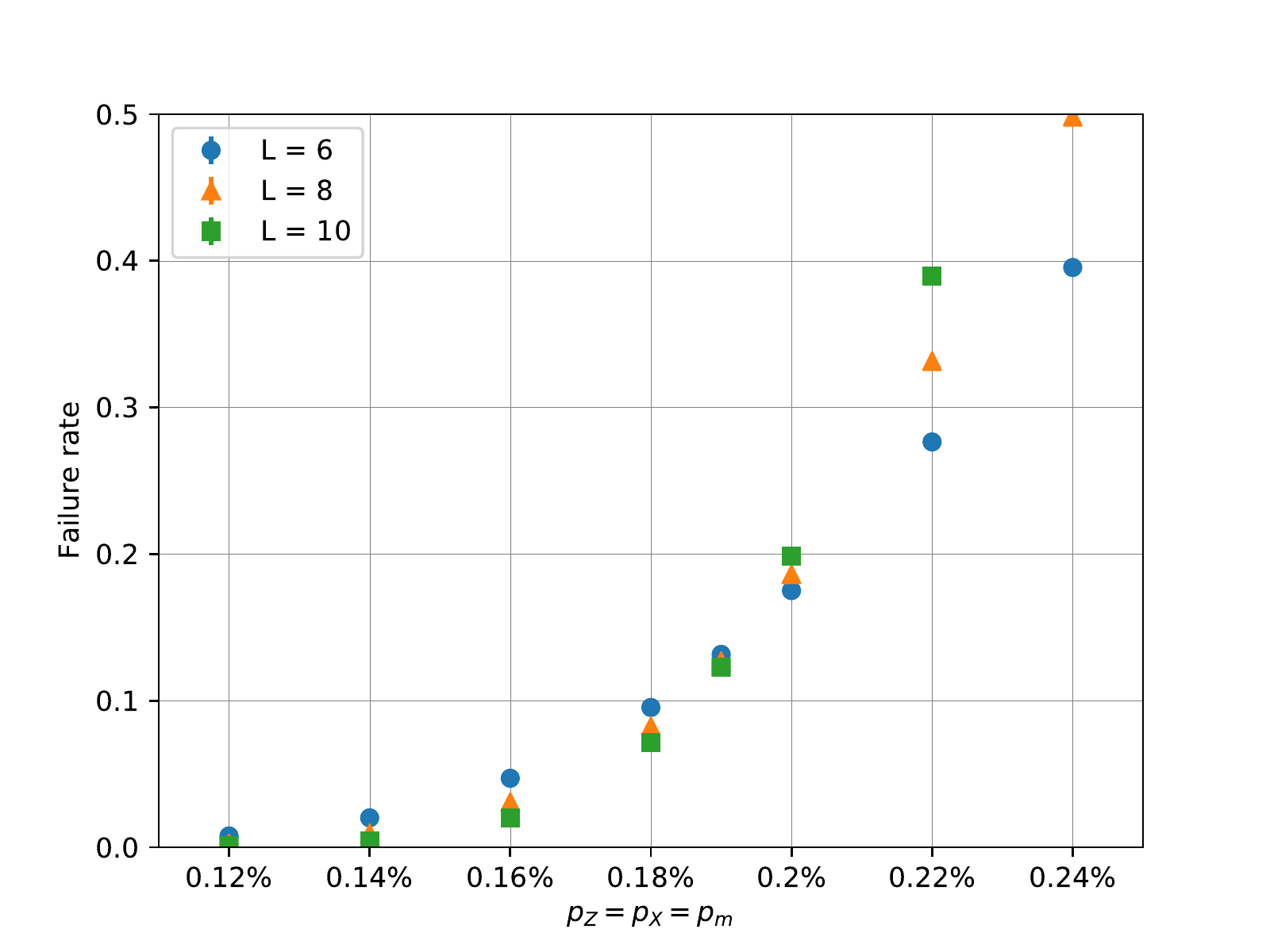}
\includegraphics[width=.49\textwidth]{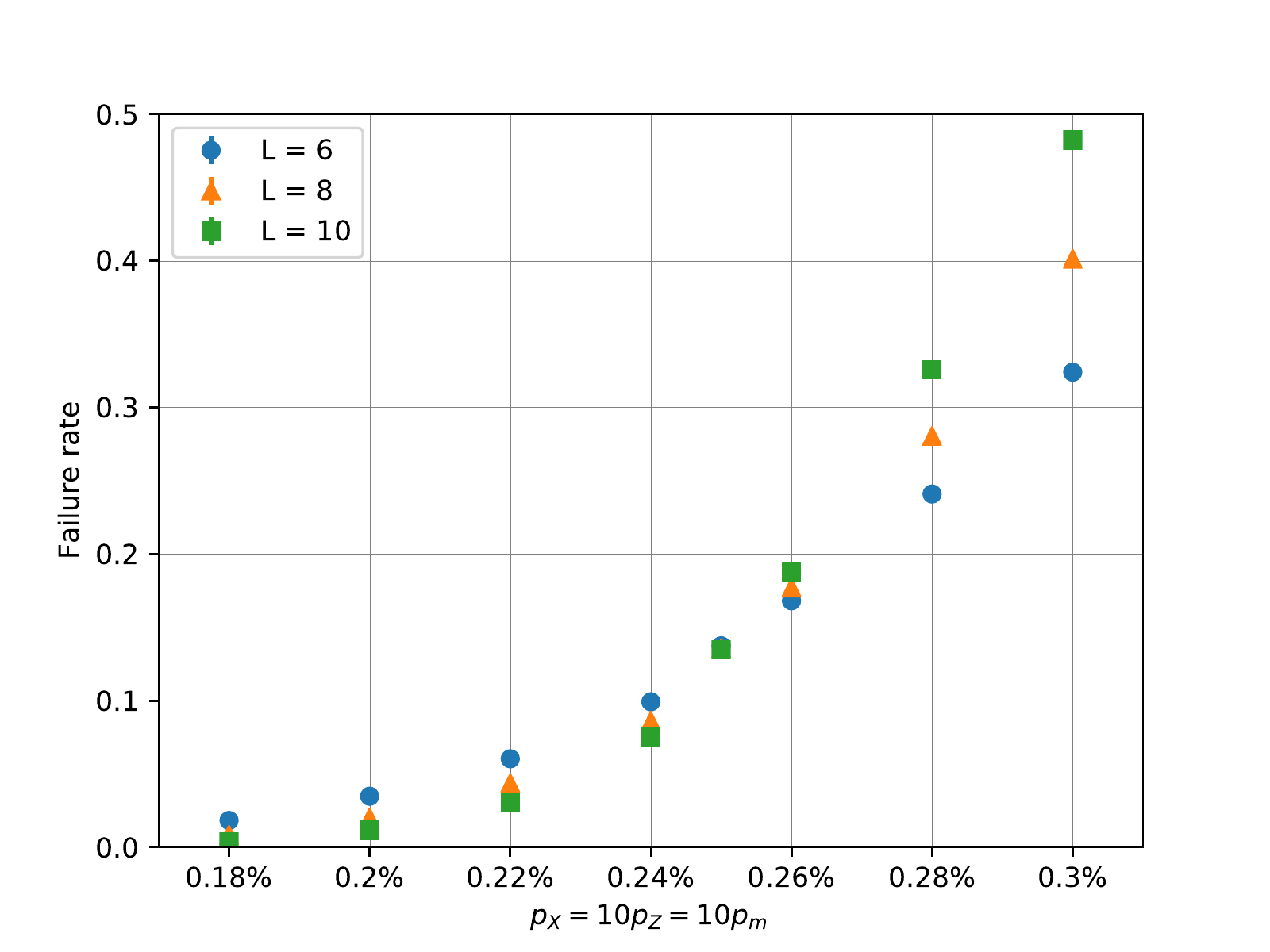}
\end{center}

\end{document}